\documentclass[11pt]{article} % use larger type; default would be 10pt

\usepackage[utf8]{inputenc} % set input encoding (not needed with XeLaTeX)

\usepackage{geometry} % to change the page dimensions
\usepackage{graphicx} % support the \includegraphics command and options

\usepackage{booktabs} % for much better looking tables
\usepackage{array} % for better arrays (eg matrices) in maths
\usepackage{paralist} % very flexible & customisable lists (eg. enumerate/itemize, etc.)
\usepackage{verbatim} % adds environment for commenting out blocks of text & for better verbatim
\usepackage{subfig} % make it possible to include more than one captioned figure/table in a single float
\usepackage{mathrsfs}
\usepackage{fancyhdr} % This should be set AFTER setting up the page geometry
\usepackage{sectsty}
\allsectionsfont{\sffamily\mdseries\upshape} % (See the fntguide.pdf for font help)
\usepackage[nottoc,notlof,notlot]{tocbibind} % Put the bibliography in the ToC
\usepackage[titles,subfigure]{tocloft} % Alter the style of the Table of Contents
\usepackage{amssymb} %maths
\usepackage{amsmath} %maths
\usepackage{mathrsfs}

\title{On the Partial-Wave Analysis of Mesonic Resonances\\Decaying to Multiparticle Final States Produced \\ by Polarized Photons}

\author{Carlos W. Salgado$^{1,2}$ and Dennis P. Weygand$^{2}$\\
{\em  Norfolk State University$^{1}$} \\ {\em and} \\
{\em The Thomas Jefferson National Accelerator Facility$^{2}$}} 
\begin{document}
\maketitle
 
 \abstract{Meson spectroscopy is going through a revival with the advent of high statistics experiments and new advances in the theoretical predictions.  The Constituent Quark Model (CQM) is finally being expanded considering more basic principles of field theory and using discrete calculations of Quantum Chromodynamics (lattice QCD). These new calculations are approaching predictive power for the spectrum of hadronic resonances  and decay modes. It will be the task of the new experiments to extract the meson spectrum from the data and compare with those predictions. The goal of this report is to describe one particular technique for extracting resonance information from multiparticle final states. The technique described here, partial wave analysis based on the helicity formalism, has been used at Brookhaven National Laboratory (BNL) using pion beams, and Jefferson Laboratory (JLab) using photon beams. In particular this report broadens this technique to include production experiments using linearly polarized real photons or quasi-real photons. This article is of a didactical nature. We describe the process of analysis, detailing assumptions and formalisms, and is directed towards people interested in starting partial wave analysis.

\tableofcontents 

\section{Introduction}

The field theory of strong interactions, Quantum Chromodynamics (QCD), allows for additional states outside the constituent quark model (CQM) in which the gluon fields can manifest externally ({\it hybrid states})~\cite{CMeyerRe}. The discovery of hybrid hadrons will provide an important test of QCD.  The signature hybrid states are {\it exotic} mesons with quantum numbers which cannot be attained by regular CQM mesons. The exotic quantum numbers prevent the mixing of these states with the conventional mesons, thus simplifying the identification of those states.  The dominant decay modes of these states were predicted to be through $S$ or $P$ waves, such as in the $b_{1}(1235) \pi$, $f_{1}(1285) \pi$ or the $\rho \pi$ decay channels~\cite{Philip,Andrei}. The lowest lying exotic are then $I^{G}J^{PC} = 1^{-} 1^{-+}$ states.
There has been evidence for $\pi_{1}(1400)$ and  $\pi_{1}(1600)$ states in pion production by the Brookhaven E852 experiment~\cite{AdamsEx,SUCEX}, and recently more evidence for the $\pi_{1}(1600)$ by the COMPASS collaboration at CERN~\cite{Nerling}.  In contrast, there are scarce data on photoproduction from old experiments~\cite{Condo,Aston,Einsenberg} and a negative result from one modern photoproduction experiment, the CLAS experiment at Jefferson Lab~\cite{Nozar}. There are various discussions in the literature as to why photo-production might be a better production mechanism for exotic mesons~\cite{Isgur1,AdamS}, but those claims still need to be confirmed experimentally. Two experiments using linearly polarized photons are planned for the upgraded Jefferson Lab accelerator at 12 GeV~\cite{CLAS12,Gluex}. They will provide high statistics and should bring greater insight into the spectrum of mesons.

In this report, we consider the analysis of multiparticle final states produced by linearly polarized real photons or quasi-real photons from electron inelastic scattering at very forward angles (low $Q^{2}$). Our goal is to identify possible short lived (strongly interacting) resonances that have decayed to the observed multiparticle final states. We are interested in mesonic resonances (B (baryon number) = 0). By "identification of a resonance" we mean: to identify an enhancement in the cross section (as well as the complex production amplitude) and to determine the quantum numbers of the resonance. The schematics of the considered reactions is shown in figure~\ref{fig:decay1}. We consider diffractive reactions , i.e. dominated by t-channel exchange, where the beam interacts peripherally with the target~\cite{Halzen}. We also regard the photon in the vector dominance model (VMD), i.e. the hadronic interactions of the photons are explained through a decomposition of the photon into vector mesons ($\rho$, $\omega$, $\phi, \cdots$) \cite{Donnachie}.

\begin{figure}[!htp]
\centerline{\includegraphics[width=12cm]{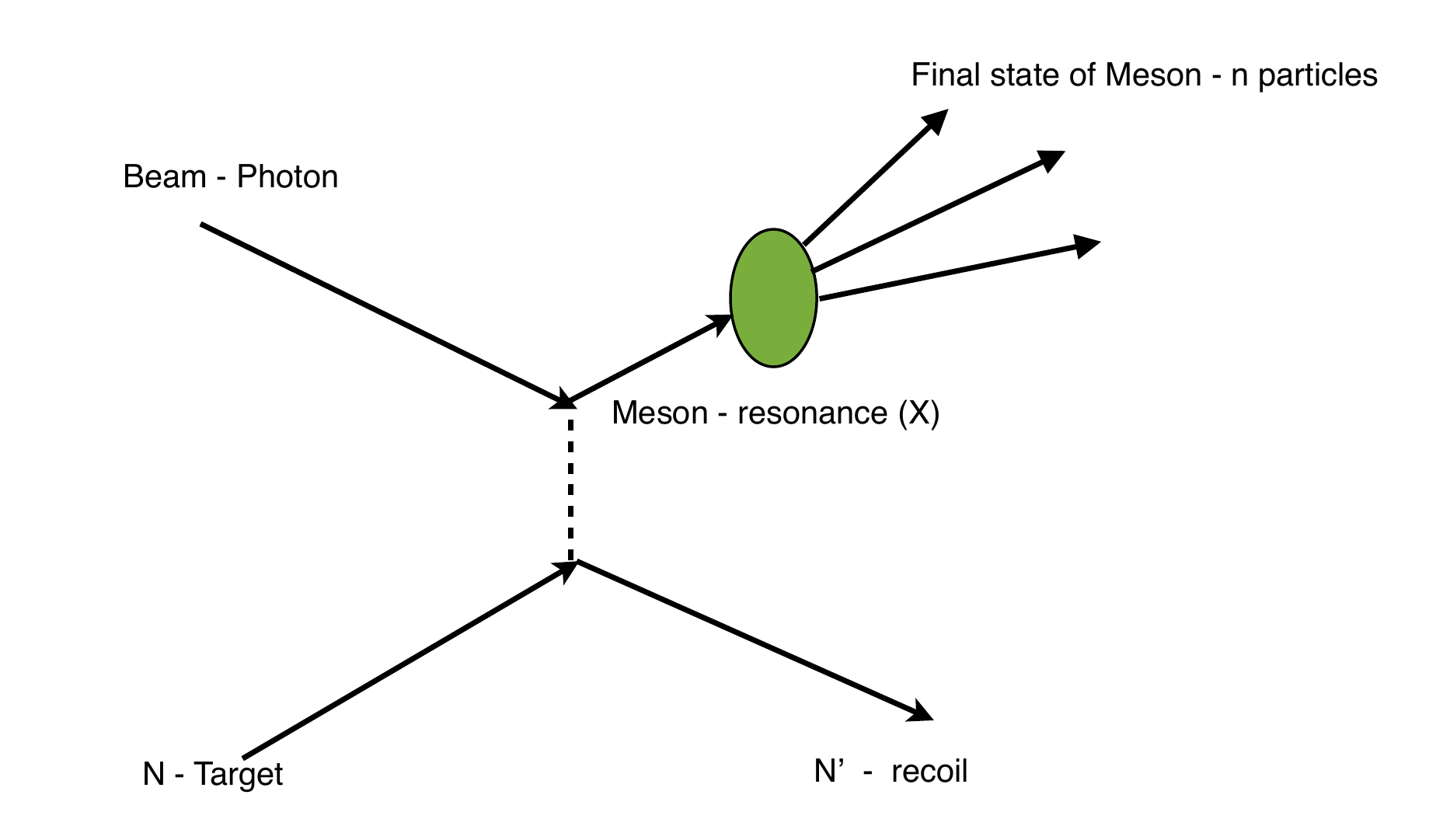}}
\caption{\label{fig:decay1}Photoproduction and decay of a mesonic resonance.}
\end{figure}

In our model, resonant states are produced by the strong interaction of the photon with the nucleon via the exchange of mesonic quantum numbers, which may be, for example, a Regge  trajectory or two or more gluons (Pomeron). The resonance's quantum numbers to be determined are: $J$ (spin), $P$ (parity), $C$ (charge conjugation), $I$ (isospin) and $G$ ($G$-parity) of the resonance~\cite{Close}. These quantum numbers are directly related to the angular distributions of the resonance's decay products. Therefore, the experimental challenge is to observe the final state particles with good acceptance, and measure their momenta with good resolution. We are also interested in measuring the resonance's production mechanisms and cross sections. However, even in a perfect $4\pi$ detector (100\% acceptance and resolution), it might be difficult to identify all final state particles coming directly from the decay of the resonance. For example, they might be coming from the breaking of the nucleon (decays of baryon resonances) or produced by dynamic effects related to radiation from the exchanged particles (Deck effect) \cite{Deck}. These effects present serious challenges to this type of analysis. 

The analysis technique described here is performed in two steps. The first is based on fitting the data to a model of the reaction that includes {\it all} possible quantum states (so-called "partial waves") decaying to the observed multiparticle final state in a restricted range of the kinematic phase-space  where cross sections are assumed to remain unchanged, therefore maintaining only angular dependences. That is, the fit is independent of the mass of the hadronic resonance ({\it mass-independent fit}). However, since it is impossible to include {\it all }possible partial waves, as in principle there may be an infinite number, several assumptions have to be made. Furthermore, different partial waves may contribute to the same final angular distribution, producing ambiguities in the solution. Therefore, we will need to choose a set of partial waves to fit the data based on physical and/or practical criteria (symmetries in a given reaction, computer time, etc). Furthermore, in the case of more than two final state particles, we will need to make important assumptions in the way the decays of the resonance are realized. In that case, we use the "isobar model" that assumes a two-body sequential decay series of the resonance.  The second step, after a dynamic distribution of intensities has been generated from the fitted values obtained in the first step (usually in mass but more generally in two variables, i.e. mass and t-Mandelstam), a fit is made to this distribution to identify resonances and extract resonance's properties (mass-dependent fit). This fit can include amplitude's phase motions. This experimental technique, based in a model that includes many physical principles, has been proved to allow the identification of many resonances. We will discuss in this report the assumptions use in the formalism and the details of its implementation in the partial wave analysis method.

Many of the principles discussed in this article, were introduced in the BNL-E852 analysis~\cite{SUC96}, and carried out by the photoproduction experiments at Jefferson Lab-CLAS~\cite{Nozar}. 

\section{Extended Likelihood Fit}
\label{sect:like}

The fit of a model to the data plays a central role in our analysis. There are several ways to obtain the best parametric fit to a set of data, and several ways to evaluate  their performance (goodness of the fit - see section 7). We will use the extended likelihood method \cite{Barlow, Orear}. The vector 
$\overrightarrow{x}$ represents the set of variables necessary to define the particular configuration of an event, and is of dimension $n$. We have measured $N$ events, each given a set of measurements represented by a vector: $\overrightarrow{x}_{i}$, where $i$ spans the set of events, i.e. $i=1,...,N$.  

Our goal will be to find a mathematical parametrization (model) that explains these observations, i.e, that is able to explain (or predict) the number of observed events for each bin. In general a model to be described by $m$ parameters, \newline \mbox{$\{a_{1},a_{2},\cdots,a_{m}\} = \overrightarrow{a}$}. We want to adjust the parameters in our model until we can best reproduce the observed data (fit). The probability of obtaining an event with the set $\overrightarrow{x}_{i}$ in our model is called $p(\overrightarrow{x}_{i},\overrightarrow{a})$.

The standard likelihood of obtaining this arrangement for $N$ measurements is the joint probability density
\begin{equation}
\mathscr{ L}= \prod_{i=1}^{N}  p(\overrightarrow{x}_{i},\overrightarrow{a})
\label{eqn:Like1}
\end{equation}
with the normalization
\begin{equation}
 \int_{\Omega} p(\overrightarrow{x},\overrightarrow{a})d^{n}\overrightarrow{x}=1
\end{equation}
where $\Omega$ represents the full phase-space.

The relaxation of this last requirement is what is called the {\it extended likelihood}. We replace $p(\overrightarrow{x},\overrightarrow{a})$ by a new function $ \mathbb{P}(\overrightarrow{x},\overrightarrow{a})$ such that

\begin{equation}
\mathbb{P}(\overrightarrow{x},\overrightarrow{a})=\mathscr{N}p(\overrightarrow{x},\overrightarrow{a})
\label{eqn:Pdef}
\end{equation}
therefore

\begin{equation}
\int_{\Omega} \mathbb{P}(\overrightarrow{x},\overrightarrow{a})d^{n}\overrightarrow{x}=\mathscr{N}.
\label{eqn:normal}
\end{equation}

The normalization $\mathscr{N}$ represents {\it the expected number of events to be observed
in the full phase-space}.

We define a new extended likelihood that will also include the probability of observing $N$ events by

\begin{equation}
\mathscr{ L}= Prob(N) \prod_{i=1}^{N}  p(\overrightarrow{x}_{i},\overrightarrow{a}). \end{equation}

Assuming a Poisson distribution for the probability of observing N events, with an expected value of $\mathscr{N}$

\begin{equation} Prob(N)=\frac{\mathscr{N}^{N}}{N!}e^{-\mathscr{N}}  \end{equation}
the extended likelihood is then

\begin{equation} \mathscr{L}=\begin{bmatrix}\frac{\mathscr{N}^{N}}{N!}e^{-\mathscr{N}}\end{bmatrix} \prod_{i=1}^{N}  p(\overrightarrow{x}_{i},\overrightarrow{a})  \end{equation}
or
\begin{equation} \mathscr{L}=\begin{bmatrix}\frac{\mathscr{N}^{N}}{N!}e^{-\mathscr{N}}\end{bmatrix} \prod_{i=1}^{N}  \frac{\mathbb{P}(\overrightarrow{x}_{i},\overrightarrow{a})}{\mathscr{N}}.  \end{equation}
Therefore
\begin{equation} \mathscr{L}=\begin{bmatrix}\frac{1}{N!}e^{-\mathscr{N}}\end{bmatrix} \prod_{i=1}^{N}   \mathbb{P}(\overrightarrow{x}_{i},\overrightarrow{a})  \end{equation}
and taking the log on both sides

\begin{equation} ln\mathscr{L}=-ln\begin{bmatrix}N!\end{bmatrix}-\mathscr{N}+\sum_{i=1}^{N}  ln\begin{bmatrix}\mathbb{P}(\overrightarrow{x}_{i},\overrightarrow{a})\end{bmatrix}.\end{equation}

Then, substituting equation~(\ref{eqn:normal}) and removing the constant term

\begin{equation} ln\mathscr{L} \propto -\int_{\Omega} \mathbb{P}(\overrightarrow{x},\overrightarrow{a})d^{n}\overrightarrow{x}+\sum_{i=1}^{N}  ln\begin{bmatrix}\mathbb{P}(\overrightarrow{x}_{i},\overrightarrow{a})\end{bmatrix}. \end{equation}

or
\begin{equation} \boxed{ ln\mathscr{L} \propto \sum_{i=1}^{N}  ln\begin{bmatrix} \mathbb{P}(\overrightarrow{x}_{i},\overrightarrow{a}) \end{bmatrix} -\mathscr{N} }
\label{eqn:like}
\end{equation} }

We will find the best parameters $\overrightarrow{a}$ for our model, maximizing the extended likelihood or equivalently minimizing the function $-ln\mathscr{L}$.
We will describe in section~\ref{sect:pwaFit} details of how we will calculate and solve this optimization problem.

The errors in the parameters are given by the square root of their variances. Let's call $a^{*}_{i}$ the fitted parameters, i.e. the values that make the function $-ln\mathscr{L}$ a minimum and find an expression for the errors \cite{Orear, Eadie}. The variances are
\begin{equation} \sigma^{2}_{ij}=E[(a_{i}-a^{*}_{i})(a_{j}-a^{*}_{j})] \end{equation}
where we also consider the correlated errors.
Let's call $w(a_{i}) \equiv  -ln\mathscr{L}$, and make a Taylor expansion around the minimum

\begin{equation}
w(a_{i}) = w(a^{*}_{i}) + \sum^{n}_{i}
\begin{vmatrix} \frac{\partial w}{\partial a_{i}}
\end{vmatrix}_{a^{*}_{i}} + \frac{1}{2} \sum^{n}_{a_{i}} \sum^{n}_{a_{j}} \mathscr{H}_{ij} \beta_{a_{i}} \beta_{a_{j}}\cdots\end{equation}
where $\beta_{a_{i}} \equiv  (a_{i}-a^{*}_{i}) $ and
\begin{equation}
\mathscr{H}_{ij} \equiv \begin{vmatrix} {\partial^{2} w \over \partial a_{i} \partial a_{j}} \end{vmatrix}_{a^{*}_{i}}
\end{equation}
is the Hessian matrix of second order partial derivatives of the negative natural logarithm of the Likelihood function respect to the parameters.

Since:
\begin{equation} \begin{vmatrix} \frac{\partial w}{\partial a_{i}} \end{vmatrix}_{a^{*}_{i}} = 0 \end{equation}
to a second order
\begin{equation} -ln\mathscr{L}(a_{i}) = w(a^{*}_{i}) + \frac{1}{2} \sum^{n}_{a_{i}} \sum^{n}_{a_{j}} \mathscr{H}_{ij} \beta_{a_{i}} \beta_{a_{j}} \end{equation}
or
\begin{equation} \mathscr{L}(a_{i}) = C e^{-\frac{1}{2} \sum^{n}_{a_{i}} \sum^{n}_{a_{j}} \mathscr{H}_{ij} \beta_{a_{i}} \beta_{a_{j}}} \end{equation}
that written in a vector-matrix notation is
\begin{equation} \mathscr{L}(\overrightarrow{a}) = C e^{-\frac{1}{2} \vec{\beta}^{T}_{\overrightarrow{a}}  \mathscr{H} \vec{\beta}_{\overrightarrow{a}}} \end{equation}
where $\overrightarrow{\beta}_{\overrightarrow{a}} = (\overrightarrow{a}-\overrightarrow{a^{*}}) $. This is the expression of a multivariate Normal Distribution \cite{Orear, Bevington}, where by normalization
\begin{equation} \int^{\infty}_{-\infty} \cdots \int^{\infty}_{-\infty} (\overrightarrow{a}-\overrightarrow{a^{*}}) e^{-\frac{1}{2} \beta^{T}_{\overrightarrow{a}}  \mathscr{H} \beta_{\overrightarrow{a}}} da_{i} \cdots da_{n} = 0. \end{equation}

Differentiating this expression we obtain
\begin{equation} \int^{\infty}_{-\infty}...\int^{\infty}_{-\infty} [I-(\overrightarrow{a}-\overrightarrow{a^{*}})^{T}(\overrightarrow{a}-\overrightarrow{a^{*}}) \mathscr{H}] e^{-\frac{1}{2} \beta^{T}_{\overrightarrow{a}}  \mathscr{H} \beta_{\overrightarrow{a}}} da_{i}...da_{n} = 0 \end{equation}
then, using that
\begin{equation} E[(\overrightarrow{a}-\overrightarrow{a^{*}})^{T}(\overrightarrow{a}-\overrightarrow{a^{*}})] = \int^{\infty}_{-\infty}...\int^{\infty}_{-\infty} (\overrightarrow{a}-\overrightarrow{a^{*}})^{T}(\overrightarrow{a}-\overrightarrow{a^{*}})  e^{-\frac{1}{2} \beta^{T}_{\overrightarrow{a}}  \mathscr{H} \beta_{\overrightarrow{a}}} da_{i}...da_{n} \end{equation}
we obtain
\begin{equation} I-E[(\overrightarrow{a}-\overrightarrow{a^{*}})^{T}(\overrightarrow{a}-\overrightarrow{a^{*}})] \mathscr{H} =0 \end{equation}
\begin{equation} E[(\overrightarrow{a}-\overrightarrow{a^{*}})^{T}(\overrightarrow{a}-\overrightarrow{a^{*}})] = \mathscr{H}^{-1} \end{equation}
therefore 

\begin{equation} \boxed{ \sigma^{2}_{ij} = E[(a_{i}-a^{*}_{i})(a_{j}-a^{*}_{j})] = E[(\overrightarrow{a}-\overrightarrow{a^{*}})^{T} (\overrightarrow{a}-\overrightarrow{a^{*}})] = [\mathscr{H}]^{-1}_{ij}. }
\label{eqn:error}
 \end{equation}

The errors of the parameters can be calculated from the inverse of the Hessian matrix evaluated at the minimum.  However, it should be noted that equation (\ref{eqn:error}) is true only if the truncated Taylor expansion is accurate, i.e. if the natural logarithm of the Likelihood can be approximated by a quadratic equation around the minimum. This approximation may be violated in PWA, and need to be checked (see section~\ref{sect:GoodFit}). The MINUIT package may use different ways of calculating errors for more general cases (i.e. using the MINUIT package MINOS, see references \cite{Eadie}, \cite{MINUIT} and \cite{MINUIT2} ).

\section{The Model }
\label{sect:Model}

It is not easy to give a general definition of hadronic resonance. From a QCD perspective, we are looking for conglomerates of quarks and gluons \cite{Close,Werle,Goldberger}. These states live for a very short time before decaying to other particles. An experimental view of a resonance is that of an enhancement in the cross section of a reaction, accompanied by an amplitude phase motion through $\pm {\pi \over 2}$ radians. Nonetheless, from the $S$-Matrix perspective, resonances are poles on the complex sheet where amplitudes are defined. In this view, it has been noted that not all peaks in the cross section are resonances, and not all resonances produce peaks in the cross sections~\cite{Goldberger,Collins}.  If resonances are {\it poles in the complex scattering matrix, that are non-observable entities}, they may not correspond to a peak in the cross section; then, how do we experimentally distinguish a resonance?  For our present search, we opt for a pragmatic definition: a resonance will be identified by an enhancement in the cross section (intensity - see section~\ref{sect:MassDep}) associated with appropriate phase motion (section~\ref{sect:Phase}). Both observations, considered together, will give us a good indication for the presence of a resonant state. 

We consider resonances and their decays mediated by the strong interaction between quarks and gluons. The Standard Model describes the behavior of the strong interaction through Quantum Chromodynamics (QCD) \cite{Close,Halzen}. However, we are not able to obtain QCD perturbative calculations for the formation and decay of hadrons at intermediate energies. Phenomenological (bag, flux-tube, Regge theory, etc.) \cite{Isgur,Barnes} and discrete (lattice QCD) models \cite{DeGrand,Dudek,Edwards} have been and are being used to make those calculations with varied success.

The analysis presented here does not invoke QCD per se,  but rather the fundamentals of quantum field theory. We base our model on Fermi's golden rule, the Feynman's rules, and the angular momentum-spin formalism of relativistic quantum mechanics \cite{Griffiths, Weinberg}, as well as the symmetries of QCD. Our goal is not to study the QCD structure of the resonance, but rather to identify the resonance. We search for its existence, measure its quantum numbers, and obtain information about its cross section and production mechanism. 

As discussed before, for practical reasons, we will also need to introduce in our model ad-hoc assumptions, as in the case of more than two final state particles which we describe via sequential 2-body decays. This is the so-called {\it isobar model}. We want to use a model that is able to incorporate known conservation laws, i.e., specific conservation laws imposed on the decay as well as production mechanisms. We will describe in this report a general overview of how to impose these constraints.

We are interested in finding possible resonances formed in the $\gamma N$ interaction. That resonance will sequentially decay to the observed multiparticle final states.  We will start by considering the reaction sketched in figure~\ref{fig:decay1}. 

%\centerline{\includegraphics[width=18cm]{ShortFigure.pdf}}
%\caption{\label{fig:decayx1}Diagram for the production of a mesonic resonance.}
%\end{figure}

We call $\tau$ the complete set of variables needed to describe the decay of the resonance- this will be reaction dependent,  but will normally include effective masses and angles. In addition, the isobar description will
include the masses and widths of the isobars.
For a number $n$ of identified particles (known mass) in the final state, we will need ($3n-4$) variables [$3n$ (spatial momenta)-$4$(momentum-energy conservation)] to identify this set in the phase space. 
The scattering cross section from a diagram with $n$ external lines depends on $(3n-10)$ variables [$4n$(unknowns)-$n$ (shell\ constraints)-$4$(momentum-energy\ constraints) - $6$(angular constraints)].

According to this, the reaction $\gamma N \rightarrow XN'$, will have two independent variables. We will take two of the Mandelstam variables, $s$ and $t$~\cite{Halzen} to identify the kinematics. The differential cross section is given then, using Fermi's golden rule, by
\begin{equation}  \frac{d\sigma}{dtds}= \sum_{ext.\;spins} \int |\mathscr{M}|^{2} d\rho(\tau) \end{equation} 
where the integral spans all the $\tau$ space, $\mathscr{M}$ is the Lorentz-invariant transition amplitude and $d\rho(\tau)$ is the Lorentz invariant phase-space element (LIPS). The spin's incoming and outgoing degrees of freedom are included in the sum over spins. The LIPS and $\mathscr{M}$ include the internal (transition) degrees of freedom. We can write~\cite{SUC71}
\begin{equation}   d\rho(\tau) \propto p_{cm}dMd\tau \end{equation} 
where $p_{cm}$ is the center-of-mass momentum, a constant in the reaction (see Appendix A) and $M$ is the mass of  X, the final mesonic system (resonance). We also assume that the cross section does not have important changes with center-of-mass energies ($\sqrt{s}$) in the energy range considered in the analysis (in practice, we limit the beam energy range).  Therefore
\begin{equation}  \frac{d\sigma}{dtdM} \propto  \sum_{ext.\;spins} \int |\mathscr{M}|^{2} d\tau
\label{eqn:fermi}
 \end{equation} 
and, if we consider small bins on $M$ and $t$ such that $\mathscr{M}$ only depends on $\tau$, we can define
\begin{equation}   I(\tau) \equiv \sum_{ext.\;spins} | \mathscr{M}|^{2}    \end{equation} 
then 
\begin{equation}  { d\sigma \over dtdM}\propto  \sum_{ext.\;spins}
\int | \mathscr{M}|^{2} d \tau = \int I(\tau) d \tau \end{equation} 
$I(\tau)$ is called the {\it intensity} and represents the probability for having a particle scattered into the
angular distribution specified by $\tau$ in the $\Delta M \Delta t$ kinematical range. This value will be associated with the probability used in the extended likelihood function discussed in section 2. $I(\tau)$ is the fundamental observed quantity.
 The complex amplitude $\mathscr{M}$ is calculated using the standard Feynman's rules \cite{Griffiths,Goldberger}.

Therefore

\begin{equation}  I(\tau) \equiv  \sum_{ext.\;spins} |\mathscr{M}|^{2} =  \sum_{ext.\;spins}  (\mathscr{M}\mathscr{M}^{*}) \end{equation} 
$\mathscr{M}$ is a representation of the scattering operator or transition operator, $\widehat{T}$, given by
\begin{equation}  \mathscr{M}= \langle out | \widehat{T} |in\rangle \end{equation} 
and then
\begin{equation}  I(\tau) \equiv  \sum_{ext.\;spins} |\mathscr{M}|^{2} =  \sum_{ext.\ spins} \langle out | \widehat{T} |in\rangle (\langle out | \widehat{T} |in\rangle)^{*} \end{equation} 
and, further
\begin{equation}  \langle out | \widehat{T} |in\rangle (\langle out | \widehat{T} |in\rangle)^{*} = \langle out | \widehat{T} |in\rangle \langle in | \widehat{T}^{\dagger}  |out\rangle.\end{equation} 

We will define the operator $|in\rangle \langle in | $, corresponding to the initial state, the {\it initial spin density matrix operator}, $\widehat{\rho_{in}}$, as
\begin{equation}  \widehat{\rho_{in}} \equiv |in\rangle \langle in |. 
\label{eqn:densdef}
\end{equation} 
Suppose that we prepare the polarization of the incoming photons and target or measure their states of polarization. The average over their polarization will be completely described by this spin density matrix. In the case of a beam of polarized photons, any polarized state can be written as a linear combination of two pure polarization states. Therefore, the general structure of this $2 \times 2$ matrix (in any particular basis defined by $|i\rangle$ and $|j\rangle$) will be
\begin{equation} \widehat{\rho_{in}}  =\sum_{i,j} \rho_{i,j} = \sum^{2}_{i,j=1} |i\rangle \langle j | .
\end{equation} 
The structure of the spin density matrix will be described in detail in section~\ref{sect:SpinDen}.

Then we have
\begin{equation}  I(\tau) =   \sum_{ext.\ spins} \sum_{i,j} \langle out | {^{i} \widehat{T}} \rho_{ij} {^{j} \widehat{T}^{\dagger} } |out\rangle .
\label{eqn:inten}
\end{equation} 
Here, in $"ext.\ spins"$ we excluded the beam and target spins, as they are described by the initial state spin density matrix. The upper-left index on the transition operators correspond to the initial state specified by the spin density matrix.
Keeping in mind the reaction represented in figure~\ref{fig:decay1},
we will assume that the transition operator can be factorized into two parts: the production (of X) and the decay operators (of X) such that:

\begin{equation}  I(\tau) \propto \sum_{ext.\ spins} \sum_{i,j}\langle out |  {^{i} \widehat{T}_{decay}} {^{i} \widehat{T}_{production}}  \rho_{ij} {^{j}\widehat{T}_{production}^{\dagger}} {^{j} \widehat{T}_{decay}^{\dagger}} |out\rangle \end{equation} 

Now we can take a complete orthogonal set of states, $|X\rangle$, such that $\sum_{X} |X\rangle\langle X| = 1$, and include them in the previous relation such that

\begin{equation}  I(\tau) \propto \sum_{ext.\ spins} \sum_{i,j} \langle out | {^{i} \widehat{T}_{d} }\sum_{X} |X\rangle\langle X|{^{i}\widehat{T}_{p}} \rho_{ij} {^{j} \widehat{T}^{\dagger}_{p}} \sum_{X'} |X'\rangle\langle X'|{^{j} {T}^{\dagger}_{d}} |out\rangle \end{equation} 

\begin{equation}  I(\tau) \propto  \sum_{ext.\ spins}\sum_{i,j} \sum_{X,X'}  \langle out | {^{i} \widehat{T}_{d}}  |X\rangle\langle X| {^{i} \widehat{T}_{p}} \rho_{ij} {^{j}\widehat{T}^{\dagger}_{p}}  |X'\rangle\langle X'| {^{j} \widehat{T}^{\dagger}_{d}}  |out\rangle. \end{equation} 

The set of states , $|X\rangle$, are called {\it partial waves}, and gives the name of {\it partial wave analysis} (PWA)  to the method presented in this report.
Each of these states can be described by a set of quantum numbers that we will collectively call $\{b\}$. This set spans all the possible intermediate states, therefore, the experimental goal of finding the quantum numbers associated with the resonance is translated to measuring the partial wave amplitudes.
We will call
\begin{equation}  \langle out | {^{i}\widehat{T}_{d}}  |X\rangle\ = {^{i}A_{b}(\tau) }
\label{eqn:decayAmp}
\end{equation} 
the {\it decay amplitude} for a given wave, $b$, which may be calculated exclusively from the $\tau$ parameters as it is going to be shown in section~\ref{subsect:DecayAmp}. 

The production amplitude contains the hadronic QCD-based interaction that we are not able to calculate, rather the production amplitudes will be considered a {\it weight}  on each partial decay amplitude of the final mix.  These weights are the parameters to be fitted to the data, and will also depend on the $k$ external spins. For example, in the case of an initial and final state nucleon (protons or neutrons), and no information about target (proton) spin, we will have $k=2 \times 2 = 4$. We have assumed here that the resonance $X$ decays to final spinless mesons. We will have
\begin{equation}  \langle X| {^{i} \widehat{T}_{p}} \rho_{ij} {^{j} \widehat{T}^{\dagger}_{p}}  |X'\rangle = {^{i} V^{k}_{b}} \rho_{ij} {^{j} V^{k*}_{b'}} \end{equation} 
being $V^{k}_{b}$ the production amplitudes. Note that the $A$'s and $V$'s are both complex numbers. Therefore
\begin{equation}  \boxed{ I(\tau) = \sum_{k} \sum_{i,j} \sum_{b,b'} \ {^{i} A_{b}(\tau)} {^{i} V^{k}_{b}} \rho_{ij} {^{j} V^{k*}_{b'}}\  {^{j} A^{*}_{b'}(\tau)}. } 
\label{eqn:intenf}
\end{equation} 
We might define the {\it resonance spin density matrices}  as
\begin{equation}  {^{i,j} \rho_{b,b'}} = \sum_{k}  {^{i} V^{k}_{b}}\rho_{ij} {^{j} V^{k *}_{b'}} 
\label{eqn:Xdens}
\end{equation} 
where $k$ represents the rank of the spin density matrix of the resonance ($X$). 
Notice that we use the same symbol ($\rho$) to name the resonance and the initial (later photon) spin density matrices. We believe that it will be clear when we use one or the other, i.e. the initial (photon) spin density matrix will run on only two or less indices.

The intensity distribution is thus given by
\begin{equation}  I(\tau) =  \sum_{i,j} \sum_{b,b'} {^{i} A_{b}(\tau)}\  {^{i,j}\rho_{b,b'}}\  {^{j} A^{*}_{b'}(\tau)}. \end{equation} 
\subsection{Decay Amplitudes}
\label{subsect:DecayAmp}

To calculate the decay amplitudes we will consider two cases: first, the resonance decaying into two particles,  and second, the resonance decaying into three or more particles.  In this latter case, we will use the {\it isobar model}~\cite{Herndon}. The isobar model assumes a series of sequential two-body decays. We consider the resonance decaying into an intermediate unstable particle (isobar) plus a stable particle (bachelor), and that all bachelors will be among the final states. The isobar will decay subsequently into other particles (children), which may also be isobars and continue the process. We assume that there are no interactions after the particles are produced through this sequential process and that all final (observed) particles are spinless. We calculate amplitudes in the spin formalism of Jacob and Wick~\cite{Wick, SUC71}.

\subsubsection{Two-Body Decays}

\begin{figure}[!htp]
\centerline{\includegraphics[width=12cm]{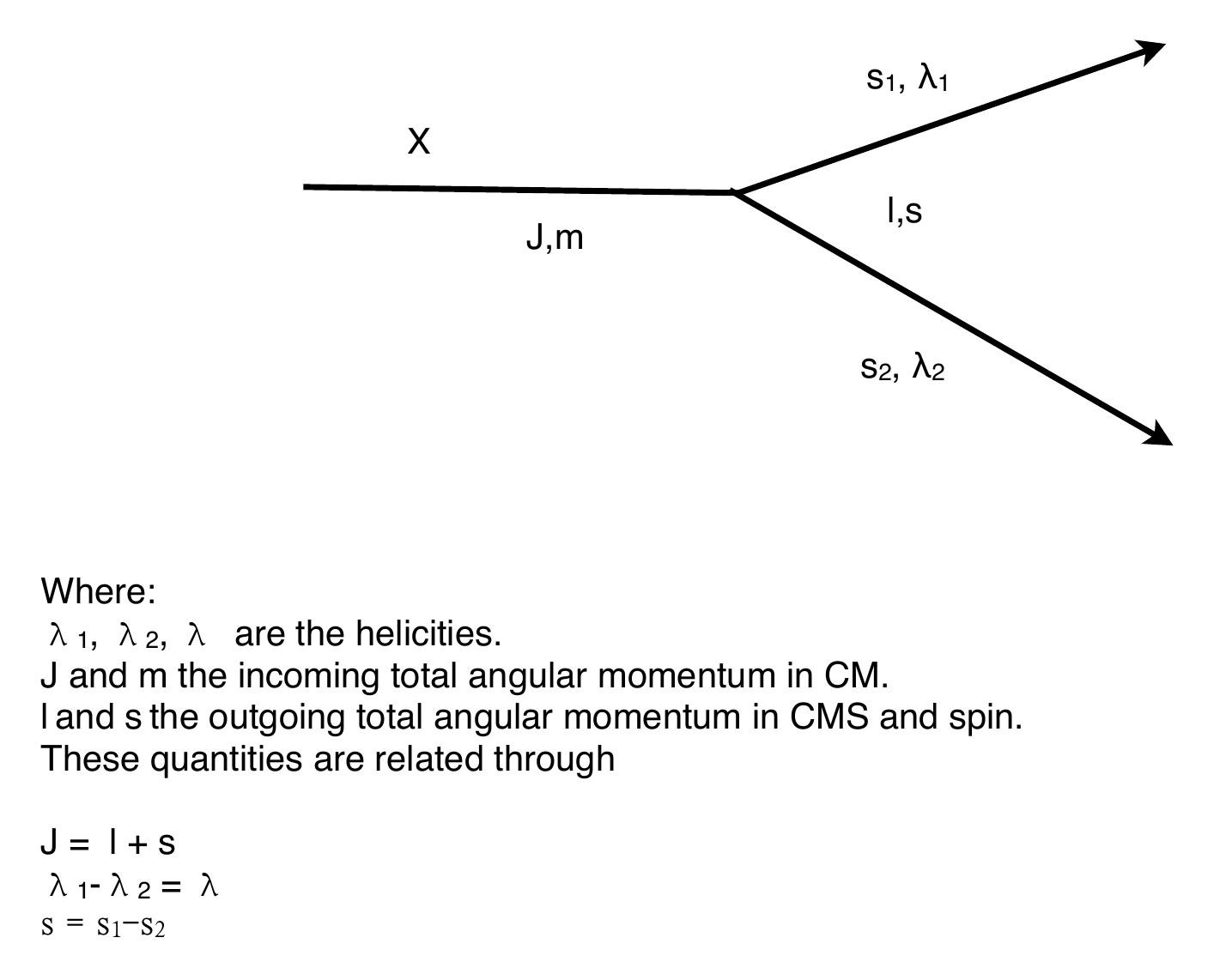}}
\caption{\label{fig:decayx} Two-body Decay.}
\end{figure}

Let's consider the case of a resonance $X$  decaying into two particles labeled as $1$ and $2$ (see figure~\ref{fig:decayx} for notation), where one of them may be an isobar with spin.
We describe the decay of $X$ in its rest frame, that is $p_{1}+p_{2}=0$, with the z in the direction of the beam; this is the Gottfried-Jackson (GJ) frame (see Appendix A). We can thus describe the kinematics with just one momentum $q(\phi,\theta)=p_{1}=-p_{2}$. 
In this case,  what we called $\tau$ to describe the final particles, will be just given by two angles 
\begin{equation}  \tau=\begin{Bmatrix}\phi_{GJ},\theta_{GJ}\end{Bmatrix} \end{equation} 
where $\phi_{GJ},\theta_{GJ}$ are the angles of one of the decay products in the Gottfried-Jackson frame. For a given mass $M$ and transfer momentum $t$, the decay amplitudes will depend only on $\tau$ (angles). The decay amplitudes were discussed before (see equation~\ref{eqn:decayAmp}) and represented by
\begin{equation}  \langle out | {^{i}\widehat{T}_{decay}}  |X\rangle\ = {^{i}A_{b}(\tau)} 
\end{equation} 
(we will omit the initial spin matrix index, $i$, in this section). 
 
The intermediate states in the expansion, $X$, are normally described in the angular momentum canonical basis labeled by their total angular momentum, $J$ and the z component $m$ ($|Jm \rangle$ orthogonal basis).
The decay products have helicities given by $\lambda_{1}$ and $ \lambda_{2}$. 

 Writing, explicitly, the form of the decay amplitude as
\begin{equation}  A_{b}(\tau) = \langle p_{1}\lambda_{1},-p_{2}\lambda_{2} | \widehat{T}_{decay}  |Jm\rangle  \end{equation} 
and since $q$ is a constant, only directions are important, then
\begin{equation}  A_{b}(\tau)  = \langle \Omega_{GJ};\lambda_{1} \lambda_{2} |\widehat{T}_{decay}|Jm \rangle \end{equation} 
where we call $\Omega_{GJ}= (\theta_{GJ},\phi_{GJ})$.
Factorizing this amplitude by explicitly introducing the outgoing angular momentum $l$ and spin $s$ into the state ket, $|Jmls \rangle$, the amplitude becomes
\begin{equation}  A_{b}(\tau) = \langle \Omega_{GJ};\lambda_{1} \lambda_{2}|Jmls \rangle\ \langle  Jmls| \widehat{T}_{decay}|Jm\rangle .
\label{eqn:amp}
\end{equation}
The left factor of (\ref{eqn:amp}) represents a change from the canonical to the helicity basis, and a rotation that takes $z$ from the beam direction to the resonance momentum direction. It will be calculated below.  

In the GJ frame, the transition decay operator, $T_{decay}$, depends only on the mass (called $w_{o}$) of the resonance, it is therefore symmetric or independent of the angles. It will be absorbed into the normalization constant. 
Let's call $a_{ls}$ to this dynamical part that contain the unknown transition amplitude
\begin{equation}  a_{ls} = \langle  Jmls| \widehat{T}_{decay}|Jm\rangle\ \end{equation} 
and introducing a complete set of  helicity states, $\sum_{\lambda^{'}_{1} \lambda^{'}_{2}} |\lambda^{'}_{1} \lambda^{'}_{2}\rangle \langle \lambda^{'}_{1} \lambda^{'}_{2} | = 1$,
\begin{equation}  A_{b}(\tau)  = \langle \Omega_{GJ};\lambda_{1} \lambda_{2}|Jmls \rangle a_{ls} = \sum_{\lambda^{'}_{1} \lambda^{'}_{2}} \langle \Omega_{GJ};\lambda_{1} \lambda_{2}|Jm \lambda^{'}_{1} \lambda^{'}_{2} \rangle \langle Jm\lambda^{'}_{1} \lambda^{'}_{2} |Jmls \rangle\ a_{ls} .\end{equation} 

We will show that

\begin{equation}  \langle \Omega_{GJ};\lambda_{1} \lambda_{2}|Jm \lambda^{'}_{1} \lambda^{'}_{2} \rangle = \sqrt{\frac{2J+1}{4 \pi}} D^{J*}_{m \lambda}(\Omega_{GJ}) \delta_{\lambda_{1} \lambda^{'}_{1}} \delta_{\lambda_{2} \lambda^{'}_2}  
\label{eqn:eqA}
 \end{equation}  
and 
\begin{equation}   \langle Jm\lambda^{'}_{1} \lambda^{'}_{2} |Jmls \rangle\ =  \sqrt{\frac{2l+1}{2J+1}} (l0s \lambda|J \lambda)(s_{1} \lambda^{'}_{1}s_{2} -\lambda^{'}_{2}|s \lambda) \  
\label{eqn:eqB}
\end{equation} 
arriving to the final expression
\begin{equation}  A_{b}(\tau)  = \sqrt{\frac{2l+1}{4 \pi}} \sum_{\lambda_{1} \lambda_{2}} D^{J*}_{m \lambda}(\Omega_{GJ}) (l0s \lambda|J \lambda)(s_{1} \lambda_{1}s_{2} -\lambda_{2}|s \lambda) a_{ls}.
\label{eqn:MAINW}
 \end{equation} 
The expressions in parenthesis represent Clebsch-Gordan coefficients (see Appendix C). Remember that $b$ represents all quantum numbers of X (J,m,l and s).

\vspace{7 mm}

Let's first prove expression (\ref{eqn:eqA}).
The state $|p \Omega_{GJ} \lambda_{1} \lambda_{2} \rangle$ can be written in a canonical angular momentum basis as  \cite{SUC71, Richman}
\begin{equation}  |p \phi \theta \lambda_{1} \lambda_{2} \rangle = \sum_{J,m} C_{Jm}(p \phi \theta \lambda_{1} \lambda_{2}) |p J m \lambda_{1} \lambda_{2} \rangle .
\label{eqn:eqA1}
\end{equation} 
We need to evaluate the coefficients $C_{Jm}$ in this expansion. We will, first,  evaluate them for the case of $\theta = \phi =0$ and then rotate the state to a general direction. We have
\begin{equation}  |p 0 0 \lambda_{1} \lambda_{2} \rangle = \sum_{J,m} C_{Jm}(p 0 0 \lambda_{1} \lambda_{2}) |p J m \lambda_{1} \lambda_{2} \rangle . \end{equation}

This state represents two particles moving in opposing directions on the z axis with helicities $\lambda_{1}$ and $\lambda_{2}$. These particles do not have angular momentum respect to z, since $L = (\hat{z} \times  \vec{p})$ =0. Therefore $|p 0 0 \lambda_{1} \lambda_{2} \rangle$ is an eigenstate of $J_{z}$ with eigenvalue $\lambda = \lambda_{1} - \lambda_{2}$ and $m=\lambda$

\begin{equation}  |p 0 0 \lambda_{1} \lambda_{2} \rangle = \sum_{J} C_{J \lambda}(p 0 0 \lambda_{1} \lambda_{2}) |p J \lambda \lambda_{1} \lambda_{2} \rangle . \end{equation} 

We now rotate this state to the angles $\phi, \theta$; by the definition of the Wigner-D functions

\begin{equation}   |p \phi \theta \lambda_{1} \lambda_{2} \rangle = R(\phi,\theta,0) |p 0 0 \lambda_{1} \lambda_{2} \rangle = \sum_{m'} D^{J}_{m' \lambda}(\phi,\theta,0) |p J m' \lambda_{1} \lambda_{2} \rangle \end{equation} 
where $R(\phi,\theta,0)$ is an active rotation (see Appendix B), leading to
\begin{equation}  |p \phi \theta \lambda_{1} \lambda_{2} \rangle = \sum_{J,m'} C_{J \lambda}(p 0 0 \lambda_{1} \lambda_{2}) D^{J}_{m' \lambda}(\phi,\theta,0) |p J m' \lambda_{1} \lambda_{2} \rangle .  
\label{eqn:eqA2}
\end{equation} 

Comparing (\ref{eqn:eqA1}) and (\ref{eqn:eqA2})

\begin{equation}  C_{Jm}(p \phi \theta \lambda_{1} \lambda_{2}) = C_{J \lambda}(p 0 0 \lambda_{1} \lambda_{2}) D^{J}_{m \lambda}(\phi,\theta,0) . \end{equation} 
With the normalization of (\ref{eqn:eqA1}), and using normalization of the Wigner-D functions \cite{SUC71, Rose}, it can be found
\begin{equation}  | C_{J \lambda}(p 0 0 \lambda_{1} \lambda_{2}) |^2 =  \frac{2J+1}{4\pi} \end{equation} 
therefore
\begin{equation}  C_{Jm}(p \phi \theta \lambda_{1} \lambda_{2}) =  \sqrt{ \frac{2J+1}{4\pi}} D^{J}_{m \lambda}(\phi,\theta,0) \end{equation} 
or
\begin{equation}  |p \Omega_{GJ} \lambda_{1} \lambda_{2} \rangle = \sum_{J,m} \sqrt{ \frac{2J+1}{4\pi}} D^{J}_{m \lambda}(\phi,\theta,0)|p J m \lambda_{1} \lambda_{2} \rangle .
\label{eqn:AAA}
\end{equation} 

For a fixed resonance mass the momentum is fixed. Taking the conjugate we have

\begin{equation}  \langle \Omega_{GJ} \lambda_{1} \lambda_{2} | = \sum_{J,m} \sqrt{ \frac{2J+1}{4\pi}} D^{J*}_{m \lambda}(\phi,\theta,0) \langle J m \lambda_{1} \lambda_{2} |  \end{equation} 
then
\begin{equation}  \langle \Omega_{GJ};\lambda_{1} \lambda_{2}|Jm \lambda^{'}_{1} \lambda^{'}_{2} \rangle = \sqrt{\frac{2J+1}{4 \pi}} D^{J*}_{m \lambda}(\Omega_{GJ}) \delta_{\lambda_{1} \lambda^{'}_{1}} \delta_{\lambda_{2} \lambda^{'}_2} . \   \end{equation}  
We will now prove expression (\ref{eqn:eqB}). We will start inverting expression (\ref{eqn:AAA}) by multiplying each side by
\begin{equation}  \int d\Omega D^{J*}_{m \lambda}(\Omega) \end{equation} 
and using the normalization of the Wigner-D functions found in Appendix B, equation (\ref{eqn:DNorm}), we obtain
\begin{equation}  |J m \lambda_{1} \lambda_{2} \rangle = \sqrt{2J+1} \int d\Omega D^{J*}_{m \lambda}(\Omega) |\phi \theta \lambda_{1} \lambda_{2} \rangle \end{equation} 

A two particles state of spins $s_{1}$ and $s_{2}$, and z projections $m_{s}=m_{s_{1}}+m_{s_{2}}$, could be obtained by two rotations on the canonical (z-projected) two particles state \cite{SUC71}, such that

\begin{equation}  |\phi \theta \lambda_{1} \lambda_{2} \rangle = \sum_{m_{s_{1}}, m_{s_{2}} } D^{s_{1}}_{m_{s_{1}}\lambda_{1}} (\phi,\theta,0) D^{s_{2}}_{m_{s_{2}} -\lambda_{2}} (\phi,\theta,0) |\phi \theta m_{s_{1}} m_{s_{2}} \rangle .
\end{equation} 
Therefore
\begin{equation}  |J m \lambda_{1} \lambda_{2} \rangle = \sqrt{2J+1} \sum_{m_{s_{1}},m_{s_{2}}} \int d\Omega D^{J*}_{m \lambda}(\Omega)  D^{s_{1}}_{m_{s_{1}}\lambda_{1}} (\phi,\theta,0) D^{s_{2}}_{m_{s_{2}}-\lambda_{2}} (\phi,\theta,0) |\phi \theta m_{s_{1}} m_{s_{2}} \rangle .
\label{eqn:BB}
\end{equation} 
Using the Wigner-D functions properties (see appendix B), we obtain
\begin{equation}  D^{s_{1}}_{m_{s_{1}}\lambda_{1}} (\phi,\theta,0) D^{s_{2}}_{m_{s_{2}}-\lambda_{2}} (\phi,\theta,0) = \sum_{s,m_{s}} (s_{1} m_{s_{1}} s_{2} m_{s_{2}} |s m_{s}) (s_{1} \lambda_{1} s_{2} -\lambda_{2} |s \lambda) D^{s}_{m_{s},\lambda} \end{equation} 
and, if we include an intermediate angular momenta set $|lm_{l} \rangle$, with $J=l \oplus s$ and $m=m_{l}+m_{s}$, we have 
\begin{equation}  D^{J*}_{m\lambda} (\phi,\theta,0) D^{s}_{m_{s}-\lambda} (\phi,\theta,0) = \sum_{l,m_{l}} \frac{\sqrt{2l+1}}{2J+1}(l m_{l} s m_{s} |J m) (l 0 s \lambda |J \lambda) Y^{l}_{m_{l}}(\Omega) .\end{equation} 

Putting all this together into (\ref{eqn:BB})

\begin{equation*}  |J m \lambda_{1} \lambda_{2} \rangle  =   \sum_{l,m_{l}} \frac{\sqrt{2l+1}}{2J+1}(l m_{l} s m_{s} |J m) (l 0 s \lambda |J \lambda)(s_{1} \lambda_{1} s_{2} -\lambda_{2} |s \lambda)  \end{equation*}
\begin{equation} \times \sum_{s,m_{s}} (s_{1} m_{s_{1}} s_{2} m_{s_{2}} |s m_{s})(l m_{l} s m_{s} |J m)  \int d\Omega Y^{l}_{m_{l}}(\Omega) |\phi \theta m_{s_{1}} m_{s_{2}} \rangle 
\label{eqn:BB2}
\end{equation} 
but we can also write
\begin{equation}  |J m l s\rangle = \sum_{s,m_{s}} (s_{1} m_{s_{1}} s_{2} m_{s_{2}} |s m_{s})(l m_{l} s m_{s} |J m)\int d\Omega \; Y^{l}_{m_{l}}(\Omega) |\phi \theta m_{s_{1}} m_{s_{2}} \rangle 
\label{eqn:BB1}
\end{equation} 
since
\begin{equation}  |J m l s\rangle = \sum_{m_{l},m_{s}} (l  m_{l}  s  m_{s} |J m) |l m_{l} s m_{s} \rangle \end{equation} 
with
\begin{equation}  |l m_{l} s m_{s} \rangle = \int d\Omega  \; Y^{l}_{m_{l}}(\Omega) |\phi \theta s m_{s} \rangle \end{equation} 
and 
\begin{equation}  |\phi \theta s m_{s} \rangle = \sum_{m_{s_{1}},m_{2}} (s_{1} m_{s_{1}} s_{2} m_{s_{2}} |s m_{s})|\phi \theta m_{s{1}} m_{s_{2}} \rangle .\end{equation} 

Substituting (\ref{eqn:BB1}) into (\ref{eqn:BB2}), we have
\begin{equation}  |J m \lambda_{1} \lambda_{2} \rangle = \sum_{l,m_{l}} \frac{\sqrt{2l+1}}{\sqrt{2J+1}} (l 0 s \lambda |J \lambda)(s_{1} \lambda_{1} s_{2} -\lambda_{2} |s \lambda) |J m l s\rangle . \end{equation} 
Taking the conjugate of this expression and applying to $|J m l s\rangle$
\begin{equation}   \langle Jm \lambda^{'}_{1} \lambda^{'}_{2} |Jmls \rangle =  \sqrt{\frac{2l+1}{2J+1}} (l0s \lambda|J \lambda)(s_{1} \lambda_{1} s_{2} -\lambda_{2}|s \lambda) \end{equation} 
obtaining the expression we wanted to prove (\ref{eqn:eqB}).

Let's repeat the main expression, for the wave $b$ ($Jmls$), the amplitude is

\begin{equation}  \boxed{ A_{b}(\tau)  = \sqrt{\frac{2l+1}{4 \pi}} F_l(p)\sum_{\lambda_{1} \lambda_{2}} D^{J*}_{m \lambda}(\Omega_{GJ}) (l0s \lambda|J \lambda)(s_{1} \lambda_{1}s_{2} -\lambda_{2}|s \lambda) a_{ls} .}
\label{eqn:twopart}
 \end{equation} 
where we introduce the factor $F_{l}(p)$, the Blatt-Weisskopf centrifugal-barrier factor, described in detail in section~\ref{sect:MassDep}. This factor takes into account the {\it centrifugal-barrier effects} caused by the angular factors on the potential. The sum on $\lambda_{1}$ and $\lambda_{2}$ is over all possible helicities of the daughters particles.

The "unknown" factor $a_{ls}$ will be included into the fitting parameters of our model ("V's") and will not be carried over our next formulas.

Consider the decay of a resonance into two spinless final particles.
Experimentally, we normally detect spinless particles , therefore this is a very common case  ( kaons, etas or pions). In this case, $\lambda = \lambda_{1}= \lambda_{2}=0$, $s = s_{1}= s_{2}=0$ and $J=l$.

Therefore
\begin{equation}  (l0s 0|J 0) = 1 \end{equation}
and
\begin{equation}  (s_{1} 0s_{2} 0|s 0) = 1. \end{equation} 

Then
\begin{equation}  A_{lm}(\tau) = F_{l}(p) \sqrt{\frac{2l+1}{4\pi}} { D^{l\ *}_{m0}(\phi,\theta,0)} 
\label{eqn:twopart}
\end{equation} 
and since
\begin{equation}  D _{m0}^{l\ *}(\phi,\theta,0) = e^{im\phi}d^{l}_{m0}(\theta) \end{equation} 
where
\begin{equation}  d^{l}_{m0}(\theta)=\sqrt{ \frac{(l-m)!}{(l+m)!} } P^{m}_{l}(cos \theta) \end{equation} 
are the Wigner small-D functions and where $P^{m}_{l}(cos \theta)$ are the Associated Legendre functions \cite{Rose}. 
Therefore,

\begin{equation}  A_{lm}(\phi , \theta) = F_{l}(p) \sqrt{ \frac{(2l+1)(l-m)!}{4 \pi (l+m)!} } P^{m}_{l}(cos \theta) e^{im\phi} = F_{l}(p) Y^{m}_{l}(\phi,\theta) \end{equation}
where $Y^{m}_{l}(\phi,\theta)$ are the spherical harmonic functions.

For example for the first three waves (l=0,1,2 or S,P,D) we have
\begin{equation}  A_{00}(\phi,\theta) =F_{0}(p) Y^{0}_{0}(\phi,\theta)= \frac{1}{\sqrt{4 \pi}}\end{equation} 
\begin{equation}  A_{11}(\phi,\theta) = F_{1}(p) Y^{1}_{1}(\phi,\theta)=-F_{1}(p) \sqrt{\frac{3}{8 \pi}} sin(\theta)e^{i\phi} \end{equation} 
\begin{equation}  A_{10}(\phi,\theta) = F_{1}(p) Y^{0}_{1}(\phi,\theta)=F_{1}(p) \sqrt{\frac{3}{4 \pi}} cos(\theta) \end{equation} 
\begin{equation}  A_{1-1}(\phi,\theta) = F_{1}(p) Y^{-1}_{1}(\phi,\theta)=-F_{1}(p) \sqrt{\frac{3}{8 \pi}} sin(\theta)e^{-i\phi} \end{equation} 

\begin{equation}  A_{22}(\phi,\theta) = F_{2}(p) Y^{2}_{2}(\phi,\theta)= F_{2}(p) \sqrt{\frac{15}{32 \pi}} sin^{2}(\theta)e^{2i\phi} \end{equation} 
\begin{equation}  A_{21}(\phi,\theta) = F_{2}(p) Y^{1}_{2}(\phi,\theta)=- F_{2}(p) \sqrt{\frac{15}{8 \pi}} sin(\theta)cos(\theta)e^{i\phi} \end{equation} 
\begin{equation}  A_{20}(\phi,\theta) = F_{2}(p) Y^{0}_{2}(\phi,\theta)= F_{2}(p) \sqrt{\frac{5}{16 \pi}} (3cos^{2}(\theta)-1) \end{equation} 
\begin{equation}  A_{2-1}(\phi,\theta) = F_{2}(p) Y^{-1}_{2}(\phi,\theta)=- F_{2}(p) \sqrt{\frac{15}{8 \pi}} sin(\theta)cos(\theta)e^{-i\phi} \end{equation} 
\begin{equation}  A_{2-2}(\phi,\theta) = F_{2}(p) Y^{-2}_{2}(\phi,\theta)= F_{2}(p) \sqrt{\frac{15}{32 \pi}} sin^{2}(\theta)e^{-2i\phi} \end{equation} 
The $F_{l}(p)$ values, the Blatt-Weisskopf centrifugal-barrier factors, are described in section~\ref{sect:MassDep}. They normally have very small variations in the (mass, t-Madelstam) range where the intensity (cross section) is calculated.

\subsubsection{Generalized Isobar Model Formalism}

Let's consider now the more general and interesting case where the final particles are three or more~\cite{SUC93}. In the isobar formalism \cite{Herndon}, we will treat the decay amplitude of the resonance as the product of successive two-body decay amplitudes

\begin{equation}  A_{b}(\tau) = A_{b'}(\tau') A_{b''}(\tau'') A_{b'''}(\tau''') ... \end{equation} 

 Let's consider a resonance decaying into a "diparticle"  (isobar) and a particle "bachelor". The isobar will decay into two children (to consider more particles the process is repeated).  The process and notation are described in figure~\ref{fig:decaya}.

\begin{figure}  
\label{fig:decaya}
\begin{center}  
\includegraphics[width=15cm]{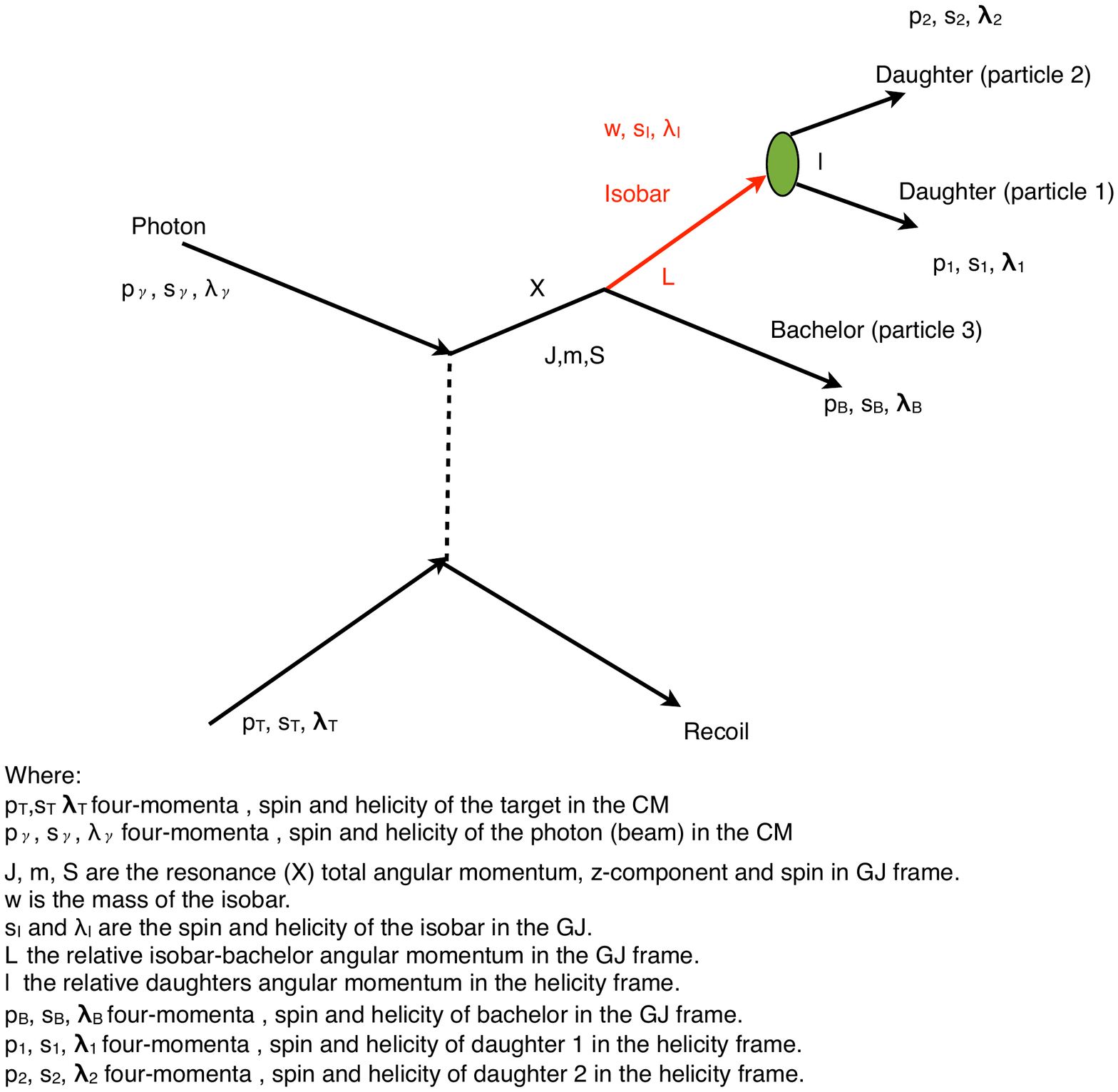} 
\caption{Isobar Model Decay.}  
\label{fig:decaya}
\end{center} 
\end{figure}

The angular description of the final three-body state in the Lab system will include three pairs of angles, but they are correlated. The degrees of freedom (uncorrelated variables describing the kinematics) will include the mass of the isobar, $w$, and the angles of the its decay products
\begin{equation}  \tau=\begin{Bmatrix}\Omega_{GJ},\Omega_{h},w\end{Bmatrix} \end{equation} 
where $\Omega_{GJ}=(\phi_{GJ},\theta_{GJ})$ and $\Omega_{h}=(\phi_{h},\theta_{h})$ are the angular descriptions in the Gottfried-Jackson and helicity frames (see Appendix A) of the isobar and its decay products respectively.
Let $l$ be the angular momentum between bachelor and isobar and $s$ the spin of the isobar (we will consider a spinless bachelor). Therefore $J=l \oplus s$.
The amplitude is then written as
\begin{equation}  A_{b}(\tau)= E^{Jls *}_{m}(\Omega_{GJ},\Omega_{h})Q_{ls}(w) .
\label{eqn:somedecay}
\end{equation} 
The amplitude has a factor that depends only on the angles, and a factor which is only dependent on mass. The mass factor comes from the propagator of the isobar, and is described in more detail in section~\ref{sect:MassDep}. The angular factor can be written, in the isobar model, as \cite{SUC93}
\begin{equation}  E_{m}^{Jls *}(\Omega,\Omega_{h}) = \langle f | \widehat{T}^{I \rightarrow D_{1}D_{2}}_{decay}  \widehat{T}^{R \rightarrow I B}_{decay} |X\rangle\   \end{equation} 
\begin{equation}  \langle f | \widehat{T}^{I \rightarrow D_{1}D_{2}}_{decay}  \widehat{T}^{R \rightarrow I B}_{decay} |X\rangle\ = \langle \Omega_{h} ; 0| \widehat{T}^{I}_{decay}  |s \lambda\rangle\ \langle \Omega_{GJ}; s \lambda| \widehat{T}^{R}_{decay}  |Jm \rangle\ \end{equation} 
where $R \rightarrow IB$ describes the decay of the resonance ($R$) into the isobar ($I$) and the bachelor ($B$ ), and $I \rightarrow D_1D_2$ is the decay of the isobar.
Using our previous result, equation (\ref{eqn:MAINW}), for each two-body decay we have
\begin{equation}  \sqrt{\frac{2l+1}{4 \pi}} \sum_{\lambda_{1} \lambda_{2}} D^{J*}_{m \lambda}(\Omega_{GJ}) (l0s \lambda|J \lambda)(s_{1} \lambda_{1}s_{2} -\lambda_{2}|s \lambda) . \end{equation} 

For the bachelor $\lambda_{2}=0$ and for the isobar $\lambda_{1}=\lambda$, therefore
\begin{equation}  (s_{1} \lambda_{1} 0 0|s \lambda) = 1 \end{equation} 
then
\begin{equation}  \langle \Omega_{GJ};s \lambda| \widehat{T}^{R}_{decay}  |Jm\rangle\ = \sqrt{\frac{2J+1}{4\pi}} { D^{J\ *}_{m\lambda}(\phi_{GJ},\theta_{GJ},0)} (l0s \lambda|J \lambda) . \end{equation}

In experiments we mostly detect final state spinless particles - pions and kaon. However, we need to emphasize that our formalism will allow to include particles with spin, or more than three particles in the final state if we just repeat the process explained here using equation (\ref{eqn:MAINW}). For clarity, we will continue with an isobar decaying into two spinless children (into a total of three particles final state)
\begin{equation}  \langle \Omega_{h} ; 0 | \widehat{T}^{I}_{decay} |s \lambda\rangle\ = \sqrt{\frac{2s+1}{4\pi}} { D^{s\ *}_{\lambda 0}(\phi_{h},\theta_{h},0)} \end{equation} 
therefore
\begin{equation}  E_{m}^{Jls *}(\Omega_{GJ},\Omega_{h})=\sqrt{(2l+1)} \sqrt{2s+1} \sum_{\lambda} D^{J *}_{m \lambda}(\phi_{GJ},\theta_{GJ},0) D^{s*}_{\lambda 0}(\phi_{h},\theta_{h},0) \langle l0s\lambda|J\lambda  \rangle . \end{equation} 

Since

\begin{equation}  D_{\lambda 0}^{s*}(\phi_{h},\theta_{h},0) = e^{i\lambda\phi_{h}}d^{s}_{\lambda 0}(\theta_{h}) \end{equation} 
and
\begin{equation}  D^{J *}_{m \lambda} (\phi_{GJ},\theta_{GJ},-\phi_{h}) = D^{J *}_{m \lambda} (\phi_{GJ},\theta_{GJ},0)  e^{-i\lambda\phi_{h}} \end{equation} 
the angular amplitude can, then, be written as
\begin{equation}  E_{m}^{Jls *}(\Omega_{GJ},\Omega_{h})=\sqrt{(2l+1)} \sqrt{2s+1} {\sum_{\lambda}} {D^{J *}_{m \lambda}}(\phi_{GJ},\theta_{GJ},\phi_{h}) d^{s}_{\lambda 0}(\theta_{h}) \langle l0s\lambda|J\lambda\rangle . \end{equation}

The mass term (as discussed in detail in section~\ref{sect:MassDep}), depends on the isobar mass,  and is given by
\begin{equation}  Q_{ls}(w) =F_{l}(p)F_{s}(q)\Psi (w) 
\label{eqn:masster}
\end{equation} 
where the $\Psi$-function is the standard relativistic Breit-Wigner form for the isobar mass distribution, $p$ is the momentum of the isobar in the GJ frame, and $q$ the momentum of the leading isobar's decay particle in the helicity frame
\begin{equation}  \Psi (w)=\frac{w_{0}\Gamma_{0}}{w^{2}_{0}-w^{2}-iw_{0}\Gamma(w)} \end{equation} 
with
\begin{equation}  \Gamma(w)=\Gamma_{0}\frac{w_{0}qF^{2}_{s}(q)}{wq_{0}F^{2}_{s}(q_{0})} \end{equation} 
$w_{0}$ and $\Gamma_{0}$ are the mass and width of the isobar, and $q_{0}$ is found such
that $\Gamma(w_{0}) = \Gamma_{0}$ and then $|\Psi(w_{0})|=1$.

The  $F_{l}(p)$ and $F_{s}(q)$ functions are the Blatt-Weisskopf centrifugal-barrier factors (see section~\ref{sect:MassDep} for details). These factors take into account the {\it centrifugal-barrier effects} caused by the angular (spin) factors on the potentials.

Adding all these components into our final form for the amplitude for a three (spinless) particles in the  final state, we obtain:

\vspace{0.5cm}
\fbox{
\addtolength{\linewidth}{-2\fboxsep}%
 \addtolength{\linewidth}{-2\fboxrule}%
 \begin{minipage}{\linewidth}
\begin{equation*}   A_{b}(\tau)= \sqrt{(2l+1)} \sqrt{2s+1}  F_{l}(p)F_{s}(q) \frac{w_{0}\Gamma_{0}}{w^{2}_{0}-w^{2}-iw_{0}\Gamma(w)} \end{equation*}
\begin{equation} \times {\sum_{\lambda}} {D^{J *}_{m \lambda}}(\phi_{GJ},\theta_{GJ},\phi_{h}) d^{s}_{\lambda 0}(\theta_{h}) \langle l0s\lambda|J\lambda\rangle .
\label{eqn:isopart}
\end{equation} 
\end{minipage}}
\vspace{0.3cm}

For more than three particles in the final state, we keep adding isobars decaying into two particles (an isobar and a bachelor) until we obtain the desired number of final state particles. Each isobar decay will introduce a term 
(\ref{eqn:somedecay}) in the amplitude, i.e. an angular decay given by equation (\ref{eqn:MAINW}) and a mass term given by quation (\ref{eqn:masster}). In our case, the mass term is of a BW form but other parametrizations are possible -  see section~\ref{sect:MassDep}.
\subsection{The Reflectivity Basis}
\label{sect:reflect}

There are important constraints, based on conservation laws, that can be imposed. In particular, the strong interaction conserves parity, that is the parity operator commutes with the scattering matrix (or transition operator). Helicity states, however, are not eigenstates of parity.  The resonance spin density matrix operator in the helicity basis is

\begin{equation}  \widehat{\rho}_{out} = \widehat{T}\widehat{\rho}_{in}\widehat{T}^{\dagger} .\end{equation} 

Since the helicity states are not eigenstates of the parity, the spin density matrix will not explicitly show any symmetry associated with parity. We will find a basis where the resonance spin density matrix shows this explicit symmetry. The basis will be constructed from the eigenstates of a new operator called {\it reflection} \cite{SUCTRU}.
Let's call $\widehat{\Pi}$, the parity operator. In the canonical representation (and in the rest system of the particle) \cite{Werle, Goldberger} we have
\begin{equation}  \widehat{\Pi} |J m \rangle = P |J m \rangle \end{equation} 
where $P =\pm 1$ are the eigenvalues.
Let's consider a particle moving with momentum $\overrightarrow{p_{z}}$ in the $z$ direction. We can get this state by boosting (see Appendix A) the state at rest
\begin{equation}  |\overrightarrow{p_{z}} J m \rangle = \mathscr{L}(\overrightarrow{p_{z}})|0; J m \rangle . \end{equation} 
For a particle moving in the z direction $J=s$ and $m=\lambda$, therefore
\begin{equation}  \widehat{\Pi} |0;s \lambda \rangle = P |0;s \lambda \rangle \end{equation}
and
\begin{equation}  |\overrightarrow{p_{z}} s \lambda \rangle = \mathscr{L}(\overrightarrow{p_{z}})|0; s \lambda \rangle .\end{equation}
Applying the parity operator
\begin{equation} \widehat{\Pi} |\overrightarrow{p_{z}} s \lambda \rangle = \widehat{\Pi} \mathscr{L}(\overrightarrow{p_{z}})|0; s \lambda \rangle \end{equation}
\begin{equation} \widehat{\Pi} |\overrightarrow{p_{z}} s \lambda \rangle = P \mathscr{L}(-\overrightarrow{p_{z}})|0; s \lambda \rangle .\end{equation}
Since, to get back from $(-\overrightarrow{p_{z}})$ to $(\overrightarrow{p_{z}})$ we need a rotation of modulo $\pi$ around the $y$ axis
\begin{equation}  \mathscr{L}(\overrightarrow{p_{z}}) = e^{-i\pi J_{y}} \mathscr{L}(-\overrightarrow{p_{z}})e^{ i\pi J_{y}} \end{equation}
and we know~\cite{Richman} that
\begin{equation}  e^{-i\pi J_{y}}|\overrightarrow{p_{z}} s \lambda \rangle = (-1)^{s-\lambda} |\overrightarrow{p_{z}} s -\lambda \rangle 
\label{eqn:rotat}
\end{equation}
we finally have
\begin{equation} \widehat{\Pi} |\overrightarrow{p_{z}} s \lambda \rangle = P (-1)^{s-\lambda} e^{i\pi J_{y}} |\overrightarrow{p_{z}} s -\lambda \rangle .\end{equation}

Since any other direction can be constructed by rotation, and the parity operator commutes with rotations (in the x-z plane), we can express the former formula in particular, in the rest frame of the resonance (GJ) with the spin quantization in the z-axis given by $m$

\begin{equation} \widehat{\Pi} |J m \rangle = P (-1)^{J-m} e^{i\pi J_{y}} | J -m \rangle .
\label{eqn:R1}
\end{equation}

 It is useful then to define the following operator, the reflection operator \cite{SUCTRU}
\begin{equation}  \widehat{\Pi}_{y}=\widehat{\Pi} e^{-i\pi J_{y}}
\label{eqn:parity}
 \end{equation} 
that involves parity and a $\pi$ angular rotation around the $y$ axis in the Gottfried-Jackson (GJ) frame. It represents a mirror {\it reflection} through the production plane (x,z).  This operator commute with the transition operator. The $y$ axis in the GJ frame is perpendicular to the production plane, therefore the transition matrix is independent of $y$, and only the $x,z$ coordinates participate in the parity transformation. Reflection commutes with the Hamiltonian.  Let's write the states $|b \rangle$ as $ |a, m\rangle$ where $a$ includes all other quantum numbers except $m$. 
Using our previous result, equation (\ref{eqn:R1}), the reflection operator acting on these states produces
\begin{equation}  \widehat{\Pi}_{y} |a, m\rangle= e^{-i\pi J_{y}} \widehat{\Pi}|a, m\rangle = e^{-i\pi J_{y}}  P(-1)^{J-m} e^{i\pi J_{y}} |a,-m\rangle = P(-1)^{(J-m)}|a,-m\rangle
\label{eqn:refle}
 \end{equation} 
where $J$ is the total angular momentum and $P$ are the parity eigenvalues.
We can build the following eigenstates of $\widehat{\Pi}_{y}$ (since the reflection changes signs on the z-projection quantum numbers, $m$, we will create eingenstates that are linear combination of both (m) signs states with adequate coefficients)
{\footnote{The sign between both terms in equation (\ref{eqn:refdef}) is arbitrary. We use the sign in the original definition of reference~\cite{SUCTRU}. Notice that $\pi$ beam experiments had used another convention adopted by reference~\cite{SUC93}.}}
\begin{equation}   |\epsilon,a,m\rangle=\begin{bmatrix} |a,m\rangle + \epsilon P (-1)^{(J-m)}|a,-m\rangle \end{bmatrix}\Theta (m) 
\label{eqn:refdef}
\end{equation} 
where
\begin{equation}  \Theta(m)=\frac{1}{\sqrt{2}}, \ if\ m >0 \end{equation} 
\begin{equation}  \Theta(m)=\frac{1}{2}, \ if\ m =0 \end{equation} 
\begin{equation}  \Theta(m)=0, \ if\ m <0 \end{equation} 
such that
\begin{equation}  \widehat{\Pi}_{y} |\epsilon,a, m\rangle= K |\epsilon,a, m\rangle .\end{equation}
We have introduced the new quantum number $\epsilon$ called {\it reflectivity}.
We now calculate K applying a second reflection on ~(\ref{eqn:refle})
\begin{equation}   \widehat{\Pi}_{y} (\widehat{\Pi}_{y} |a, m\rangle)= \widehat{\Pi}_{y} (P (-1)^{(J-m)}|a,-m\rangle)= (P (-1)^{(J-m)})\widehat{\Pi}_{y}|a,-m\rangle \end{equation} 
then
\begin{equation}  \widehat{\Pi}_{y}^{2} |a, m\rangle= (P (-1)^{(J-m)})(P (-1)^{(J+m)}|a, m\rangle)=P^{2}(-1)^{2J}|a, m\rangle=(-1)^{2J}|a, m\rangle 
\label{eqn:eigen}
\end{equation} 
and
\begin{equation}  \widehat{\Pi}_{y}^{2} |a, -m\rangle = (-1)^{2J}|a, -m\rangle .
\end{equation} 
Therefore
\begin{equation}  \widehat{\Pi}_{y}^{2} |\epsilon, a, m\rangle = (-1)^{2J}|\epsilon, a, m\rangle
\end{equation} 
and
\begin{equation}  K^{2}=(-1)^{2J}. \end{equation} 

Normalizing the eigenstates such that:
\begin{equation}  \langle \epsilon',a',m'|\epsilon,a,m\rangle = \begin{bmatrix}\langle a',m'|a, m\rangle +\epsilon^{*}\epsilon P^{2}(-1)^{2(J-m)} \langle a',-m'|a,-m\rangle\end{bmatrix}\Theta^{2}(m) \end{equation} 
\begin{equation} [1 + \epsilon^{*}\epsilon] \times \frac{1}{2} = 1 \end{equation}
then
\begin{equation}  \epsilon^{*}\epsilon=1. \end{equation} 

The eigenvalues of the reflection operator can then be written as
\begin{equation} K = \epsilon (-1)^{2J} \end{equation}
where
\begin{equation}  \epsilon = \pm 1\  for\ J=0,1,2...\ (bosons) \end{equation} 
\begin{equation}  \epsilon = \pm i\  for\ J=1/2,3/2,...\ (fermions) \end{equation} 

Therefore, we write
\begin{equation}  \widehat{\Pi}_{y}  |\epsilon,a, m \rangle= \epsilon(-1)^{2J} |\epsilon,a, m\rangle . \end{equation} 
For bosons (ie. mesons):
\begin{equation}  \widehat{\Pi}_{y}  |\epsilon,a, m\rangle= \epsilon |\epsilon,a, m\rangle .
\label{eqn:defi}
\end{equation} 

Notice that since each state defined in the reflectivity basis includes $m$ and $-m$, in this basis the projections of the spin on the quantization axis, $m$, are replaced by $|m|$ (absolute value) and the reflectivity $\epsilon$ (that includes the sign). 

The inverse of this basis change is given by \cite{SUCTRU}

\begin{equation}  |a, m\rangle = \sum_{\epsilon = \pm 1} \begin{bmatrix} |\epsilon,a, m\rangle \Theta(m) +\epsilon^{*}P(-1)^{J+m}|\epsilon,a,-m\rangle \Theta(-m)\end{bmatrix} .
\end{equation}

We have now all the elements to obtain our goal: to make the production explicitly parity conserving by writing the resonance spin density matrix in the reflectivity basis. We will find the relation between the resonance density matrix (\ref{eqn:Xdens}) in the helicity basis and in the new reflectivity basis.  Using the definition of the states in the reflectivity basis and equation (\ref{eqn:refdef}) applied to each side of the spin density matrix, we obtain
\begin{equation*} 
^{\epsilon\epsilon'}\rho^{aa'}_{mm'} = \langle \epsilon',a',m'|\ {^{i,j}\rho^{aa'}_{mm'}}|\epsilon,a,m\rangle 
\end{equation*}
\begin{equation*} 
 \langle \epsilon',a',m'|\ {^{i,j}\rho^{aa'}_{mm'}}|\epsilon,a,m\rangle =
 [{^{i,j}\rho^{aa'}_{mm'}}+\epsilon^{*}\epsilon' PP'(-1)^{(m-m')} (-1)^{(J-J')}\ {^{i,j}\rho^{aa'}_{-m-m'} } \end{equation*}
\begin{equation} +\epsilon^{*} P(-1)^{(J-m)}\ {^{i,j}\rho^{aa'}_{-mm'}} +\epsilon'P'(-1)^{(J'-m')}\ {^{i,j}\rho^{aa'}_{m-m'} }] \Theta(m)\Theta(m') .
\label{eqn:longe}
\end{equation} 

If parity is conserved, using equation (\ref{eqn:R1}) 
\begin{equation}  {^{i,j}\rho^{aa'}_{mm'}} = \widehat{\Pi}\  {^{i,j}\rho^{aa'}_{mm'}}  \widehat{\Pi}^{\dagger} = PP' (-1)^{J-J'} (-1)^{m-m'} e^{i\pi J_{y}}\  {^{i,j}\rho^{aa'}_{-m-m'}}\ e^{-i\pi J'_{y}} \end{equation}
but, from equation (\ref{eqn:Xdens}), writing $\rho_{ij} =  | i \rangle \langle j |$, and using equation (\ref{eqn:rotat}), we have
\begin{equation} e^{i\pi J_{y}}\  {^{i,j}\rho^{aa'}_{-m-m'}}\ e^{-i\pi J'_{y}} = {^{i} V^{k}_{b}} e^{i\pi J_{y}}\ 
| i \rangle \langle j | e^{-i\pi J'_{y}} {^{j} V^{k *}_{b'}} = (-1)^{(j-i)}\ {^{i,j}\rho^{aa'}_{-m-m'}}
\end{equation}
and, therefore, the resonance density matrix will have the following symmetries
\begin{equation}  {^{ij}\rho^{aa'}_{mm'}}= \pm PP'(-1)^{(m-m')} (-1)^{(J-J')}\ {^{ij}\rho^{aa'}_{-m-m'} }
\label{eqn:fistre}
\end{equation}
where the positive sign is for $i=j$ and the negative for $i \neq j$ (remembering that $i,j =1,2$). 
Notice that the case of diagonal initial (photon) spin density matrix, ($i=j$), correspond to definite (pure) states or unpolarized photons.

It can be seen, after some algebra, that introducing the positive version of equation (\ref{eqn:fistre}) into equation (\ref{eqn:longe}), the right-hand side of (\ref{eqn:longe}) will vanish, $^{\epsilon\epsilon'}\rho^{aa'}_{mm'}=0$, for $\epsilon \neq \epsilon'$. Therefore, the resonance density matrix is non-zero only if  $\epsilon = \epsilon'$. The spin density matrix became block-diagonal for the case of unpolarized photons (or definite spin particles). In the other case, if  ($i \neq j$),  introducing the negative version of equation (\ref{eqn:fistre}) into equation (\ref{eqn:longe}), the right-hand side of (\ref{eqn:longe}) will vanish for $\epsilon = \epsilon'$, therefore, for this case the resonance spin density matrix contains only block off-diagonal elements. The interfering elements of the resonance spin density matrix come from the off-diagonal elements of the photon spin density matrix. 

In general (i.e polarized photon beams) the expression for the intensity will have diagonal and off-diagonal elements.  Equation (\ref{eqn:intenf}), in the reflectivity basis is then
\begin{equation} \boxed{   I(\tau) =  \sum_{k} \sum_{\epsilon,\epsilon'} \sum_{b,b'} {^{\epsilon}A_{b}}\ {^{\epsilon}V^{k}_{b}}\   \rho_{\epsilon,\epsilon'} {^{\epsilon'}V^{k *}_{b'}}\  {^{\epsilon'}A^{*}_{b'}(\tau)} } 
\label{eqn:totall}
\end{equation} 
where $b,b'$ represent all the partial waves quantum numbers.

For the case of $i=j$, i.e. unpolarized photons or definite spin particles (pions), the matrix becomes block diagonal in the reflectivity basis:

\begin{equation}  ^{\epsilon}\rho^{aa'}_{mm'}=\begin{pmatrix} ^{(+)}\rho^{aa'}_{mm'} & 0 \\ 0 & ^{(-)}\rho^{aa'}_{mm'} \end{pmatrix} . \end{equation}

In this case, an advantage of using the reflectivity basis is that, practically,  reduces by a factor of two the rank of the resonance spin density matrix. Then,
\begin{equation}  {^{\epsilon}\rho_{b,b'}} = \sum_{k} {^{\epsilon}V^{k}_{b}}\   {^{\epsilon}\rho_{\gamma}} {^{\epsilon}V^{k *}_{b'}} \end{equation} 
therefore
\begin{equation}  I(\tau) =  \sum_{\epsilon} \sum_{b,b'} {^{\epsilon}A_{b}(\tau)}\   {^{\epsilon}\rho_{b,b'}} {^{\epsilon}A^{*}_{b'}(\tau)} 
\end{equation} 
or
\begin{equation}   I(\tau) =  \sum_{k \epsilon} \sum_{b,b'} {^{\epsilon}A_{b}(\tau)}^{\epsilon}\ {^{\epsilon}V^{k}_{b}}\   {^{\epsilon}\rho_{\gamma}} {^{\epsilon}V^{k *}_{b'}} {^\epsilon}A^{*}_{b'}(\tau) 
\label{eqn:totall}
\end{equation} 
or
\begin{equation}  I(\tau) = \sum_{k} \sum_{b,b'} [{^{+}\rho_{b,b'}} {^{+}A_{b}(\tau)}^{+}A^{*}_{b'}(\tau)+ {^{-}\rho_{b,b'}} {^{-}A_{b}(\tau)}^{-}A^{*}_{b'}(\tau)] . \end{equation} 

For unpolarized photons ${^{+}\rho_{\gamma}} = {^{-}\rho_{\gamma}} = \frac{1}{2}$, therefore

\begin{equation}  I(\tau) = \frac{1}{2} \sum_{k} \sum_{b,b'} [{^{+}V^{k}_{b}} {^{+}V^{k*}_{b'}}\ {^{+}A_{b}(\tau)}\ {^{+}A^{*}_{b'}(\tau)}+ {^{-}V^{k}_{b}}{^{-}V^{k*}_{b'}}\ {^{-}A_{b}(\tau)}\ {^{-}A^{*}_{b'}(\tau)}] . \end{equation}

The sum involves non-interfering terms of the amplitudes when expressed in the reflectivity basis. The absence of the interfering terms of different reflectivities is a direct consequence of parity conservation.

We have seen, equation (\ref{eqn:isopart}), that the decay amplitudes are given by a combination of Wigner-D functions and Clebsch-Gordan coefficients. Only the Wigner-D functions will be affected by the change to the reflectivity basis, therefore to evaluate the amplitudes in this new basis we need to show how the Wigner D-functions, $D^{J *}_{m \lambda}(\phi_{GJ},\theta_{GJ},\phi_{h})$, are affected by the reflectivity operator, i.e. how can we write those functions on the reflectivity basis?
We will start from

\begin{equation}   |\epsilon,a,m\rangle=\begin{bmatrix} |a,m\rangle + \epsilon P (-1)^{(J -m)}|a, -m\rangle \end{bmatrix}\Theta (m) .\end{equation} 
The Wigner D-functions in the reflectivity basis follow a similar relation
\begin{equation}  ^\epsilon D _{m\lambda}^{J *}(\phi_{GJ},\theta_{GJ},\phi_{h})=\Theta(m)\begin{bmatrix} D _{m \lambda}^{J *}(\phi_{GJ},\theta_{GJ},\phi_{h})+\epsilon P(-1)^{J-m} D _{-m\lambda}^{J\ *}(\phi_{GJ},\theta_{GJ},\phi_{h})\end{bmatrix} . \end{equation} 

%Using that 
%
%\begin{equation}  D^{J\ *}_{m\lambda} (\phi_{GJ},\theta_{GJ},\phi_{h})=(-1)^{m-\lambda} D^{J}_{-m-\lambda}
%(\phi_{GJ},\theta_{GJ},\phi_{h}) 
%\label{eqn:duax}
%\end{equation} 
%
%we obtain
%
%\begin{equation*}  ^\epsilon D _{m\lambda}^{J *}(\phi_{GJ},\theta_{GJ},\phi_{h})=\Theta(m) \end{equation*}
%\begin{equation} \times \begin{bmatrix} D _{m \lambda}^{J *}(\phi_{GJ},\theta_{GJ},\phi_{h})+\epsilon P %(-1)^{J-m} (-1)^{-m-\lambda} D _{m-\lambda}^{J}(\phi_{GJ},\theta_{GJ},\phi_{h})\end{bmatrix} \end{equation} 
%

For resonances with $m = 0$, we have
%
%\begin{equation}  D_{0 \lambda}^{J\ *}(\phi_{GJ},\theta_{GJ},\phi_{h}) = (-1)^{\lambda}D _{0 -\lambda}^{J}
%(\phi_{GJ},\theta_{GJ},\phi_{h}) \end{equation} 
%
% we have
%
\begin{equation}  ^\epsilon D _{0\lambda}^{J *}(\phi_{GJ},\theta_{GJ},\phi_{h})=\frac{1}{2}\begin{bmatrix}1+\epsilon P (-1)^{J}\end{bmatrix}  D _{0 \lambda}^{J *}(\phi_{GJ},\theta_{GJ},\phi_{h})
\label{eqn:natu}
\end{equation}
therefore, for $m=0$ only one value of $\epsilon$ is possible.  Only values of $\epsilon P (-1)^{J}=1$ produce non-zero states. For even (odd) total angular momentum, we have $\epsilon = P$ ($\epsilon = -P$).

Let's consider the case of a resonance decaying into a spinless (pseudo-scalar)  bachelor plus an isobar decaying into two spinless (pseudo-scalar) particles. Using that for this case $P=(-1)^{J+1}$ and that

\begin{equation}  D^{J\ *}_{m\lambda} (\phi_{GJ},\theta_{GJ},\phi_{h})=(-1)^{m-\lambda} D^{J}_{-m-\lambda}
(\phi_{GJ},\theta_{GJ},\phi_{h}) 
\label{eqn:duax}
\end{equation} 

we have

\begin{equation*}  ^\epsilon D _{m\lambda}^{J *}(\phi_{GJ},\theta_{GJ},\phi_{h})=\Theta(m) \end{equation*}
\begin{equation} \times \begin{bmatrix} D _{m \lambda}^{J *}(\phi,\theta,0)-\epsilon (-1)^{2J}  (-1)^{-2m} (-1)^{-\lambda} D _{m-\lambda}^{J}(\phi_{GJ},\theta_{GJ},\phi_{h})\end{bmatrix} \end{equation} 
therefore, for bosons we have
\begin{equation}  ^\epsilon D _{m\lambda}^{J *}(\phi_{GJ},\theta_{GJ},\phi_{h})=\Theta(m)\begin{bmatrix} D _{m \lambda}^{J *}(\phi_{GJ},\theta_{GJ},\phi_{h})-\epsilon (-1)^{\lambda} D _{m-\lambda}^{J}(\phi_{GJ},\theta_{GJ},\phi_{h})\end{bmatrix}. 
\label{eqn:dref}
\end{equation}

For resonances with $m = 0$, since
\begin{equation}  D_{0 \lambda}^{J\ *}(\phi_{GJ},\theta_{GJ},\phi_{h}) = (-1)^{\lambda}D _{0 -\lambda}^{J}(\phi_{GJ},\theta_{GJ},\phi_{h}) \end{equation} 
using equation (\ref{eqn:dref}), we have
\begin{equation}  ^\epsilon D _{0\lambda}^{J *}(\phi_{GJ},\theta_{GJ},\phi_{h})=\frac{1}{2}\begin{bmatrix}1-\epsilon\end{bmatrix}  D _{0 \lambda}^{J *}(\phi_{GJ},\theta_{GJ},\phi_{h})\end{equation}
therefore, for $m=0$ and an isobar decaying into two spinless pseudo-scalars, there is only one possible state with $\epsilon = -1$.

Let's consider the case of a resonance decaying into two spinless final state particles. 
Taken the GJ subindexes from the notation  and taken $\lambda$ = 0 in equation (\ref{eqn:dref})

\begin{equation}   ^\epsilon D _{m0}^{l *}(\phi,\theta,0)=\Theta(m)\begin{bmatrix} D _{m0}^{l *}(\phi,\theta,0)-\epsilon D _{m0}^{l}(\phi,\theta,0)\end{bmatrix} 
\label{eqn:dofref}
\end{equation} 
since:
\begin{equation}  D _{m0}^{l *}(\phi,\theta,0) = d _{m0}^{l}(\theta) e^{im\phi} \end{equation}
and
\begin{equation}  D _{m0}^{l}(\phi,\theta,0) = d _{m0}^{l}(\theta) e^{-im\phi} \end{equation}
for $\epsilon = -1$, we have
\begin{equation}   ^{(-)} D _{m0}^{l *}(\phi,\theta,0)=\Theta(m) d _{m0}^{l}(\theta)\begin{bmatrix}e^{-im\phi} + e^{im\phi}\end{bmatrix} = 2\Theta(m) d _{m0}^{l}(\theta) cos (m\phi) \end{equation} 
 that is real, and for $\epsilon = +1$, we have
\begin{equation}   ^{(+)} D _{m0}^{l *}(\phi,\theta,0)=\Theta(m) d _{m0}^{l}(\theta)\begin{bmatrix}e^{im\phi} - e^{-im\phi}\end{bmatrix} = 2i\Theta(m) d _{m0}^{l}(\theta) sin (m\phi) \end{equation} 
that is imaginary.
Using equation (\ref{eqn:twopart}) (without the BW factors for this argument)

\begin{equation}
^{(-)} A_{lm} (\phi,\theta) = \sqrt{\frac{2l+1}{4\pi}} 2\Theta(m) d _{m0}^{l}(\theta) cos (m\phi)
\end{equation}

\begin{equation}
 ^{(+)} A_{lm} (\phi,\theta) = \sqrt{\frac{2l+1}{4\pi}} 2i\Theta(m) d _{m0}^{l}(\theta) sin (m\phi)
\label{eqn:spinless}
\end{equation}
that shows, again, that for m=0 only the $^{(-)} A_{l0} (\phi,\theta)$ are non-zero.

A state is said to have natural parity if $P=(-1)^{J}$, while is said to have unnatural parity if $P=-(-1)^{J}$. We can recast this definition by introducing the {\it naturality} of the exchanged particle, $\mathcal{N}$, as

\begin{equation} \mathcal{N} = P \times (-1)^{J} .\end{equation}

Therefore, naturality is $\mathcal{N}=+1$ (natural) for $J^{P}=0^{+},1^{-},2^{+}, \cdots$ and $\mathcal{N}=-1$ (unnatural) for $J^{P}=0^{-},1^{+},2^{-}, \cdots$.

Let's recall equation (\ref{eqn:inten})

\begin{equation}  I(\tau) =   \sum_{ext.\ spins} \sum_{i,j} \langle f | \widehat{T} \ \rho_{i,j} \widehat{T}^{\dagger} |f\rangle
\label{eqn:total} 
\end{equation} 
and separate the transition operator in two terms~\cite{Sch70}, one corresponding to the exchange of a natural particle, $\widehat{T}^{N}$ and another, $\widehat{T}^{U}$, corresponding to the exchange of an unnatural particle. Therefore
\begin{equation} \widehat{T}= \begin{pmatrix} \widehat{T}^{N} \\ \widehat{T}^{U} \end{pmatrix}. \end{equation}

The photon spin density matrix in the reflectivity basis, calculated in section~\ref{sect:SpinDen}, has the form
\begin{equation} 
\rho _{i,j} \left(\mathscr{P},\alpha \right)={\raise0.7ex\hbox{$ 1 $}\!\mathord{\left/{\vphantom{1 2}}\right.\kern-\nulldelimiterspace}\!\lower0.7ex\hbox{$ 2 $}} \left(\begin{array}{cc} {1+\mathscr{P}\cos 2\alpha } & {i\mathscr{P}\sin 2\alpha } \\ {-i\mathscr{P}\sin 2\alpha } & {1-\mathscr{P}\cos 2\alpha } \end{array}\right) 
\end{equation} 
where $\mathscr{P}$ is the partial polarization and $\alpha$ the angle between the electric field direction and the production plane. Including these values in equation (\ref{eqn:total})
\begin{equation}  I(\tau) =   \sum_{ext.\ spins} \langle f |  \begin{pmatrix} \widehat{T}^{N} \\ \widehat{T}^{U} \end{pmatrix}   {\raise0.7ex\hbox{$ 1 $}\!\mathord{\left/{\vphantom{1 2}}\right.\kern-\nulldelimiterspace}\!\lower0.7ex\hbox{$ 2 $}} \left(\begin{array}{cc} {1+\mathscr{P}\cos 2\alpha } & {i\mathscr{P}\sin 2\alpha } \\ {-i\mathscr{P}\sin 2\alpha } & {1-\mathscr{P}\cos 2\alpha } \end{array}\right)  \begin{pmatrix} \widehat{T}^{N \dagger} & \widehat{T}^{U \dagger} \end{pmatrix} |f\rangle 
\end{equation} 
therefore
\begin{equation*}
 I(\tau) \propto (1+\mathscr{P}cos (2 \alpha) ) | \widehat{T}^{N}|^{2} +(i\mathscr{P}sin(2 \alpha) ) \widehat{T}^{N} \widehat{T}^{U \dagger} \end{equation*}
\begin{equation} +(-i\mathscr{P}sin( 2\alpha) ) \widehat{T}^{U} \widehat{T}^{N \dagger}+(1-\mathscr{P}cos( 2\alpha )) | \widehat{T}^{U}|^{2} . \end{equation}
The interference terms between natural and unnatural parity exchanges vanished in the limit of high energies (beam energies of above 5 GeV)~\cite{Sch70, Cohen68}, therefore
\begin{equation}
 I(\tau) \propto (1+\mathscr{P}cos (2 \alpha) ) | \widehat{T}^{N}|^{2} +(1-\mathscr{P}cos( 2\alpha )) | \widehat{T}^{U}|^{2} .
\label{eqn:mix}
\end{equation}

Consider a fully linearly polarized photon beam, $\mathscr{P}=1$, if the photon polarization is perpendicular to the production plane, $\alpha=0$, we have
\begin{equation}
 I(\tau) \propto  | \widehat{T}^{N}|^{2}  \end{equation}
and , if the photon polarization is parallel to the production plane, $\alpha=\frac{\pi}{2}$, we have
\begin{equation} I(\tau) \propto  | \widehat{T}^{U}|^{2} . \end{equation}

This result is called the {\it Stichel theorem}~\cite{Stichel,Bajpai} which states that only natural (unnatural) parity exchange contributes to the polarized cross-section, when the photon polarization is perpendicular (parallel) to the production plane. In the general case, through equation (\ref{eqn:mix}), knowing the photon polarization, $\mathscr{P}$ and $\alpha$, we have accesses to the exchange particle naturality.

The naturality of the exchanged particle is related to the reflectivity of the produced resonance. From equation (\ref{eqn:eigen}), we find that

\begin{equation}  \widehat{\Pi}_{y}^{2} |a, m\rangle= \big{[} \mp P(-1)^{J} \big{]}^{2} |a, m\rangle 
\end{equation} 
that, for bosons, is also
\begin{equation}  \widehat{\Pi}_{y}^{2} |a, m\rangle= \epsilon^{2} |a, m\rangle 
\end{equation} 
then
\begin{equation}  \epsilon = P(-1)^{J} \end{equation} 
the reflectivity coincides with the naturality of the resonance. Reflection is a conserved quantum number, since both rotation and parity are conserved. Therefore, the product of the initial beam reflectivity and the exchange particle reflectivity must equal the reflectivity of the resonance (and then, so do the naturalities):

\begin{equation} \epsilon_{beam} \times \epsilon_{ex} = \epsilon_{R} .\end{equation}

For a pion beam where ($J=0$, $m=0$ and $P=-1$), there is only one reflectivity value, $\epsilon = -1$. Therefore, for a positive resonance's reflectivity the exchange particle belongs to an unnatural parity Regge trajectory (i.e. a pion), and for negative resonance's reflectivity the exchange particle belong to a natural parity Regge trajectory (i.e. a $\rho$).

 For a photon beam the two values of the reflectivity are possible, therefore, the reflectivity of the resonance and the naturality of the exchange are in principle not directly related (see section~\ref{sect:SpinDen}).

The photon spin density matrix in the reflectivity basis, represents a general mix state of the photon  (see section~\ref{sect:SpinDen}). However, for full polarization $\mathscr{P}=1$, there are two states corresponding to eigestates of reflectivity. For $\alpha = 0$ ($\epsilon=+1$) and for  $\alpha = \frac{\pi}{2}$ ($\epsilon=-1$ ), as was demonstrated by the Stichel theorem. Therefore, using linearly polarized photons at those explicit configurations we could determine the naturality of the exchange particle. In the case of pion exchange (or other Regge unnatural trajectory particle) the reflectivity of the resonance is opposite to that of the photon. In the case of $\rho$ exchange (or other Regge natural trajectory particle), the reflectivity of the resonance and the photon will be the same.

\section{Spin Density Matrices of Linearly Polarized Photons and Virtual Photons}
\label{sect:SpinDen}
\subsection{Photoproduction}

Consider a real photon beam prepared in a linearly polarized state. Since the photon wave is transverse (Lorentz condition), any polarization state will be in a plane perpendicular to the direction of the photon momentum. Therefore, the polarization of real photons will have two possible pure spin states (let's call them: up and down, $+1$ or $-1$ ). In a classical view, the $+1$ will coincide with the electric field direction. Any polarization direction can be represented by the superposition of these two orthogonal (pure) states contained in the transverse plane. Let's take these two states to be $|P_{1} \rangle $, in the direction of a pure state ( the direction of the electric vector of the incoming photon) and $|P_{2} \rangle $ orthogonal to the $|P_{1} \rangle $ state, as the basis.

Consider the reaction 
\begin{equation} \gamma N\to XN'\end{equation}  
where $N,N'$ are nucleons and $X$ is a mesonic resonance. Recall from section~\ref{sect:Model} the form of the scattering amplitude  $\mathscr{M}$, represented by the transition operator, $\widehat{T}$, given by
\begin{equation}  \mathscr{M}= \langle out | \widehat{T} |in\rangle \end{equation} 
and then
\begin{equation}  I(\tau) =  \sum_{ext.-spins} |\mathscr{M}|^{2} =  \sum_{ext.-spins} \langle out | \widehat{T} \widehat{\rho_{in}} \widehat{T}^{\dagger}  |out\rangle = \sum_{ext.-spins} \sum_{i,j}\langle out | \widehat{T} \ \rho_{i,j} \widehat{T}^{\dagger}  |out\rangle .
\label{eqn:matscat}
\end{equation}

The operator $|in\rangle \langle in | $ was defined as the initial spin density matrix operator, $\widehat{\rho_{in}}$, where the indices run over initial spins

\begin{equation}  \widehat{\rho_{in}} = |in \rangle \langle in | = \sum_{i,j} \rho_{i,j} .\end{equation}

If we include the spin information of the target in the sum over external spins, then $ \widehat{\rho_{in}}$ includes only the beam spin information and it is defined as {\it the spin density matrix of the incoming photon} $ \widehat{\rho_{\gamma}}$. 

In the $|P_{1} \rangle ,|P_{2} \rangle $ basis, the spin density matrix of a mixed polarization state (superposition of two pure polarization states), can be written in the formal notation, where W${}_{1}$ and W${}_{2}$ are the weights for each state~\cite{Blum}
 
\begin{equation} \label{GrindEQ__2_} 
\widehat{\rho_{\gamma}} = \sum_{i,j}  \rho_{i,j} =W_{1} |P_{1} \rangle \langle P_{1} |+W_{2} |P_{2} \rangle \langle P_{2} | .
\end{equation} 

Consider $N$ beam photons. The meaning of (\ref{GrindEQ__2_}) is that, when the beam polarization is measured, we will find $N{}_{1}$${}_{ }$ photons polarized in the state with amplitude  $\langle P_{1} |\widehat{\rho_{\gamma}}|P_{1} \rangle $, and $N{}_{2}$ in the state with amplitude $\langle P_{2} |\widehat{\rho_{\gamma}} |P_{2} \rangle $ such that $N{}_{1}$ + $N{}_{2}$${}_{ }$ =$N$. Let's assume that $N_{1} \geq N_{2}$, i.e. the index one corresponding to the $\overrightarrow{OX}$ axis is assigned to the maximum number of photons (assuming the opposite will produce a change of signs in the formulas with no physical consequences).
We define the {\it partial polarization} (or {\it degree of polarization}), $\mathscr{P}$, such that

\begin{equation} \label{GrindEQ__3_} 
\mathscr{P}=\frac{N_{1} -N{}_{2} }{N} . 
\end{equation} 

Notice that $0 \leq \mathscr{P} \leq 1$, corresponding to the magnitude of the {\it polarization vector}, 
$\overrightarrow{\mathscr{P}}$, that is generally defined (in the helicity basis)~\cite{Sch70} by the identity

\begin{equation} \widehat{\rho_{\gamma}}  = \frac{1}{2} I + \frac{1}{2} \overrightarrow{\mathscr{P}} \cdot \sigma\end{equation}

where $I$ is the unit matrix ($2 \times 2$), and the $\sigma_{i}$ are the three Pauli matrices. In this expression we write the photon spin density matrix in a complete set from the space of $2 \times 2$ hermitian matrices. 

If all photons are found in the $\langle P_{1} |\widehat{\rho_{\gamma}} |P_{1} \rangle $  state, $N{}_{1}=N$ then $\mathscr{P} =1$ (full polarization), if all beam particles are found equally distributed between  $\langle P_{1} |\widehat{\rho_{\gamma}} |P_{1} \rangle $ and $\langle P_{2} |\widehat{\rho_{\gamma}} |P_{2} \rangle $ (no polarization), $\mathscr{P} =0$.

We can now calculate the weights in \eqref{GrindEQ__2_} using the interpretation of probabilities as frequencies
\begin{equation}  W_{1} =\frac{N_{1} }{N} \ and\ W_{2} =\frac{N_{2} }{N} \end{equation} 
and solving the system
\begin{equation} \begin{array}{l} {N=N_{1} +N_{2} } \\ {} \\ {\mathscr{P}N=N_{1} -N_{2} } \end{array}\end{equation}  
we obtain
\begin{equation}    W_{1} =\frac{N_{1} }{N} =\frac{\left(1+\mathscr{P}\right)}{2} \  and\  W_{2} =\frac{N_{2} }{N} =\frac{\left(1-\mathscr{P}\right)}{2} \end{equation} 
therefore
\begin{equation} \rho_{\gamma} =\frac{(1+\mathscr{P})}{2} |P_{1} \rangle \langle P_{1} |+\frac{(1-\mathscr{P})}{2} |P_{2} \rangle \langle P_{2} |\end{equation}  
and in matrix form:
\begin{equation} \label{GrindEQ__4_} 
\rho_{i,j} \left(\mathscr{P}\right)={\raise0.7ex\hbox{$ 1 $}\!\mathord{\left/{\vphantom{1 2}}\right.\kern-\nulldelimiterspace}\!\lower0.7ex\hbox{$ 2 $}} \left(\begin{array}{cc} {1+\mathscr{P}} & {0} \\ {0} & {1-\mathscr{P}} \end{array}\right) .
\end{equation} 

We will now calculate this matrix in three different bases: the Gottfried-Jackson (GJ) basis, the helicity basis, and finally in the reflectivity basis.

The Gottfried-Jackson (GJ) frame is defined in Appendix A. To transform the matrix from the ($|P_{1} \rangle $ and $|P_{2} \rangle $) basis, as calculated in \eqref{GrindEQ__4_}, to the GJ basis ($|x\rangle $ and $|y\rangle $) we need to perform a rotation about the z axis (beam) of the form

\begin{equation} \label{GrindEQ__5_} 
\begin{array}{l} {|x\rangle =\cos \alpha |P_{1} \rangle +\sin \alpha |P_{2} \rangle } \\ {} \\ {|y\rangle =-\sin \alpha |P_{1} \rangle +\cos \alpha |P_{2} \rangle } \end{array} 
\end{equation} 
where $\alpha $ is the angle between (from) the polarization vector $|P_{1} \rangle$ and the production plane, which is the $x$ axis of the GJ frame.

In the GJ basis, the new matrix is
\begin{equation} \label{GrindEQ__51_} 
\rho_{x,y} \left(\mathscr{P}\right)= \left(\begin{array}{cc} |x \rangle \langle x|& |x \rangle \langle y| \\ |y \rangle \langle x| & |y \rangle \langle y| \end{array}\right) .
\end{equation} 
We can calculate each element of the matrix using \eqref{GrindEQ__5_} 
\begin{equation} |x\rangle \langle x|= (\cos \alpha |P_{1} \rangle +\sin \alpha |P_{2} \rangle )(\cos \alpha \langle P_{1} |+\sin \alpha \langle P_{2} |) \end{equation}
or
\begin{equation} {|x\rangle \langle x| =\cos ^{2} \alpha |P_{1} \rangle \langle P_{1} |+\sin \alpha \cos \alpha |P_{1} \rangle \langle P_{2} |+\sin \alpha \cos \alpha |P_{2} \rangle \langle P_{1} |+\sin ^{2} \alpha |P_{2} \rangle \langle P_{2} |} \end{equation}
and using the values obtained in \eqref{GrindEQ__4_}
 \begin{equation} |x\rangle \langle x| =\frac{(1+\mathscr{P})}{2} \cos ^{2} \alpha +\frac{(1-\mathscr{P})}{2} \sin ^{2} \alpha . \end{equation} 

In the same way

\begin{equation} |y\rangle \langle y| = (-\sin \alpha |P_{1} \rangle +\cos \alpha |P_{2} \rangle ) (-\sin \alpha \langle P_{1} |+\cos \alpha \langle P_{2} |)  \end{equation}
or
\begin{equation} |y\rangle \langle y|=\sin ^{2} \alpha |P_{1} \rangle \langle P_{1} |-\sin \alpha \cos \alpha |P_{1} \rangle \langle P_{2} |-\sin \alpha \cos \alpha |P_{2} \rangle \langle P_{1} |+\cos ^{2} \alpha |P_{2} \rangle \langle P_{2} |\end{equation}  
therefore
\begin{equation} |y\rangle \langle y| = \frac{(1+\mathscr{P})}{2} \sin ^{2} \alpha +\frac{(1-\mathscr{P})}{2} \cos ^{2} \alpha . \end{equation} 

The off-diagonal elements are (both elements are the same)

\begin{equation} \begin{array}{l} {|x\rangle \langle y|=\left[\left(\cos \alpha |P_{1} \rangle +\sin \alpha |P_{2} \rangle \right)\left(-\sin \alpha \langle P_{1} |+\cos \alpha \langle P_{2} |\right)\right]} \\ {=-\sin \alpha \cos \alpha |P_{1} \rangle \langle P_{1} |+\cos ^{2} \alpha |P_{1} \rangle \langle P_{2} |-\sin ^{2} \alpha |P_{2} \rangle \langle P_{1} |+\sin \alpha \cos \alpha |P_{2} \rangle \langle P_{2} |}  \end{array}\end{equation}  
therefore
\begin{equation} |x\rangle \langle y|= -\frac{(1+\mathscr{P})}{2} \sin \alpha \cos \alpha +\frac{(1-\mathscr{P})}{2} \sin \alpha \cos \alpha .\end{equation}

In matrix form

\begin{equation*} \rho _{GJ} \left(\mathscr{P},\alpha \right)= \end{equation*}
\begin{equation} \left(\begin{array}{cc} {\frac{\left(1+\mathscr{P}\right)}{2} \cos ^{2} \alpha +\frac{\left(1-\mathscr{P}\right)}{2} \sin ^{2} \alpha } & {-\frac{\left(1+\mathscr{P}\right)}{2} \sin \alpha \cos \alpha +\frac{\left(1-\mathscr{P}\right)}{2} \cos \alpha \sin \alpha } \\ {-\frac{(1+\mathscr{P})}{2} \cos \alpha \sin \alpha +\frac{\left(1-\mathscr{P}\right)}{2} \sin \alpha \cos \alpha } & {\frac{\left(1+\mathscr{P}\right)}{2} \sin ^{2} \alpha +\frac{\left(1-\mathscr{P}\right)}{2} \cos ^{2} \alpha } \end{array}\right) . \end{equation}  

After some algebra, the spin density matrix of the photon in the GJ basis becomes:

\begin{equation} \label{GrindEQ__6_} 
\rho _{GJ} \left(\mathscr{P},\alpha \right)={\raise0.7ex\hbox{$ 1 $}\!\mathord{\left/{\vphantom{1 2}}\right.\kern-\nulldelimiterspace}\!\lower0.7ex\hbox{$ 2 $}} \left(\begin{array}{cc} {1+\mathscr{P}\cos 2\alpha } & {-\mathscr{P}\sin 2\alpha } \\ {-\mathscr{P}\sin 2\alpha } & {1-\mathscr{P}\cos 2\alpha } \end{array}\right) .
\end{equation}

Next, we transform the matrix to the helicity basis.
We start by using the relations between the GJ basis and the helicity basis given by \cite{Halzen}

\begin{equation} \label{GrindEQ__7_} 
\begin{array}{l} {|\lambda =+1\rangle =-\frac{1}{\sqrt{2} } (|x\rangle +i|y\rangle )} \\ {} \\ {|\lambda =-1\rangle =\frac{1}{\sqrt{2} } (|x\rangle -i|y\rangle )} \end{array} 
\end{equation} 
and using the relations in \eqref{GrindEQ__5_}, we obtain         
\begin{equation}                       \begin{array}{l} {|\lambda =+1\rangle =-\frac{1}{\sqrt{2} } \left[\left(\cos \alpha |P_{1} \rangle +\sin \alpha |P_{2} \rangle \right)+i\left(-\sin \alpha |P_{1} \rangle +\cos \alpha |P_{2} \rangle \right)\right]} \\ {=-\frac{1}{\sqrt{2} } \left[\left(\cos \alpha -i\sin \alpha \right)|P_{1} \rangle +\left(\sin \alpha +i\cos \alpha \right)|P_{2} \rangle \right]}  \end{array} 
\end{equation} 
therefore
\begin{equation}   |\lambda =+1\rangle = -\frac{1}{\sqrt{2} } \left(e^{-i\alpha } |P_{1} \rangle +ie^{-i\alpha } |P_{2} \rangle \right) . 
\label{eqn:HELp}
\end{equation}
The other state of helicity is
\begin{equation}                       \begin{array}{l} {|\lambda =-1\rangle =\frac{1}{\sqrt{2} } \left[\left(\cos \alpha |P_{1} \rangle +\sin \alpha |P_{2} \rangle \right)-i\left(-\sin \alpha |P_{1} \rangle +\cos \alpha |P_{2} \rangle \right)\right]} \\ {=\frac{1}{\sqrt{2} } \left[\left(\cos \alpha +i\sin \alpha \right)|P_{1} \rangle +\left(\sin \alpha -i\cos \alpha \right)|P_{2} \rangle \right]}  \end{array} 
\end{equation}
therefore
\begin{equation}  |\lambda =-1\rangle =\frac{1}{\sqrt{2} } \left(e^{i\alpha } |P_{1} \rangle -ie^{i\alpha } |P_{2} \rangle \right) . 
\label{eqn:HELm}
\end{equation}

Then, we can calculate the elements of the new matrix:
\begin{equation} \begin{array}{l} {|\lambda =+1\rangle \langle \lambda =+1|=\left[(-\frac{1}{\sqrt{2} } )(e^{-i\alpha } |P_{1} \rangle +ie^{-i\alpha } |P_{2} \rangle )(-\frac{1}{\sqrt{2} } )(e^{i\alpha } \langle P_{1} |-ie^{i\alpha } \langle P_{2} |)\right]} \\ = {\raise0.7ex\hbox{$ 1 $}\!\mathord{\left/{\vphantom{1 2}}\right.\kern-\nulldelimiterspace}\!\lower0.7ex\hbox{$ 2 $}} \left(|P_{1} \rangle \langle P_{1} |+|P_{2} \rangle \langle P_{2} |\right)  \end{array}\end{equation}  
and using the relations in \eqref{GrindEQ__5_} we obtain 
\begin{equation} |\lambda =+1\rangle \langle \lambda =+1| ={\raise0.7ex\hbox{$ 1 $}\!\mathord{\left/{\vphantom{1 2}}\right.\kern-\nulldelimiterspace}\!\lower0.7ex\hbox{$ 2 $}} \left(1/2+(\mathscr{P}/2)+1/2-(\mathscr{P}/2)\right)=1/2 .\end{equation}
We can find the other diagonal element 
\begin{equation} \begin{array}{l} {|\lambda =-1\rangle \langle \lambda =-1|=\left[(\frac{1}{\sqrt{2} } )(e^{i\alpha } |P_{1} \rangle -ie^{i\alpha } |P_{2} \rangle )(\frac{1}{\sqrt{2} } )(e^{-i\alpha } \langle P_{1} |+ie^{-i\alpha } \langle P_{2} |)\right]} \\ = {\raise0.7ex\hbox{$ 1 $}\!\mathord{\left/{\vphantom{1 2}}\right.\kern-\nulldelimiterspace}\!\lower0.7ex\hbox{$ 2 $}} \left(|P_{1} \rangle \langle P_{1} |+|P_{2} \rangle \langle P_{2} |\right)  \end{array}\end{equation}  
and using the relations in \eqref{GrindEQ__5_} we obtain
\begin{equation} |\lambda =-1\rangle \langle \lambda =-1| ={\raise0.7ex\hbox{$ 1 $}\!\mathord{\left/{\vphantom{1 2}}\right.\kern-\nulldelimiterspace}\!\lower0.7ex\hbox{$ 2 $}} \left(1/2+(\mathscr{P}/2)+1/2-(\mathscr{P}/2)\right)=1/2 .\end{equation}

The off-diagonal terms are:
\begin{equation} \begin{array}{l} {|\lambda =-1\rangle \langle \lambda =+1|=\left[\frac{1}{\sqrt{2} } (e^{i\alpha } |P_{1} \rangle -ie^{i\alpha } |P_{2} \rangle )(-\frac{1}{\sqrt{2} } )(e^{i\alpha } \langle P_{1} |-ie^{i\alpha } \langle P_{2} |)\right]} \\ {=-{\raise0.7ex\hbox{$ 1 $}\!\mathord{\left/{\vphantom{1 2}}\right.\kern-\nulldelimiterspace}\!\lower0.7ex\hbox{$ 2 $}} \left(e^{2i\alpha } |P_{1} \rangle \langle P_{1} |-e^{2i\alpha } |P_{2} \rangle \langle P_{2} |\right)} \\ {=-{\raise0.7ex\hbox{$ 1 $}\!\mathord{\left/{\vphantom{1 2}}\right.\kern-\nulldelimiterspace}\!\lower0.7ex\hbox{$ 2 $}} e^{2i\alpha } \left(1/2+(\mathscr{P}/2)-1/2+(\mathscr{P}/2)\right)= -(\mathscr{P}/2) e^{2i\alpha } } \end{array}\end{equation}  
and
\begin{equation} \begin{array}{l} {|\lambda =+1\rangle \langle \lambda =-1|=\left[-\frac{1}{\sqrt{2} } (e^{-i\alpha } |P_{1} \rangle +ie^{-i\alpha } |P_{2} \rangle )(\frac{1}{\sqrt{2} } )(e^{-i\alpha } \langle P_{1} |+ie^{-i\alpha } \langle P_{2} |)\right]} \\ {=-{\raise0.7ex\hbox{$ 1 $}\!\mathord{\left/{\vphantom{1 2}}\right.\kern-\nulldelimiterspace}\!\lower0.7ex\hbox{$ 2 $}} \left(e^{-2i\alpha } |P_{1} \rangle \langle P_{1} |-e^{-2i\alpha } |P_{2} \rangle \langle P_{2} |\right)} \\ {=-{\raise0.7ex\hbox{$ 1 $}\!\mathord{\left/{\vphantom{1 2}}\right.\kern-\nulldelimiterspace}\!\lower0.7ex\hbox{$ 2 $}} e^{-2i\alpha } \left(1/2+(\mathscr{P}/2)-1/2+(\mathscr{P}/2)\right)=-(\mathscr{P}/2) e^{-2i\alpha } }. \end{array}\end{equation}

The spin density matrix of the photon in the helicity basis is then
\begin{equation} \label{GrindEQ__9_} 
\rho _{\lambda \lambda '} \left(\mathscr{P},\alpha \right)={\raise0.7ex\hbox{$ 1 $}\!\mathord{\left/{\vphantom{1 2}}\right.\kern-\nulldelimiterspace}\!\lower0.7ex\hbox{$ 2 $}} \left(\begin{array}{cc} {1} & {-\mathscr{P}e^{-2i\alpha } } \\ {-\mathscr{P}e^{2i\alpha } } & {1} \end{array}\right) 
\end{equation} 
(in agreement with reference~\cite{Sch70}).
\\
Let's consider an example of its application to calculate the intensity in the helicity frame. Using equation (\ref{eqn:intenf}), but considering only rank one (k=1) we have that the intensity is

\begin{equation}  I(\tau) = \sum_{i,j=+,-} \sum_{b,b'} {^{i}A_{b}(\tau)} {^{i}V_{b}(\tau)} \rho_{i,j}{^{j}V^{*}_{b'}(\tau)}  {^{j}A^{*}_{b'}(\tau)} .
\end{equation} 
Using the notation $|L\rangle = |\lambda=+1\rangle$ (for left-handed) and $|R\rangle = |\lambda=-1\rangle$ (for right handed) states and considering the spin density matrix in its operator form we have
\begin{equation}  \sum_{R,L} \rho _{i,j} \left(\mathscr{P},\alpha \right)= \frac{1}{2} [|R\rangle\langle R|-\mathscr{P}e^{-2i\alpha } |R\rangle\langle L|-\mathscr{P}e^{2i\alpha } |L\rangle\langle R|+|L\rangle\langle L|] \end{equation} 
therefore
\begin{equation*}  I(\tau)= \frac{1}{2} \sum_{b} [\langle V_{b}A_{b}(\tau)|R\rangle\langle R|V_{b}A_{b}(\tau) \rangle -\mathscr{P}e^{-2i\alpha } \langle V_{b}A_{b}(\tau) |R\rangle\langle L|V_{b}A_{b}(\tau) \rangle \end{equation*} 
\begin{equation}-\mathscr{P}e^{2i\alpha } \langle V_{b}A_{b}(\tau)|L\rangle\langle R| V_{b}A_{b}(\tau) \rangle + \langle V_{b}A_{b}(\tau)|L\rangle\langle L| V_{b}A_{b}(\tau) \rangle ] . \end{equation} 
After some algebra this expression can be written as
\begin{equation*}   I(\tau) = \frac{1-\mathscr{P}}{4} \begin{vmatrix} \sum_{b} \langle V_{b}A_{b}(\tau)|R\rangle + e^{2i\alpha } \sum_{b} \langle V_{b}A_{b}(\tau)|L\rangle \end{vmatrix}^{2} \end{equation*}
\begin{equation} +\frac{1+\mathscr{P}}{4}\begin{vmatrix} \sum_{b} \langle V_{b}A_{b}(\tau)|R \rangle-e^{2i\alpha } \sum_{b} \langle V_{b} A_{b}(\tau)|L\rangle \end{vmatrix}^{2} .\end{equation}

To calculate the spin density matrix in the reflectivity basis, we turn to the relations of the reflectivity basis with the helicity and the GJ bases [6,7].

We have that
\begin{equation}
 \label{GrindEQ__10_}
 {|\epsilon a\lambda \rangle =\left[|a\lambda \rangle +\epsilon P (-1)^{j-\lambda } |a-\lambda \rangle \right]\Theta (\lambda )}
 \end{equation}
 where $P$ is the parity of particle "a", and 
 \begin{eqnarray}
\Theta(\lambda) & = & {1 \over \sqrt{2}} \hspace{2mm} \mbox{for} \hspace{1.3mm} \lambda > 0 \\
\Theta(\lambda) & = & {1 \over 2}  \hspace{2mm} \mbox {for}\hspace{1.3mm} \lambda = 0 \\
\Theta(\lambda) & = & 0   \hspace{2mm} \mbox{ for } \hspace{1.3mm}\lambda <  0
\end{eqnarray}
the eigenvalue of reflectivity for $\lambda $=0 is $P (-1)^{J}$.

For a real photon $P =-1$, $J=1$ and $\lambda =+1,-1$,therefore

\noindent 
\begin{equation} \label{GrindEQ__11_} 
|\epsilon \lambda \rangle =\frac{1}{\sqrt{2} } \left[|\lambda \rangle -\epsilon (-1)^{1-\lambda } |-\lambda \rangle \right] 
\end{equation} 
then (the reflectivity eigenvalues for a photon are $\epsilon =\pm 1$ ).
\begin{equation} \label{GrindEQ__12_} 
\begin{array}{l} {|\epsilon =+1,\lambda =+1\rangle =\frac{1}{\sqrt{2} } (|\lambda =+1\rangle -|\lambda =-1\rangle )} \\ {} \\ {|\epsilon =-1,\lambda =+1\rangle =\frac{1}{\sqrt{2} } (|\lambda =+1\rangle +|\lambda =-1\rangle )} \end{array} .
\end{equation} 

Using the relations in \eqref{GrindEQ__7_}, we obtain 
\begin{equation} \label{GrindEQ__13_} 
\begin{array}{l} {|\epsilon =+1\rangle =\frac{1}{\sqrt{2} } \left[-\frac{1}{\sqrt{2} } |x\rangle -\frac{1}{\sqrt{2} } i|y\rangle -\frac{1}{\sqrt{2} } |x\rangle +\frac{1}{\sqrt{2} } i|y\rangle \right]=-|x\rangle } \\ {} \\ {|\epsilon =-1\rangle =\frac{1}{\sqrt{2} } \left[-\frac{1}{\sqrt{2} } |x\rangle -\frac{1}{\sqrt{2} } i|y\rangle +\frac{1}{\sqrt{2} } |x\rangle -\frac{1}{\sqrt{2} } i|y\rangle \right]=-i|y\rangle } \end{array} 
\end{equation} 
Therefore, we find that, using values in \eqref{GrindEQ__6_}
\noindent 
\begin{equation} \begin{array}{l} {|\epsilon =+1\rangle \langle \epsilon =+1|=|x\rangle \langle x|={\raise0.7ex\hbox{$ 1 $}\!\mathord{\left/{\vphantom{1 2}}\right.\kern-\nulldelimiterspace}\!\lower0.7ex\hbox{$ 2 $}} \left(1+\mathscr{P}\cos 2\alpha \right)} \\ {} \\ {|\epsilon =-1\rangle \langle \epsilon =-1|=-i|y\rangle (i\langle y|)=|y\rangle \langle y|={\raise0.7ex\hbox{$ 1 $}\!\mathord{\left/{\vphantom{1 2}}\right.\kern-\nulldelimiterspace}\!\lower0.7ex\hbox{$ 2 $}} \left(1-\mathscr{P}\cos 2\alpha \right)} \\ {} \\ {|\epsilon =+1\rangle \langle \epsilon =-1|=-|x\rangle (i\langle y|)=-i|x\rangle \langle y|={\raise0.7ex\hbox{$ 1 $}\!\mathord{\left/{\vphantom{1 2}}\right.\kern-\nulldelimiterspace}\!\lower0.7ex\hbox{$ 2 $}} i\left(\mathscr{P}\sin 2\alpha \right)} \\ {} \\ {|\epsilon =-1\rangle \langle \epsilon =+1|=-i|y\rangle (-\langle x|)=-i|y\rangle \langle x|=-{\raise0.7ex\hbox{$ 1 $}\!\mathord{\left/{\vphantom{1 2}}\right.\kern-\nulldelimiterspace}\!\lower0.7ex\hbox{$ 2 $}} i\left(\mathscr{P}\sin 2\alpha \right)} \end{array}\end{equation}  

After some algebra, we obtain the spin density matrix of the photon in the reflectivity basis:

\begin{equation} 
\boxed{ \rho _{\epsilon \epsilon '} \left(\mathscr{P},\alpha \right)={\raise0.7ex\hbox{$ 1 $}\!\mathord{\left/{\vphantom{1 2}}\right.\kern-\nulldelimiterspace}\!\lower0.7ex\hbox{$ 2 $}} \left(\begin{array}{cc} {1+\mathscr{P}\cos 2\alpha } & {i\mathscr{P}\sin 2\alpha } \\ {-i\mathscr{P}\sin 2\alpha } & {1-\mathscr{P}\cos 2\alpha }  \end{array}\right) . }
\end{equation}

\subsection{Virtual Photoproduction}

\begin{figure}[!htp]
\centerline{\includegraphics[width=12cm]{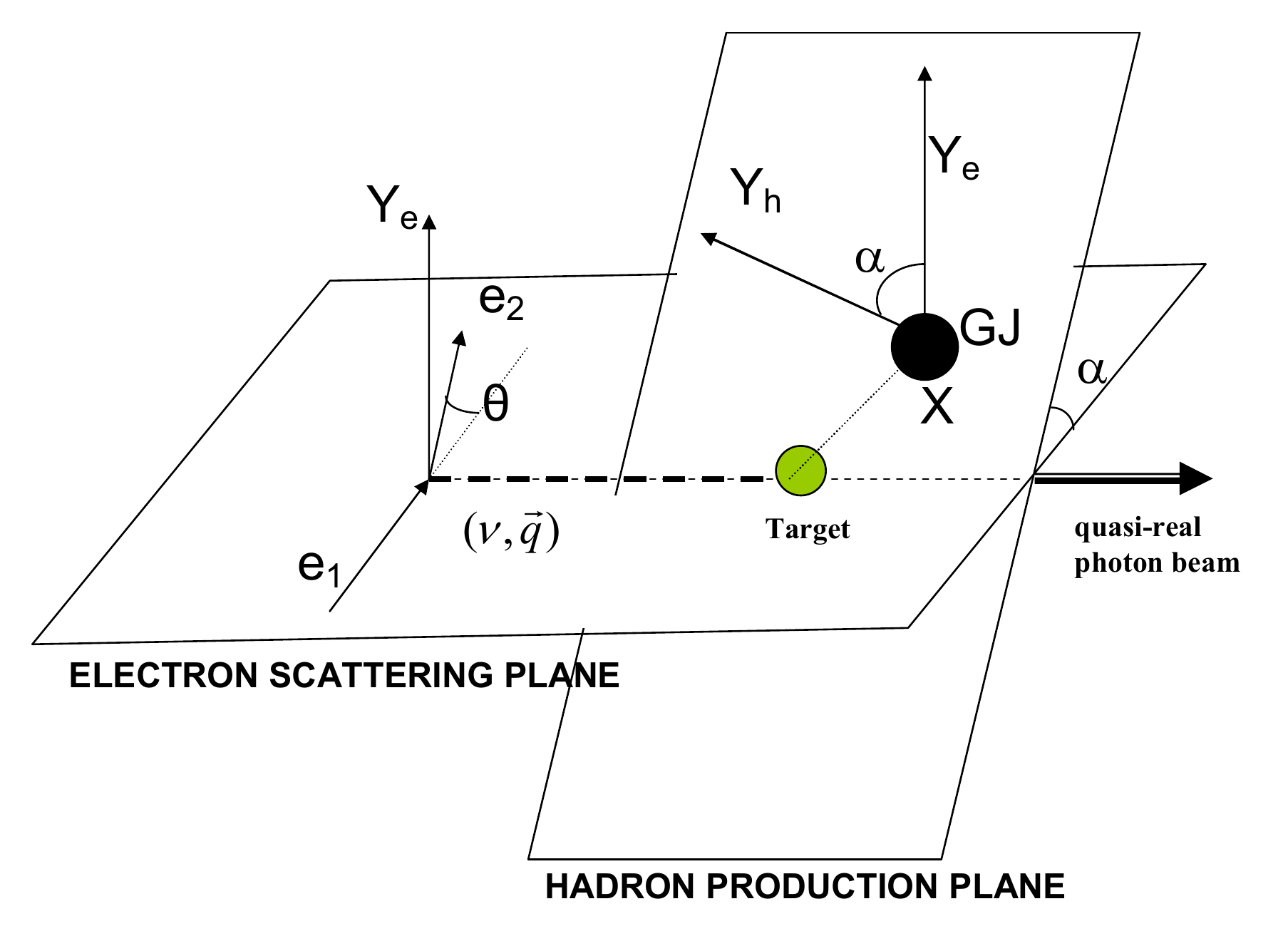}}
\caption{\label{fig:electron}Electron scattering at small $Q^{2}$.}
\end{figure}

Electron scattering can be regarded as the interaction of a virtual photon with the target \cite{Domby}, as seen in figure~\ref{fig:electron}. The electron radiates a virtual photon (off mass shell) of energy $\nu$, where $\nu=E-E'$, being $E$ the energy of the incoming electron (beam) and $E'$ the energy of the scattered electron. The momentum of the virtual photon is $\overrightarrow{q}$ such that $\overrightarrow{q}=\overrightarrow{p}-\overrightarrow{p'}$, where $\overrightarrow{p}$ and  $\overrightarrow{p'}$ are the momenta of the incoming and scattered electrons. The scattered electron makes an angle $\theta$ with respect to the incoming electron. We define
\begin{equation} Q^{2} = \overrightarrow{q} \cdot \overrightarrow{q} - \nu^{2} = -q^{2} \end{equation}
and 
\begin{equation} W^{2} = (\nu + M)^{2} -  \overrightarrow{q} \cdot \overrightarrow{q} \end{equation}
where $M$ is the mass of the nucleon and $q$ is the virtual photon four momentum. $W$ is the total energy of the virtual photon-nucleon system.
In this case, the Lorentz condition is not satisfied, and a longitudinal component of the polarization in the direction of motion of the virtual photon is possible.

It is common to exploit the concept of a virtual photon beam by defining a virtual photon flux and cross sections. This also allows a direct comparison with real photoproduction. We can write (for $Q^{2} >>$ electron mass) the electron scattering cross section as~\cite{Halzen}
\begin{equation} \frac{d^{2} \sigma}{d \Omega dE'} = \Gamma \lbrack \sigma_{T}(Q^{2},\nu) +\mathscr{P}\sigma_{L}(Q^{2},\nu) \rbrack 
\label{eqn:elec}
\end{equation}
where $\sigma_{T}$ and $\sigma_{L}$ are the total cross sections for transverse and longitudinal virtual photons respectively, $\Gamma$ is the flux of transverse photons (defined below) and $\mathscr{P}$ is a parameter characterizing the polarization of the virtual photon.  The flux of transverse photons is~\cite{Halzen}
\begin{equation} \Gamma = \frac{\alpha}{4 \pi^{2}} \frac{E'}{E} \frac{W^{2}-M^{2}}{MQ^{2}} \frac{1}{1-\mathscr{P}} \end{equation}
$\alpha$ is the fine-structure constant and the value of $\mathscr{P}$ will be obtained below. Note that although a flux and a cross section were defined in equation (\ref{eqn:elec}), only the product (cross section of the electron scattering) can be measured. Notice that a virtual photon can not be fully polarized ($\mathscr{P}=1$).

We might also apply the standard QED Feynman rules to obtain the invariant electron scattering amplitude as given by \cite{Domby}
\begin{equation}  {\mathscr{M}} \propto \frac{e^{2} }{Q^{2} } \langle N'|J_{\nu } |N\rangle \langle l_{2} |j_{\mu } |l_{1} \rangle \end{equation}  
where $J_{\nu }$ and $j_{\mu }$ are the current densities of the nucleon and target respectively. The terms included in the cross section, the squared invariant amplitude, can be then represented by two tensors, one corresponding to the hadronic (current) vertex, ${\rm T} _{\mu \nu }$, and another to the electromagnetic (current) vertex, $L_{\mu \nu }$ ($\nu$ and $\mu$ span over 0,1,2,3 the Minkowski space), such that \cite{Domby}

\begin{equation} \left|{\mathscr{M}}\right|^{2} =\frac{2e^{4} }{Q^{2} } {\rm T} _{\mu \nu } L_{\mu \nu } . \end{equation}  

Comparing with the photoproduction result of equation (\ref{eqn:matscat}), we can identify $L_{\mu \nu } $ with the spin density matrix of the virtual photon,  this tensor is given by \cite{Domby,Sch73}

\begin{equation} L_{\mu \nu } ={\raise0.7ex\hbox{$ 1 $}\!\mathord{\left/{\vphantom{1 2Q^{2} }}\right.\kern-\nulldelimiterspace}\!\lower0.7ex\hbox{$ 2Q^{2}  $}} \left(\sum _{spins}\langle l_{2} |j_{\nu } |l_{1} \rangle ^{*} \langle l_{2} |j_{\mu } |l_{1} \rangle  \right)=\frac{1}{2} Tr\left[\gamma \cdot l_{1} \gamma _{\nu } \gamma \cdot l_{2} \gamma _{\mu } \right] . \end{equation}  

Calculating the trace, the elements of the tensor are \cite{Domby,Sch73}

\begin{equation} \label{GrindEQ__15_} 
L_{\mu \nu } ={\raise0.7ex\hbox{$ 1 $}\!\mathord{\left/{\vphantom{1 Q^{2} }}\right.\kern-\nulldelimiterspace}\!\lower0.7ex\hbox{$ Q^{2}  $}} \left(l_{1\nu } l_{2\mu } +l_{1\mu } l_{2\nu } +\frac{1}{2} lQ^{2} \delta _{\nu \varpi } \right) .
\end{equation} 

After some algebraic manipulation and defining, by comparison with the real photon case, the partial polarization $\mathscr{P}$ such that

\begin{equation} \label{GrindEQ__16_} 
\frac{L_{11} }{L_{22} } \equiv \frac{(1+\mathscr{P})}{(1-\mathscr{P})}  
\end{equation} 
we can define the {\it virtual photon spin density matrix} \cite{Domby} by taken only the space-like part of $L_{\nu \mu }$ (a $3 \times 3$ matrix) such that
\begin{equation} \label{GrindEQ__17_} 
\sum_{i,j} \rho _{ij} =(1-\mathscr{P})L_{\nu \mu }  
\end{equation} 
with $i,j = 1,2,3$.
The elements of this matrix (in the frame defined in figure~\ref{fig:electron}, where $XZ$ is the scattering plane, $Z$ the direction of the virtual photon) can be calculated using definition \eqref{GrindEQ__16_} and the values in \eqref{GrindEQ__15_} (see references~\cite{Domby} and~\cite{Sch73}). The matrix is 
\begin{equation} \label{GrindEQ__18_} 
\rho _{X,Y,Z} =\left(\begin{array}{ccc} {\frac{1}{2} (1+\mathscr{P})} & {0} & {-\left[\frac{1}{2} \mathscr{P}_{L} (1+\mathscr{P})\right]^{{\raise0.7ex\hbox{$ 1 $}\!\mathord{\left/{\vphantom{1 2}}\right.\kern-\nulldelimiterspace}\!\lower0.7ex\hbox{$ 2 $}} } } \\ {0} & {\frac{1}{2} (1-\mathscr{P})} & {0} \\ {-\left[\frac{1}{2} \mathscr{P}_{L} (1+\mathscr{P})\right]^{{\raise0.7ex\hbox{$ 1 $}\!\mathord{\left/{\vphantom{1 2}}\right.\kern-\nulldelimiterspace}\!\lower0.7ex\hbox{$ 2 $}} } } & {0} & {\mathscr{P}_{L} } \end{array}\right) 
\end{equation} 
where $\mathscr{P}{}_{L}$, the longitudinal partial polarization, is defined by

\begin{equation} \label{GrindEQ__19_} 
\mathscr{P}_{L} \equiv \frac{Q^{2} }{\nu ^{2} } \mathscr{P} .
\end{equation} 

The value of $\mathscr{P}$ can be calculated from the kinematics of the scattering as \cite{Sch73}

\begin{equation} \label{GrindEQ__20_} 
\mathscr{P}=\left[1+2\frac{(Q^{2} +\nu ^{2} )}{Q^{2} } \tan ^{2} {\theta \over 2}\right]^{-1}  .
\end{equation} 

A transformation to the GJ frame will only affect the $XY$ components, as $Z$ remains unchanged, and in the direction of the virtual photon, {\it not} the electron beam direction.

In a similar way as it was calculated for photoproduction, the spin density matrix of the virtual photon in the GJ frame is:
\begin{equation} \label{GrindEQ__21_} 
\rho _{x,y,z} (\mathscr{P},\alpha ,Q^{2} ,\nu )=\left(\begin{array}{ccc} {\frac{1}{2} (1+\mathscr{P}\cos 2\alpha )} & {-\frac{1}{2} \mathscr{P}\sin 2\alpha } & {-\left[\frac{1}{2} \mathscr{P}_{L} (1+\mathscr{P})\right]^{{\raise0.7ex\hbox{$ 1 $}\!\mathord{\left/{\vphantom{1 2}}\right.\kern-\nulldelimiterspace}\!\lower0.7ex\hbox{$ 2 $}} } } \\ {-\frac{1}{2} \mathscr{P}\sin 2\alpha } & {\frac{1}{2} (1-\mathscr{P}\cos 2\alpha )} & {0} \\ {-\left[\frac{1}{2} \mathscr{P}_{L} (1+\mathscr{P})\right]^{{\raise0.7ex\hbox{$ 1 $}\!\mathord{\left/{\vphantom{1 2}}\right.\kern-\nulldelimiterspace}\!\lower0.7ex\hbox{$ 2 $}} } } & {0} & {\mathscr{P}_{L} } \end{array}\right) 
\end{equation} 
where $\alpha $ is the angle between the scattering and production plane.

We now find the values of the matrix in the helicity frame. The virtual photon will have three helicity states in the GJ frame~\cite{Halzen}

\begin{equation} \label{GrindEQ__22_} 
\begin{array}{l} {|\lambda =+1\rangle =-\frac{1}{\sqrt{2} } (|x\rangle +i|y\rangle )} \\ {} \\ {|\lambda =0\rangle =|z\rangle } \\ {} \\ {|\lambda =-1\rangle =\frac{1}{\sqrt{2} } (|x\rangle -i|y\rangle )} \end{array} 
\end{equation} 

Using our previous results, equations (\ref{eqn:HELp}) and (\ref{eqn:HELm}),  the relation from helicity to the basis called $|P_{1}\rangle$, $|P_{2}\rangle$ and $|P_{3}\rangle$, where $|P_{3}\rangle$ is in the virtual photon direction and $|P_{3}\rangle$, $|P_{1}\rangle$ in the scattering plane (see figure  ~\ref{fig:electron}) is

\begin{equation} \begin{array}{l} {|\lambda =+1\rangle =-\frac{1}{\sqrt{2} } \left(e^{-i\alpha } |P_{1} \rangle +ie^{-i\alpha } |P_{2} \rangle \right)} \\ {} \\ {|\lambda =0\rangle =|P_{3} \rangle } \\ {} \\ {|\lambda =-1\rangle =\frac{1}{\sqrt{2} } \left(e^{i\alpha } |P_{1} \rangle -ie^{i\alpha } |P_{2} \rangle \right)} \end{array}\end{equation}  

The elements of the matrix, calculated following the some procedure we used in the real photon case, are

\begin{equation}
\begin{array}{l} {|\lambda =+1\rangle \langle \lambda =+1|=1/2} \\ {} \\ {|\lambda =-1\rangle \langle \lambda =-1|=1/2} \\ {} \\ {|\lambda =-1\rangle \langle \lambda =+1|= -{\mathscr{P} \over 2} e^{2 i \alpha} } \\ {} \\  {|\lambda =+1\rangle \langle \lambda =-1|= -{\mathscr{P} \over 2} e^{-2 i \alpha} } \end{array} \end{equation}  

The longitudinal elements are

\begin{equation} \begin{array}{l} {|\lambda =+1\rangle \langle \lambda =0|=\left[(-\frac{1}{\sqrt{2} } )(e^{-i\alpha } |P_{1} \rangle +ie^{-i\alpha } |P_{2} \rangle )(\langle P_{3} |)\right]} \\ {=(-\frac{1}{\sqrt{2} } )\left(e^{-i\alpha } |P_{1} \rangle \langle P_{3} |+ie^{-i\alpha } |P_{2} \rangle \langle P_{3} |\right)=-\frac{1}{\sqrt{2} } \left[\frac{1}{2} \mathscr{P}_{L} (1+\mathscr{P})\right]^{{\raise0.7ex\hbox{$ 1 $}\!\mathord{\left/{\vphantom{1 2}}\right.\kern-\nulldelimiterspace}\!\lower0.7ex\hbox{$ 2 $}} } e^{-i\alpha } } \\ {=-\frac{1}{2} \left[\mathscr{P}_{L} (1+\mathscr{P})\right]^{{\raise0.7ex\hbox{$ 1 $}\!\mathord{\left/{\vphantom{1 2}}\right.\kern-\nulldelimiterspace}\!\lower0.7ex\hbox{$ 2 $}} } e^{-i\alpha } } \\ {} \\ {|\lambda =0\rangle \langle \lambda =0|=|P_{3} \rangle \langle P_{3} |=\mathscr{P}_{L} } \end{array}\end{equation}

The resulting spin density matrix of the virtual photon in the helicity frame is

\begin{equation} \label{GrindEQ__23_} 
\rho _{\lambda \lambda '} (\mathscr{P},\alpha ,Q^{2} ,\nu )=\frac{1}{2} \left(\begin{array}{ccc} {1} & {-\left[\mathscr{P}_{L} (1+\mathscr{P})\right]^{{\raise0.7ex\hbox{$ 1 $}\!\mathord{\left/{\vphantom{1 2}}\right.\kern-\nulldelimiterspace}\!\lower0.7ex\hbox{$ 2 $}} } e^{-i\alpha } } & {-\mathscr{P}e^{-2i\alpha } } \\ {-\left[\mathscr{P}_{L} (1+\mathscr{P})\right]^{{\raise0.7ex\hbox{$ 1 $}\!\mathord{\left/{\vphantom{1 2}}\right.\kern-\nulldelimiterspace}\!\lower0.7ex\hbox{$ 2 $}} } e^{i\alpha } } & {2\mathscr{P}_{L} } & {0} \\ {-\mathscr{P}e^{2i\alpha } } & {0} & {1} \end{array}\right) 
\end{equation} 
(in agreement with reference \cite{Sch73}).

We now transform this matrix to the reflectivity basis. We start by using \eqref{GrindEQ__10_}

\begin{equation} |\varepsilon, \lambda \rangle =\frac{1}{\sqrt{2} } \left[|\lambda \rangle -\varepsilon (-1)^{1-\lambda } |-\lambda \rangle \right]\end{equation}  
For the virtual photon $P$ =-1, $J$ =1 and $\lambda $=+1, 0,-1$;$ therefore

\begin{equation} \label{GrindEQ__24_} 
\begin{array}{l} {|\varepsilon =+1,\lambda =+1\rangle =\frac{1}{\sqrt{2} } (|\lambda =+1\rangle -|\lambda =-1\rangle )} \\ {} \\ {|\varepsilon =-1,\lambda =+1\rangle =\frac{1}{\sqrt{2} } (|\lambda =+1\rangle +|\lambda =-1\rangle )} \\ {} \\ {|\varepsilon =-1,\lambda =0\rangle =|\lambda =0\rangle } \end{array} 
\end{equation} 
Using equations (\ref{GrindEQ__22_}), we have then
\begin{equation} \label{GrindEQ__25_} 
\begin{array}{l} {|\varepsilon =+1,\lambda =+1\rangle =-|x\rangle } \\ {} \\ {|\varepsilon =-1,\lambda =+1\rangle =-i|y\rangle } \\ {} \\ {|\varepsilon =-1,\lambda =0\rangle =|z\rangle } \end{array} 
\end{equation} 
 and
\begin{equation} \begin{array}{l} {|\varepsilon =+1,+1\rangle \langle \varepsilon =+1,+1|=|x\rangle \langle x|={\raise0.7ex\hbox{$ 1 $}\!\mathord{\left/{\vphantom{1 2}}\right.\kern-\nulldelimiterspace}\!\lower0.7ex\hbox{$ 2 $}} \left(1+\mathscr{P}\cos 2\alpha \right)} \\ {} \\ {|\varepsilon =-1,+1\rangle \langle \varepsilon =-1,+1|=-i|y\rangle (i\langle y|)=|y\rangle \langle y|={\raise0.7ex\hbox{$ 1 $}\!\mathord{\left/{\vphantom{1 2}}\right.\kern-\nulldelimiterspace}\!\lower0.7ex\hbox{$ 2 $}} \left(1-\mathscr{P}\cos 2\alpha \right)} \\ {} \\ {|\varepsilon =+1,+1\rangle \langle \varepsilon =-1,+1|=-|x\rangle (i\langle y|)=-i|x\rangle \langle y|={\raise0.7ex\hbox{$ 1 $}\!\mathord{\left/{\vphantom{1 2}}\right.\kern-\nulldelimiterspace}\!\lower0.7ex\hbox{$ 2 $}} i\left(\mathscr{P}\sin 2\alpha \right)} \\ {} \\ {|\varepsilon =-1,+1\rangle \langle \varepsilon =+1,+1|=-i|y\rangle (-\langle x|)=-i|y\rangle \langle x|=-{\raise0.7ex\hbox{$ 1 $}\!\mathord{\left/{\vphantom{1 2}}\right.\kern-\nulldelimiterspace}\!\lower0.7ex\hbox{$ 2 $}} i\left(\mathscr{P}\sin 2\alpha \right)} \\ {} \\ {|\varepsilon =+1,+1\rangle \langle \varepsilon =-1,0|=-|x\rangle \langle z|=-\left[\frac{1}{2} \mathscr{P}_{L} \left(1+\mathscr{P}\right)\right]^{{\raise0.7ex\hbox{$ 1 $}\!\mathord{\left/{\vphantom{1 2}}\right.\kern-\nulldelimiterspace}\!\lower0.7ex\hbox{$ 2 $}} } } \\ {} \\ {|\varepsilon =-1,0\rangle \langle \varepsilon =-1,0|=|z\rangle \langle z|=\mathscr{P}_{L} } \end{array}\end{equation}  
or in matrix form, the spin density matrix of the virtual photon in the reflectivity basis is
\begin{equation} \label{GrindEQ__26_} 
\rho _{\varepsilon \varepsilon '} (\mathscr{P},\alpha ,Q^{2} ,\nu )=\left(\begin{array}{ccc} {\frac{1}{2} (1+\mathscr{P}\cos 2\alpha )} & {-\left[\frac{1}{2} \mathscr{P}_{L} (1+\mathscr{P})\right]^{{\raise0.7ex\hbox{$ 1 $}\!\mathord{\left/{\vphantom{1 2}}\right.\kern-\nulldelimiterspace}\!\lower0.7ex\hbox{$ 2 $}} } } & {i\frac{\mathscr{P}}{2} \sin 2\alpha } \\ {-\left[\frac{1}{2} \mathscr{P}_{L} (1+\mathscr{P})\right]^{{\raise0.7ex\hbox{$ 1 $}\!\mathord{\left/{\vphantom{1 2}}\right.\kern-\nulldelimiterspace}\!\lower0.7ex\hbox{$ 2 $}} } } & {\mathscr{P}_{L} } & {0} \\ {-i\frac{\mathscr{P}}{2} \sin 2\alpha } & {0} & {\frac{1}{2} (1-\mathscr{P}\cos 2\alpha )} \end{array}\right) 
\end{equation}

\subsection{Very-low Q2 Regime}

For cases where $m^{2} <<Q^{2} <<\nu ^{2} $, we can take $\mathscr{P}_{L} \cong 0$ ($m$ is the electron mass). For example, for the planned forward tagger facility at CLAS12~\cite{CLAS12}, these values will be about: $(0.5 \cdot 10^{-3} )^{2} <<0.16<<64$. In that case, the "virtual photon'' spin density matrices are identical to the real photon spin density matrices. Since
\begin{equation} \mathscr{P}_{L}=\frac{Q^{2}}{\nu^{2}} \mathscr{P} \approx 0 .\end{equation}
Paraphrasing reference \cite{Domby}: "\dots experiments using unpolarized leptons {\it are equivalent} in the small Q${}^{2}$ limit to those using partially linearly polarized photons. Furthermore, the (partial) polarization $\mathscr{P}$ will be known very accurately.'' The polarization vector is given by $\mathscr{P}$ and the angle $\alpha $  between the scattering plane and the production plane. The value of $\mathscr{P}$ is given by \eqref{GrindEQ__20_}:
\begin{equation}  \mathscr{P}=\left[1+2\frac{(Q^{2} +\nu ^{2} )}{Q^{2} } \tan ^{2} \frac{1}{2} \theta \right]^{-1} \approx \left[1+\frac{\nu ^{2} }{2Q^{2} } \theta ^{2} \right]^{-1} \end{equation}  
and since
\begin{equation} Q^{2} \cong 4E_{1} E_{2} \sin ^{2} \frac{1}{2} \theta \approx E_{1} E_{2} \theta ^{2} \end{equation}  
then
\begin{equation}  \mathscr{P}\approx \left[1+\frac{(E_{1} -E_{2} )^{2} }{2E_{1} E_{2} } \right]^{-1} \approx \frac{2E_{1} E_{2} }{E_{1} ^{2} +E_{2} ^{2} }.\end{equation}  
Therefore, $\mathscr{P}$ depends only on the beam and scattered electron energies.
The angle $\alpha$, between the electron scattering and hadron production planes, is calculated from the measured momenta of the electron beam, $\overrightarrow{e}$, the scattered electron, $\overrightarrow{e'}$ and all final particle's,  $\overrightarrow{p_{i}}$. We have that the normal to the electron scattering plane is

\begin{equation}
\overrightarrow{n_{e}} =\frac{ \overrightarrow{e} \times \overrightarrow{e'} } {\big{|} \overrightarrow{e} \times \overrightarrow{e'} \big{|} }
\end{equation}

and the normal to the hadron production plane is

\begin{equation}
\overrightarrow{n_{h}} =\frac{ \overrightarrow{\gamma^{*}} \times \overrightarrow{X} } {\big{|} \overrightarrow{\gamma^{*}} \times \overrightarrow{X} \big{|} }
\end{equation}

where $\overrightarrow{\gamma^{*}} = \overrightarrow{e'} - \overrightarrow{e}$ and $\overrightarrow{X} = \sum_{i} \overrightarrow{p_{i}}$. Therefore

\begin{equation}
\alpha = arccos(\overrightarrow{n_{e}} \cdot \overrightarrow{n_{h}})
\end{equation}

Notice that the errors on the determination of $\mathscr{P}$ depend only on the determination of the energies of the incoming electron beam and outgoing scattered electron, therefore, they should be small. However, the determination of the angle $\alpha$ depends on the isolation of the resonance decay products and, therefore, the errors are affected by baryon contamination and other problems (see section \ref{sect:Baryon}).

\section{Partial Waves}

We now write an expression for the intensity, $I(\tau)$. Recall from (\ref{eqn:totall}) that

\begin{equation}  I(\tau) = \sum_{k} \sum_{\epsilon,\epsilon'}  \sum_{b,b'} \  {^{\epsilon}V^{k}_{b}}\  {^{\epsilon}A_{b}(\tau)} \rho_{\epsilon,\epsilon'} {^{\epsilon'}V^{k*}_{b'}}\  {^{\epsilon'}A^{*}_{b'}(\tau)}. \end{equation}

We have organized the indices such that $k$ are the external or non-interfering indices. The states with different reflectivity,  off-diagonal elements of the spin density matrix, are present in the case of polarized beams. For unpolarized beams, both reflectivities do not interfere and $\epsilon = \epsilon'$. The index $b$ contains the characterization of each partial wave (intermediate states) and $\tau$ the angular distribution of the final states. The particular indices of $b$ depend of the type of reaction and final states.
However, it normally contains
\begin{equation} b=\begin{Bmatrix}I^{G}J^{PC}mL[w_{0},\Gamma_{0}]ls\end{Bmatrix} \end{equation}  
where\\

$I$ : isospin of the resonance.\\

$G$: $G$-parity of the resonance.\\
This is a generalization of the $C$-parity (see below). Since $C$-parity is only a "good" quantum number (eigenvalue) for neutral particles, whereas $G$-parity is valid for all charges and is defined by

\begin{equation} G = C e^{i\pi I_{2}} \end{equation}
where $I_{2}$ is the second component of the isospin.\\

$J$: total spin of the resonance.\\

$P$: parity of the resonance.\\

$C$: charge conjugation (or $C$-parity) of the resonance.\\

$C$ is related to $G$ through

\begin{equation} C = G \times (-1)^{I} 
\label{eqn:gpar}
\end{equation}

$m$: spin projection of the resonance about the beam axis (z-direction).\\

$L$: is the orbital angular momentum between the isobar and the bachelor particle.\\

${[w_{0},\Gamma_{0}]}$  the central mass and width of the isobar particle.\\

$l$: the angular momentum between the two daughters of the isobar (spin of isobar).\\

$s$: spin of isobar (in most cases $l=s$, as we have spinless final particles).\\

The total set of numbers identifying a wave will also include the rank ($k$) and the reflectivity ($\epsilon$), we treated them independently just for pedagogical reasons.

In the isobar model, the mass and width of the isobar(s) ($w_{0}, \Gamma_{0}$) are taken from the Particle Data Group values~\cite{PDG}. These parameters will be included (together with the angular) in what we called $\tau$, such that

\begin{equation}  \tau=\begin{Bmatrix}\Omega_{GJ},\Omega_{h},w_{0},\Gamma_{0}\end{Bmatrix}. \end{equation} 

In each vertex of the decay chain we can impose all strong interaction conservation laws.
For each vertex we have

\begin{equation} G = G_{1} \times G_{2} \end{equation}
\begin{equation} P = P_{1} \times P_{2} \times (-1)^{L} \end{equation}
and the usual rules for the isospin
\begin{equation} \bigg{|}I_{1}-I_{2}\bigg{|} \leq I \leq I_{1}+I_{2} 
\label{eqn:isospin}
\end{equation}
and the angular momentum
\begin{equation} \bigg{|}L-s\bigg{|} \leq J \leq L+s .\end{equation}

The final multiparticle state will normally establish restrictions in the total values of $I^{G}$. For example, let's consider the system $\pi^{+} \pi^{-} \pi^{0}$. This system has a $G$-parity of $-1$, since 
 $G=G(\pi^{+}) G(\pi^{-})G( \pi^{0})=(-1)(-1)(-1)=-1$. There are no known resonances with $I=2$ or more, therefore, for this argument let's restrict $I \leq 1$. Because of equation (\ref{eqn:gpar}), and since it is a neutral system (an eigenstate of $C$), we have for this system that both  $I^{G}=0^{-}$ (isoscalar), $C=-1$ and $I^{G}=1^{-}$ (isovector), $C=+1$ are allowed. However, for the charged final system $\pi^{+} \pi^{+} \pi^{-}$, using equation (\ref{eqn:isospin}) and working backward from the possible isobars/bachelors, we find that only resonances with $I^{G}=1^{-}$ are possible. In a similar way we can analyze other channels.

The spin z-projections, $m$, are normally restricted to $m \leq 2$ since in peripheral photoproduction the helicity change at the baryon vertex can only produce $m=1$ (helicity flip) or $m=0$ (helicity non-flip), and for the photon $m \leq 1$.

Since the partial waves basis must be finite, the question to be answered in PWA is which partial waves to include in a fit?
For a two body decay, we start with two independent quantum numbers for the resonance: $J$ and $m$. Therefore, the number of possible resonant (initial) states is  $\sum_{J} (2J+1)$. If the final decay particles have spins $s_{1}$ and $s_{2}$, we will have $\sum_{J,s_{1},s_{2}} (2J+1)(2s_{1}+1)(2s_{2}+1)$ number of final states.  For spinless final particles, $J=l$, the number of waves is $\sum_{l} (2l+1)$. In the reflectivity basis $m \geq 0$, but there are two reflectivities for each wave, therefore we obtain the same number of waves.

For three or more final state particles, we need to establish the number of isobars first. We normally establish the isobars from the data  (and previous experimental results). In the case of three final particles (one isobar), we plot the masses of any set of two final states and look for known resonances. For example, for the $\pi\pi\pi$ final state we normally find $\pi \pi$ resonances in the $\rho$, $f_0$, and $f_{2}$ mesons.
For each partial wave $\{b\}$ we can calculate decay amplitudes from the formulas discussed in section~\ref{subsect:DecayAmp}. The equation (\ref{eqn:somedecay}) contains a factor $Q_{ls}$ that stems from the isobar propagator. This factor should be, in general, a dynamical parameter depending on the isobar and daughter masses and the interfering and overlapping resonances. However, our model will adapt only well-defined resonant contributions given by the Breit-Wigner formalism (see section~\ref{sect:MassDep}). 

The photon spin density matrix, $^{\epsilon}\rho_{\gamma}$ is calculated with the formulas discussed in section~\ref{sect:SpinDen}. The $^{\epsilon}V_{b}^{k}$ are the parameters of the model (equivalent to $\overrightarrow{a}$ in section~\ref{sect:like}). Therefore,  for each wave, we will have $k \times 2 \times 2$ fitted parameters in our model. A factor of two appears because the $V's$ are complex numbers, and another factor of two corresponding to both reflectivities.

To choose the number of waves to include in the fit is a delicate and reaction-dependent decision. We will give some guiding principles. It is practice to start with a large number of waves and reduce the number of waves to only the waves contributing to the fit. However, it is important to check several combinations as one wave could become important in a different combination. For each reaction there are relations between the quantum numbers of the resonance and the final states that limited the number of possible waves to include in the set. Those constraints are normally related to vanishing CG coefficients or conservation laws. Finally, the statistical significance of the contribution of each wave can
be analyzed by the relative value of the ln(Likelihood) function (see section~\ref{sect:GoodFit}).

Recalling the form of the photon spin density matrix in the reflectivity basis

\begin{equation}    \widehat{\rho_{\gamma}} = \begin{pmatrix}\rho_{11} & \rho_{1-1} \\ \rho_{-11} & \rho_{-1-1}\end{pmatrix} = {\raise0.7ex\hbox{$ 1 $}\!\mathord{\left/{\vphantom{1 2}}\right.\kern-\nulldelimiterspace}\!\lower0.7ex\hbox{$ 2 $}} \left(\begin{array}{cc} {1+\mathscr{P}\cos 2\alpha } & {i\mathscr{P}\sin 2\alpha } \\ {-i\mathscr{P}\sin 2\alpha } & {1-\mathscr{P}\cos 2\alpha } \end{array}\right) \end{equation} 
we have
\begin{equation*}
 I(\tau)=\sum_{k} \sum_{b,b'} \bigg{[} {^{1}}V_{b}^{k}\  {^{1}}A_{b}(\tau)\rho_{11}  {^{1}}V_{b'}^{k *}\  {^{1}A^{*}_{b'}(\tau)} + {^{1}}V_{b}^{k}\  {^{1}}A_{b}(\tau)\rho_{1-1}  {^{-1}}V_{b'}^{k *}\  {^{-1}A^{*}_{b'}(\tau)} \end{equation*}
\begin{equation}
 +{^{-1}}V_{b}^{k}\  {^{-1}}A_{b}(\tau)\rho_{-11}  {^{1}}V_{b'}^{k *}\  {^{1}A^{*}_{b'}(\tau)}+{^{-1}}V_{b}^{k}\  {^{-1}}A_{b}(\tau)\rho_{-1-1}  {^{-1}}V_{b'}^{k *}\  {^{-1}A^{*}_{b'}(\tau)} \bigg{]}
 \end{equation}
 or
 \begin{equation*}
 I(\tau)=\sum_{k} \bigg{[} \rho_{11} \sum_{b,b'} {^{1}}V_{b}^{k}\  {^{1}}A_{b}(\tau)  {^{1}}V_{b'}^{k *}\  {^{1}A^{*}_{b'}(\tau)} + \rho_{1-1} \sum_{b,b'}  {^{1}}V_{b}^{k}\  {^{1}}A_{b}(\tau) {^{-1}}V_{b'}^{k *}\  {^{-1}A^{*}_{b'}(\tau)} \end{equation*}
\begin{equation}
 +\rho_{-11}\sum_{b,b'}  {^{-1}}V_{b}^{k}\  {^{-1}}A_{b}(\tau)  {^{1}}V_{b'}^{k *}\  {^{1}A^{*}_{b'}(\tau)}+\rho_{-1-1} \sum_{b,b'} {^{-1}}V_{b}^{k}\  {^{-1}}A_{b}(\tau)  {^{-1}}V_{b'}^{k *}\  {^{-1}A^{*}_{b'}(\tau)} \bigg{]}
 \end{equation}
 or
 \begin{equation*}
 I(\tau)=  \sum_{k}  \bigg{[} \rho_{11} \bigg{|}\sum_{b} {^{1}}V_{b}^{k}\  {^{1}}A_{b}(\tau)\bigg{|}^{2} + \rho_{1-1} \sum_{b,b'}  {^{1}}V_{b}^{k}\  {^{1}}A_{b}(\tau) {^{-1}}V_{b'}^{k *}\  {^{-1}A^{*}_{b'}(\tau)} \end{equation*}
\begin{equation}
 +\rho_{-11}\sum_{b,b'}  {^{-1}}V_{b}^{k}\  {^{-1}}A_{b}(\tau)  {^{1}}V_{b'}^{k *}\  {^{1}A^{*}_{b'}(\tau)}+\rho_{-1-1} \bigg{|} \sum_{b} {^{-1}}V_{b}^{k}\  {^{-1}}A_{b}(\tau)\bigg{|}^{2} \bigg{]}.
 \end{equation} 

Let's call $\rho_{nd}=\frac{i}{2}\mathscr{P}sin2\alpha=\rho_{1-1}=-\rho_{-11}$ (non-diagonal elements), then

\begin{equation*}
 I(\tau)= \sum_{k} \bigg{[} \rho_{11} \bigg{|}\sum_{b} {^{1}}V_{b}^{k}\  {^{1}}A_{b}(\tau)\bigg{|}^{2} + \rho_{nd} \sum_{b,b'}[  {^{1}}V_{b}^{k}\  {^{1}}A_{b}(\tau) {^{-1}}V_{b'}^{k *}\  {^{-1}A^{*}_{b'}(\tau)} \end{equation*}
\begin{equation}
 -{^{-1}}V_{b}^{k}\  {^{-1}}A_{b}(\tau)  {^{1}}V_{b'}^{k *}\  {^{1}A^{*}_{b'}(\tau)}]+\rho_{-1-1} \bigg{|} \sum_{b} {^{-1}}V_{b}^{k}\  {^{-1}}A_{b}(\tau)\bigg{|}^{2} \bigg{]}
 \end{equation} 
 calling $C_{bb'}={^{1}}V_{b}^{k}\  {^{1}}A_{b}(\tau) {^{-1}}V_{b'}^{k *}\  {^{-1}A^{*}_{b'}(\tau)}$ we have
  \begin{equation*}
 I(\tau)= \sum_{k}  \bigg{[} \rho_{11} \bigg{|}\sum_{b} {^{1}}V_{b}^{k}\  {^{1}}A_{b}(\tau)\bigg{|}^{2} + \rho_{nd} \sum_{b,b'}[  C_{bb'} -C^{*}_{bb'}] \end{equation*}
\begin{equation}+\rho_{-1-1} \bigg{|} \sum_{b} {^{-1}}V_{b}^{k}\  {^{-1}}A_{b}(\tau)\bigg{|}^{2} \bigg{]}
 \end{equation} 
 and since $C_{bb'}-C^{*}_{bb'}=i \times 2 \times Im(C_{bb'})$
 \begin{equation*}
 I(\tau)= \sum_{k}  \bigg{[} \rho_{11} \bigg{|}\sum_{b} {^{1}}V_{b}^{k}\  {^{1}}A_{b}(\tau)\bigg{|}^{2} + 2i\rho_{nd}\sum_{b,b'} Im(C_{bb'}) \end{equation*}
\begin{equation} +\rho_{-1-1} \bigg{|} \sum_{b} {^{-1}}V_{b}^{k}\  {^{-1}}A_{b}(\tau)\bigg{|}^{2} \bigg{]}.
 \end{equation}   
 
 Notice that the non-diagonal elements of the photon spin density matrix in the reflectivity basis ($\rho_{nd}$) are imaginary making the intensity real.

Substituting back the values of $\rho$

\begin{equation*}
 I(\tau)= \frac{1}{2} \sum_{k}  \bigg{[} (1+\mathscr{P}cos2\alpha) \bigg{|}\sum_{b} {^{1}}V_{b}^{k}\  {^{1}}A_{b}(\tau)\bigg{|}^{2} - 2\mathscr{P}sin2\alpha \sum_{b,b'} Im(C_{bb'}) \end{equation*}
\begin{equation} +(1-\mathscr{P}cos2\alpha)\bigg{|} \sum_{b} {^{-1}}V_{b}^{k}\  {^{-1}}A_{b}(\tau)\bigg{|}^{2} \bigg{]}.
 \end{equation}   

For no polarization, $\mathscr{P}=0$, we have:

\begin{equation}
 I(\tau)= \frac{1}{2} \sum_{k}  \bigg{[} \bigg{|}\sum_{b} {^{1}}V_{b}^{k}\  {^{1}}A_{b}(\tau)\bigg{|}^{2} +\bigg{|} \sum_{b} {^{-1}}V_{b}^{k}\  {^{-1}}A_{b}(\tau)\bigg{|}^{2} \bigg{]}
\label{eqn:nopo}
\end{equation} 
In this case the amplitudes in both reflectivities do not interfere.

Let's consider the $\phi$ dependence for the decay of two spinless particles. If the beam and the target are not polarized, there is no particular spatial direction for the reaction (other than the direction of the beam) and then, there is no $\phi$ dependence on the intensity. There is no $\phi$ dependence for $m=0$. For other values of $m$, we need to include equations (\ref{eqn:spinless}) into the intensity equation (\ref{eqn:nopo})

\begin{equation*}
 I(\tau) \propto \frac{1}{2} \sum_{k}  \bigg{[} \bigg{|}\sum_{lm} {^{1}}V_{b}^{k} \sqrt{\frac{2l+1}{4\pi}} 2i\theta(m) d _{m0}^{l}(\theta) sin(m \phi)\bigg{|}^{2} \end{equation*}
\begin{equation} +\bigg{|} \sum_{lm} {^{-1}}V_{b}^{k} \sqrt{\frac{2l+1}{4\pi}} 2\theta(m) d _{m0}^{l}(\theta) cos(m \phi)\bigg{|}^{2} \bigg{]}.
\end{equation} 
This expression can only be independent of $\phi$ if ${^{1}}V_{b}^{k}$ = ${^{-1}}V_{b}^{k}$. We have
\begin{equation}
 I(\tau) \propto \frac{1}{2} \sum_{k} \sum_{lm} \frac{2l+1}{4\pi}  \bigg{|}d _{m0}^{l}(\theta) {^{1}} V_{lm}^{k}\bigg{|}^{2} \bigg{[} \bigg{|}sin(m \phi)\bigg{|}^{2} +\bigg{|} cos(m \phi)\bigg{|}^{2} \bigg{]}
\end{equation}
or
\begin{equation}
 I(\tau) \propto \frac{1}{2} \sum_{k} \sum_{lm} \frac{2l+1}{4\pi}  \bigg{|}d _{m0}^{l}(\theta) {^{1}} V_{lm}^{k}\bigg{|}^{2} 
\end{equation}
therefore, {\it for unpolarized beams (plus target) the positive and negative reflectivity states are expected to have equal contributions to the intensity}. Without linearly polarization beams (circular polarization had also $\phi$ symmetry) we do not have enough information to separate ${^{1}}V_{b}^{k}$ and ${^{-1}}V_{b}^{k}$. 
\section{(Mass-independent) Fit to the Model}
\label{sect:pwaFit}
\indent {     } From our discussion in section~\ref{sect:like}, the log of the extended likelihood function can be written as

\begin{equation}  ln\mathscr{L} = \sum_{i=1}^{N}  ln\begin{bmatrix} \mathbb{P}(\overrightarrow{x}_{i},\overrightarrow{a}) \end{bmatrix} -\mathscr{N} .
\label{eqn:likefun}
\end{equation} 
Through the model developed in last sections we can calculate the probability for having a particle scattered into the angular distribution specified by $\tau$ in the phase-space defined by $\Delta M \Delta t$

\begin{equation}  I(\tau)= \sum_{k} \sum_{\epsilon,\epsilon'}  \sum_{b,b'}    {^{\epsilon}A_{b}(\tau) } \ {^{\epsilon}V^{k}_{b}} \rho_{\epsilon,\epsilon'}  {^{\epsilon'}V^{k *}_{b'}}\  {^{\epsilon'}A^{*}_{b'}}(\tau).
\label{eqn:Ivalue}
 \end{equation} 
We make the association, see equation~(\ref{eqn:Pdef}),

\begin{equation}   \mathbb{P}(\overrightarrow{x}_{i},\overrightarrow{a}) \equiv I(\tau_{i}) \end{equation}

therefore, the normalized probability for the $\Delta M \Delta t$ bin is

\begin{equation} p(\overrightarrow{x}_{i},\overrightarrow{a}) = \frac{I(\tau_{i})}{ \mathscr{N }}  = \frac{I(\tau_{i})}{ \int I(\tau) \eta (\tau) d \tau} . \end{equation}

 where in this formula $\eta (\tau)$ is an acceptance that will be defined below.

The value of $\mathscr{N}$, the average number of events expected to be observed in the total phase-space defined by $\Delta M \Delta t$, is calculated numerically through a Monte Carlo (MC) full simulation of the detector and a (flat) phase-space generator of the reaction. In many cases, due to limited statistics, the binning is done only in $M$, therefore a model for the $t$ cross section dependence is introduced in the MC. It is common to use a distribution, inspired by the Regge theory~\cite{Collins}, of the form $e^{-bt}$, where the value of $b$ is extracted by a fit to the the data $Mandelstam-t$ distribution (see section \ref{sect:varia} for other possibilities).
The value of $\mathscr{N}$ is
\begin{equation}  \mathscr{N }= \frac{1}{N_{g}} \sum^{N_{g}}_{i} I(\tau_{i})\eta(\tau_{i}) \end{equation} 
$N_{g}$ is the total number of events generated in the MC. The function $\eta(\tau)$ represents the acceptance (resolution is taken to be perfect, only acceptance is considered here - see section 15 for a discussion on the resolution effects). A MC simulation of the detector will provide the values of this function that are

\begin{equation}  \eta(\tau)=1 \ \mbox{if the event is accepted}\end{equation} 
\begin{equation}  \eta(\tau)=0 \ \mbox{ if the event is  not accepted} \end{equation} 
then
\begin{equation}  \mathscr{N }= \frac{1}{N_{g}} \sum^{N_{a}}_{i} I(\tau_{i}) \end{equation} 
where $N_{a}$ is the total number of accepted events. Let's introduce
\begin{equation}  \eta_{x}= \frac{N_{a}}{N_{g}} \end{equation} 
as the total fraction of events accepted, or total acceptance, then
\begin{equation}  \mathscr{N }= \eta_ {x} \frac{1}{N_{a}} \sum^{N_{a}}_{i} I(\tau_{i}) \end{equation} 
therefore
\begin{equation}  \mathscr{N } =  \eta_ {x} \frac{1}{N_{a}} \sum^{N_{a}}_{i} \sum_{k} \sum_{\epsilon,\epsilon'}  \sum_{b,b'}   {^{\epsilon}A_{b}(\tau_{i})}\ {^{\epsilon}  V^{k}_{b}}\ \rho_{\epsilon,\epsilon'} {^{\epsilon'}V^{k *}_{b'}}   \ {^{\epsilon'}A^{*}_{b'}(\tau_{i})} .
\label{eqn:nofacto}
\end{equation} 

If we assume that {\it all events are produced from the same vertex and by the same mechanism} (t-channel diffraction), the $^{\epsilon}V^{k}_{b}$ parameters are independent of the event number (i.e. they have the same structure for all events), the production parameters can be factored out of the event loop, giving

\begin{equation}  \mathscr{N } =  \eta_ {x}  \sum_{k} \sum_{\epsilon,\epsilon'}  \sum_{b,b'} {^{\epsilon} V^{k}_{b}} \  {^{\epsilon'}V^{k *}_{b'}} \frac{1}{N_{a}} \sum^{N_{a}}_{i}{^{\epsilon}A_{b}(\tau_{i})} \ \rho_{\epsilon,\epsilon'}{^{\epsilon'}A^{*}_{b'}(\tau_{i})}. \end{equation} 
Calling 
\begin{equation} {^{\epsilon,\epsilon'}}\Psi_{b,b'}^{x} =  \frac{1}{N_{a}} \sum^{N_{a}}_{i} {^{\epsilon}A_{b}(\tau_{i})} \  \rho_{\epsilon,\epsilon'}{^{\epsilon'}A^{*}_{b'}}(\tau_{i}) \end{equation}
 the {\it accepted normalization integral}, we obtain
\begin{equation}  \mathscr{N } =  \eta_ {x}  \sum_{k} \sum_{\epsilon,\epsilon'} \sum_{b,b'} {^{\epsilon} V^{k}_{b}} \  {^{\epsilon'}V^{k *}_{b'}}\  {^{\epsilon,\epsilon'}}\Psi_{b,b'}^{x} .
\label{eqn:facto}
\end{equation} 
Notice that this integral needs to be calculated only once during the minimization process, saving computer resources.

Including equations (\ref{eqn:Ivalue}) and (\ref{eqn:facto}) into the likelihood function, equation (\ref{eqn:likefun}),  we have

\vspace{0.4cm}
\fbox{
\addtolength{\linewidth}{-8.\fboxsep}%
\addtolength{\linewidth}{-8.\fboxrule}%
\begin{minipage}{\linewidth}
\begin{equation*}  -ln\mathscr{L} \propto -\sum_{i=1}^{N}  ln\begin{bmatrix}\sum_{k} \sum_{\epsilon,\epsilon'} \sum_{b,b'}  {^{\epsilon}A_{b}(\tau_{i})} {^{\epsilon} V^{k}_{b}} \rho_{\epsilon,\epsilon'}  {^{\epsilon'}V^{k *}_{b'}}\  {^{\epsilon'}A^{*}_{b'}(\tau_{i})}\end{bmatrix} \end{equation*}
\begin{equation}  +  \eta_ {x}  \sum_{k} \sum_{\epsilon,\epsilon'} \sum_{b,b'} {^{\epsilon} V^{k}_{b}} \  {^{\epsilon'}V^{k *}_{b'}}\    {^{\epsilon,\epsilon'}}\Psi_{b,b'}^{x} . \end{equation}
\end{minipage}}
\vspace{0.4cm}

This is the function to be minimized to obtain the $ {^{\epsilon} V^{k}_{b}}$ values~\cite{Cummings}.
To find the {\it true} or predicted number of events in the $\Delta M \Delta t$ bin, which we will call $N_{true}$,
we take
\begin{equation}  N_{true} = \frac{1}{N_{g}} \sum^{N_{g}}_{i} I(\tau_{i}) \end{equation} 
where we will use equation (\ref{eqn:Ivalue}) with the fitted $ {^{\epsilon} V^{k}_{b}}$ values.
 Then
\begin{equation}  N_{true} = \frac{1}{N_{g}}\sum^{N_{g}}_{i} \sum_{k} \sum_{\epsilon,\epsilon'} \sum_{b,b'} {^{\epsilon}A_{b}(\tau)} \ {^{\epsilon} V^{k}_{b}} \rho_{\epsilon,\epsilon'}  {^{\epsilon'}V^{k *}_{b'}} \   {^{\epsilon'}A^{*}_{b'}(\tau)}\end{equation} 
and calling
\begin{equation} {^{\epsilon,\epsilon'}}\Psi^{r}_{b,b'} =  \frac{1}{N_{g}} \sum^{N_{g}}_{i} {^{\epsilon}A_{b}(\tau_{i})}  \rho_{\epsilon,\epsilon'}  {^{\epsilon'}A^{*}_{b'}}(\tau_{i}) \end{equation}
the {\it raw} normalization integral. Notice that these integrals (and also the accepted) are, in general, complex numbers and that they are represented by hermitian matrices. This can be shown using the fact that $\widehat{\rho}_{\gamma}$ is also hermitian. Then

\begin{equation}  N_{true} =  \sum_{k} \sum_{\epsilon,\epsilon'} \sum_{b,b'} {^{\epsilon} V^{k}_{b}} \  {^{\epsilon'}V^{k *}_{b'}}\  {^{\epsilon,\epsilon'}} \Psi_{b,b'}^{r}. 
\label{eqn:trueN}
\end{equation} 

If we include all quantum numbers in one index defining a wave, $\alpha = (b, \epsilon, k)$, then in a more abbreviated notation

\begin{equation}  N_{true} =  \sum_{\alpha,\alpha'} V_{\alpha} V^{*}_{\alpha'}  \Psi_{\alpha,\alpha'}^{r}
\end{equation} 

and the yield for each partial wave is

\begin{equation}  N_{\alpha ,true} =  V_{\alpha} V^{*}_{\alpha}  \Psi_{\alpha,\alpha}^{r} =  \big{|} V_{\alpha} \big{|}^{2} \Psi_{\alpha,\alpha}^{r} .
\label{eqn:yield}
\end{equation} 

If the model includes amplitudes related to other vertices or non-$t$-channel production mechanism (for example the Deck effect, Baryon contaminations, etc.) the factorization used in (\ref{eqn:facto}) is not always possible, and the accepted and raw normalization integrals will not factor. This has a very important effect in the time expended in the minimization process, as the normalization integrals need to be recomputed at each minimization step. The use of GPUs or other computing advances could greatly improve this aspect of the fitting process since we need to include directly equation (\ref{eqn:nofacto}) into the likelihood and calculate the normalization at each step.
After we obtain the $ {^{\epsilon} V^{k}_{b}}$ values, we are able to generate MC events through our partial wave model and {\it predicted} many distributions of data properties (i.e., angular distributions, t-distributions, etc.) to compare directly with data. This method is detailed in Appendix D. These comparisons allow the verification of the fit (see section \ref{sect:GoodFit}). To make the predictions we use the values of $I(\tau; {^{\epsilon} V^{k}_{b}})$ to weight a generated (raw), phase-space, sample and then apply a detector simulation to produce a sample of observed events.

\begin{figure}[!h]
\centerline{\includegraphics[width=10cm]{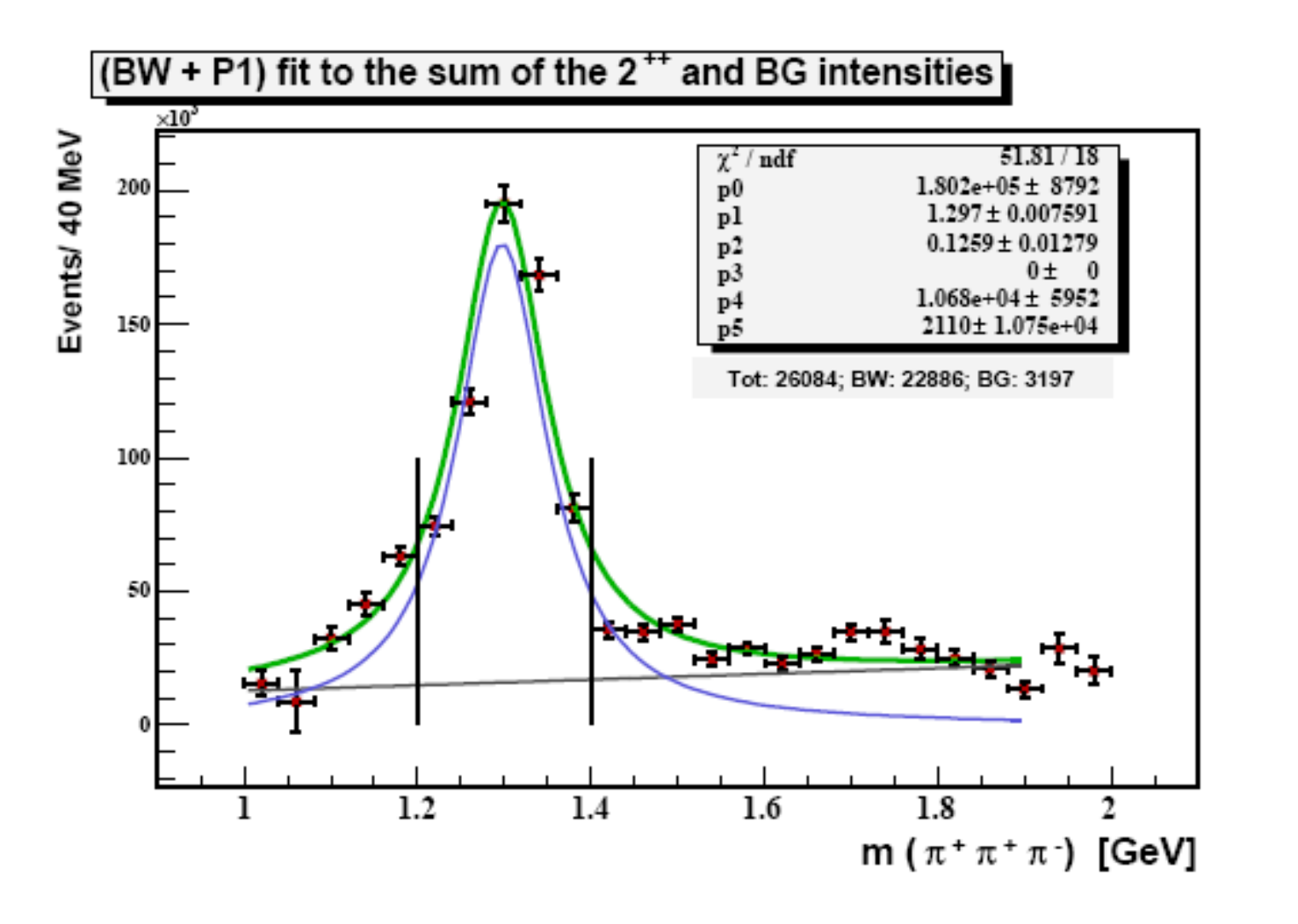}}
\caption{\label{fig:minaresult} Predicted intensities adding all $J^{P C}$ = $2^{++}$ waves in the PWA of  $\gamma p \rightarrow \pi^{+} \pi^{+} \pi^{-} n$. The known $a_{2}(1320)$ resonance is clearly visible. The fit to the obtained mass distribution was done using a Breit-Wigner plus a polynomial background (green curve)~\cite{Nozar2}. The blue curve represents the BW portion and the black the polynomial background portion of the total. }
\end{figure}

As an example of PWA fits, let’s look at one of the results obtained on the $\gamma p \rightarrow \pi^{+} \pi^{+} \pi^{-} n$  analysis of CLAS-g6c run data \cite{Nozar,Nozar2}. Figure \ref{fig:minaresult} shows the predicted intensities going to $J^{P C}$ = $2^{++}$ waves. The figure shows results of $25$ independent fits performed on data binned in masses of $40$ MeV widths (data was integrated on Mandelstam-t).  Each fit was done using $12$ waves in total, in rank $2$. Since each wave has two real parameters there were $12 \times 2 \times 2 = 48$ parameters in the PWA fit. The \mbox{ $J^{P C}$ = $2^{++}$}  represents the addition of $4$ waves for $m^{\epsilon}=1^{\pm}$ and $2^{\pm}$.  A $\rho (770)$ was used as the isobar ($l=2$) . A clear enhancement is observed in the region of the known $a_{2}(1320)$ resonance. Also shown in the figure, there is a fit to the mass distribution using a Breit-Wigner function plus a polynomial background (see section \ref{sect:MassDep}). The values obtained for the mass and width ($1297 \pm 126$  MeV) were consistent with those previously measured~\cite{PDG}.

\section{Validation of the Fits}
\label{sect:GoodFit}

Let's consider three statistical problems associated with our analysis:
\begin{enumerate}
\item Is the fit of our model to the data appropriate? (goodness-of-fit), 
\item How does our model (wave set) compare with other models (other wave sets)? (hypothesis-tests)
\item What is the best way of fitting the model to the data? (estimation) \cite{Eadie}.
\end{enumerate}

We address the third point first. We use the extended Likelihood method. We chose this function as the estimator. An estimator should have four desirable properties: consistency, unbiasedness, efficiency, and robustness.  Consistency, referring to the property of an estimator that must converge to the true value as the number of observations increases, is the most important of all. An estimator is biased if its expected value differs from the true value for any number of events, and it is unbiased otherwise. The maximum-likelihood is a biased estimator~\cite{Eadie}. However, the consistency is more important than unbiasedness since, if the estimator is consistent, it will be always become unbiased for a large number of events. Therefore, with the Likelihood method, it is imperative to have a "large number" of observations in relation to the number of parameters we are fitting. What we mean by "large"  here is difficult to quantify, as it depends on the channel being studied, the set of waves being fitted (not just the number of parameters), and other experimental quantities (as the acceptance of the detector). In every analysis we must evaluate the required number of events studying the fit's stability (as discussed below). 

Maximizing the Likelihood [or equivalently minimizing the -ln(Likelihood)] we can obtain several roots. It can be shown that one of those roots is a consistent estimator \cite{Eadie,Brandt}, but one obvious problem is to identify which of the roots has this property. Furthermore, as the number of observations increases, it is possible that the relative strength of the different maxima of the likelihood change. This is the fundamental uncertainty about finding the true value and it is always present in any finite sample. 

The calculated value of the maximum likelihood, $\mathscr{L}$, bears no statistical significance. To test our hypothesis against another similar hypothesis, we compare the statistical goodness of different PWA fits to the same data by different wave sets using the Likelihood Ratio (LR) test~\cite{Eadie}. In the LR test, a relatively more complex PWA fit, $F_1$ (testing hypothesis $H_1$), is compared with a simpler fit, $F_0$ (testing hypothesis $H_0$), with $F_1$ containing an additional $r$ parameters over $F_0$. This comparison is used to determine if $F_1$ fits the data set significantly better. The LR test is only valid if used to compare hierarchically nested models. That is, the more complex model must differ from the simple model only by the addition of a few extra parameters (normally an extra wave). Addition of extra parameters always results in an equal or higher likelihood value, 
$\mathscr{L}_{1} \geq \mathscr{L}_{0}$.

The ratio of the likelihood values obtained from fits, $F_0$ and $F_1$ is defined as

\begin{equation}  \lambda = \frac{\mathscr{L}_{0}(\overrightarrow{x}_{i},\overrightarrow{a})}{\mathscr{L}_{1}(\overrightarrow{x}_{i},\overrightarrow{a}+\overrightarrow{r})} \end{equation} 

The square of this ratio in log form is the LR statistic
\begin{equation}  LR = -2ln(\lambda) = -2[ln(\mathscr{L}_{0})-ln(\mathscr{L}_{1})] \end{equation} 

The LR statistic is approximately distributed as $\chi^{2}(r)$. In likelihood fits, a change in the $ln(\mathscr{L})$ of 0.5 corresponds to one-standard-deviation error~\cite{Orear}. To determine if the difference in likelihood values among the two fits is statistically significant, the number of degrees of freedom must be taken into account. In the LR, the additional degrees of freedom in fit $F_0$ are the number of additional parameters in the more complex fit, $r$. In PWA, $r$ corresponds to the extra waves included in fit $F_1$. Using this information, we can determine the critical value, CV, of the test statistic for a given significance level, p , (e.g. p = $99$\%) from the standard statistical tables. As a general rule, a high level of significance is required to retain the additional wave(s) and accept hypothesis $H_1$ \cite{Eadie} over $H_0$.
If the LR value is less than or equal to CV, $\mathscr{L}_{1} $ does not fit the data significantly better than $\mathscr{L}_{0}$, and we then infer that the addition of the wave(s) with the $r$ new parameters is not statistically meaningful (given our power to detect such differences). However if LR is greater than CV, the fits differ statistically significantly at the p (e.g. $99$\%) level.

Since in our PWA formalism, we perform an independent fit for each $M$ and $t$ bin, we have then LR values for each of the those bins. Therefore we can perform an analysis on the stability of LR from the different fits in different bins. We expect a smooth transition among bins. There are however some caveats on using the LR test (or any other statistical test) in a PWA formalism to specifically distinguish between hypothesis obtained by adding more parameters. As a general rule using more parameters may fit the data better and make a statistical test unusable. We need to apply other complementary criteria to test the robustness and stability of the fits. As discussed before, we also compare predicted angular distributions and other kinematical distributions to the data.

Since the multi-dimensional Likelihood function can have more than one minimum in the region of interest and may not be quadratic near the minimum, it may be hard to ascertain if we indeed are in the absolute minimum ("minimum minimorum") and, furthermore, if the errors have reasonable meaning . Practically, the way to proceed is to produce many fits with as many different wave sets and {\it feel} your way through the parameter space. Granted, this sounds more an art that a scientific procedure. New methods have recently been used to make this process more automatic and mathematically sound by the Compass collaboration~\cite{Compass, MacKay}. There has been also work in evaluating several goodness-of-fit criteria using likelihood fits in Dalitz analysis that may be easily carried over to PWA~\cite{MWill}.

The errors on the predicted intensities come from the Likelihood fits, as were discussed in section~\ref{sect:like}. These were only statistical errors. But there are other systematic components to the errors.  The systematic errors come mostly from the truncation on the rank and on the number of partial waves used for the fits. An important source of systematic errors comes from the choice of the final set of partial waves used in the fits. Normally, our final partial wave set will include the smallest partial wave set that will fit the data.

The total errors on the predicted intensities include a combination of statistical and systematic errors.  To estimate the systematic errors, we compare different sets, ($j=1,\dots,N$), of intensities, $I_{j}$, and calculate their variance \cite{Bevington}
\begin{equation} \sigma^{2}_{sys} = {1 \over N}\sum^{N}_{j=1} (\hat{I}-I_{j})^{2} \end{equation}
where $\hat{I}$ is the average intensity of the superset of $N$ wave sets. The statistical errors come from the Likelihood fit. The errors on the estimated parameters ($V$) are calculated using the inverse of the Hessian matrix as it was shown in section \ref{sect:like} (or in other possible ways by the MINUIT package \cite{MINUIT}). The statistical errors in the predicted $N_{true}$ values are calculated using the variance obtained by propagating errors through equations (\ref{eqn:trueN}) and (\ref{eqn:yield}).

However, the V's are complex numbers. Let's include all wave numbers in one index ($\alpha$), as we did in equation (\ref{eqn:yield}). We write $V_{\alpha} = v_{\alpha R} + i v_{\alpha I}$, where $v_{\alpha R(I)}$ are the real (imaginary) part of $V_{\alpha}$ and, $\alpha=1, \cdots, n$, where $n$ is the total number of partial waves. Therefore, we can rewrite equation  (\ref{eqn:trueN}) as
\begin{equation}  N_{true} =  \sum_{\alpha,\alpha'}^{n} { (v_{\alpha R} + i\ v_{\alpha I})} \  { (v_{\alpha' R} - i\ v_{\alpha'I})}\ \Psi_{\alpha,\alpha'}^{r} 
\label{eqn:newtrue}
\end{equation} 

and equation  (\ref{eqn:yield}) as

\begin{equation}  N_{\alpha} =  { (v_{\alpha R} + i\ v_{\alpha I})} \  { (v_{\alpha R} - i\ v_{\alpha I})}\ \Psi_{\alpha,\alpha}^{r} =  { (v_{\alpha R}^{2}+ v_{\alpha I}^{2})} \Psi_{\alpha,\alpha}^{r}
\label{eqn:newtrue}
\end{equation} 

By propagation of errors from $ v_{\alpha R(I)}$ to 
$N_{true}$, the variance of $N_{true}$ is given by~\cite{Hall, Bevington, Orear}

\begin{equation}
\sigma^{2}_{stat} = \mathscr{J} \cdot \mathscr{C} \cdot \mathscr{J}^{T}
\end{equation}
where $\mathscr{C} $ is the covariance or, more general, the error matrix (normally produced by MINUIT) and $\mathscr{J}$ is the Jacobian. The Vs are typically correlated, therefore, the error matrix is not diagonal. $\mathscr{C} $ is a symmetric $(2n) \times (2n)$ matrix. The form of the matrix is

\begin{equation} 
\mathscr{C} =\begin{bmatrix}  \sigma^{2}(v_{1R}) &  \sigma(v_{1R},v_{1I})  & \cdots & \sigma (v_{1R},v_{nR}) & \sigma (v_{1R},v_{nI}) \\ 
 \sigma(v_{1I},v_{1R}) & \sigma^{2}(v_{1I})  & \cdots & \sigma (v_{1I},v_{nR}) & \sigma (v_{1I},v_{nI}) \\
\cdots & \cdots & \cdots & \cdots & \cdots  \\
\cdots & \cdots & \cdots & \cdots & \cdots  \\
\sigma(v_{nR},v_{1R}) &  \sigma(v_{nR},v_{1I})  & \cdots & \sigma^{2}(v_{nR}) & \sigma (v_{nR},v_{nI}) \\
\sigma(v_{nI},v_{1R}) &  \sigma(v_{nI},v_{1I})  & \cdots & \sigma (v_{nI},v_{nR}) & \sigma^{2}(v_{nI}) 
\label{eqn:errormat}
 \end{bmatrix} 
\end{equation} 
where the diagonal terms, $\sigma^{2}(v_{\alpha R(I)})$, represent the standard variance of the $v_{\alpha R(I)}$ values, and the terms $\sigma (v_{\alpha R(I)}, v_{\alpha I(R)})$ off diagonal represent the covariance between $v_{\alpha R(I)}, v_{\alpha I(R)}$.
$\mathscr{J}$ is given by the $\alpha  \times 2n$ vector

\begin{equation} 
\mathscr{J} = \begin{bmatrix} \cdots & \frac{\partial N_{true}}{\partial v_{\alpha R}} & \frac{\partial N_{true}}{\partial v_{\alpha I}} & \cdots 

\end{bmatrix}
\end{equation}

By taken partial derivatives of equation (\ref{eqn:newtrue}), and taken into account that the $ \Psi^{r}_{\alpha ,\alpha'}$ are symmetric, we have

\begin{equation}
\frac{\partial N_{true}} {\partial \  v_{\alpha R(I)}} =  2 \times\sum^{n}_{j=1}  \Psi^{r}_{\alpha ,j} v_{jR(I)}  .
\end{equation}

The total error in the predicted intensity ($N_{true}$) is then obtained from

\begin{equation} \sigma^{2}_{N_{true}} = \sigma^{2}_{sys} + \sigma^{2}_{stat} \end{equation}
The analysis of these errors, i.e. looking for abnormally large contributions,  from different wave sets and fits provides another way to assess the quality of the fit. 
\section{Mass Dependent Fit}
\label{sect:MassDep}

After performing mass-independent fits in each bin of $M$ (or $M$ and $t$) we obtained the predicted distribution of $N_{true}(M)$ for each partial wave. We are ready to analyze these mass distributions. We will first look for regions of enhancement (peaks or valleys) in the distributions and fit a theoretical based distribution to obtain the resonance properties (mass and width). We use the relativistic Breit-Wigner (BW) prescription, with corrections, as explained below. The BW distribution represents just an approximation for the mass distribution (as naively derived below). In general, the properties of the resonances should be obtained from the poles on the complex amplitudes of the S-Matrix Feynman expansion~\cite{Eden}. These poles (and thresholds) are in the complex plane (Riemann surfaces)  and only their projected real axis values can be evaluated experimentally. In the case of multiple poles with same quantum numbers and/or poles far from the real axis, the axis projections can deviate from the BW distribution. The shape of these distributions are also influence by the QCD dynamics. Effective field theories, i.e Chiral Perturbation Theory has been  combined with the dispersion relations to obtain better parametrization of the mass distributions~\cite{Pelaez}. Resonances are now being defined by the PDG as poles in the complex plane and their properties are given as complex numbers~\cite{PDG}. In the model described in this report we used the BW as first approximation, new parametrizations may be included in places where the BW is used. The BW parametrization had worked well in the past when more isolated and narrow meson resonances were studied. Future developments for high statistics experiments covering broader and less defined resonances will may use a more sophisticated parametrization as theory dictates. 

The space propagators for unstable particles (of mass M) produced in a $ab \rightarrow bc$ type reaction are Green's functions satisfying the Klein-Gordon equation \cite{Weinberg}. If we do not consider particles with spin (it turns out that the pole behavior of the propagator is similar for the Dirac equation, i.e. particles with spin), this means that there are functions G($\bar{x}$) such that
\begin{equation}  (\frac{\partial^{2}}{\partial t^{2}} - \nabla^{2}+M^{2})G(\overrightarrow{x},\overrightarrow{y})=-\delta({\overrightarrow{x}-\overrightarrow{y}}) \end{equation} 
$\overrightarrow{x}$ and $\overrightarrow{y}$ are the initial and final points of the trajectory of the particle in the Minkowski space.

For the solution of this equation we write the Fourier transform of the Green function into the momentum space
\begin{equation}  G(\overrightarrow{x},\overrightarrow{y})=\frac{1}{(2\pi)^{4}}\int d^{4}p \frac{e^{-ip(x-y)}}{p^{2}-M^{2} \pm i\epsilon} 
\label{eqn:green}
\end{equation} 
the term $i\epsilon$ is introduced to make the integral defined over the two mass poles. We can give a physical significance to this term if we look at
\begin{equation}  M^{2}+i\epsilon = M(M+i(\epsilon/M)) \end{equation} 
then, we can think about $\epsilon$ as an imaginary contribution to the mass. From the time dependence of the wave function we can also observe
\begin{equation}  e^{-iE t} \rightarrow e^{-i(M+i\epsilon/M) t} = e^{-iMt}e^{-(-\epsilon/M) t} \end{equation} 
therefore, we can associate the term $\epsilon$ with a time decay constant that is associated with a width, $\Gamma$, such that

\begin{equation}   \Gamma = -\epsilon/M.
\label{eqn:gamma}
\end{equation} 

We can then write the energy/mass amplitude distribution as
\begin{equation}  \Psi(w) = \frac{\Psi_{0}}{(w_o^2-w^2-i w_o\Gamma)} \end{equation} 
where we change our notation to make: $w_0 = M $ the mass of the resonance, $w^2 = p^2=s$, four-momentum square or mass squared.

Since
\begin{equation}  \int_{-\infty}^\infty \begin{vmatrix}\Psi(w)\end{vmatrix}^{2}dw=1 \end{equation} 
and $\Gamma(w_o)=\Gamma_0$ (width of the resonance), we obtain
\begin{equation} \boxed{ \Psi(w)=\frac{w_o\Gamma_o}{(w_o^2-w^2-iw_o\Gamma(w))} .} 
\label{eqn:BWE}
\end{equation} 

This is the well known relativistic Breit-Wigner (BW) mass distribution amplitude~\cite{Blatt}. In the case where $\Gamma_o << w_o$, therefore $w  \sim w_o$ and $\Gamma \sim \Gamma_o$; we obtain the non-relativistic case

\begin{equation}  \Psi(w)=\frac{\Gamma_o}{\frac{(w_o-w)(w_o+w)}{w_o}-i\Gamma_o} ~\sim \frac{\Gamma_o}{\frac{(w_o-w)(2w_o)}{w_o}-i\Gamma_o}  \sim  \frac{\Gamma_{o}/2} {w_o-w-i\Gamma_{o}/2}. \end{equation}  

We use in our analysis the relativistic formula.
Introducing the phase shift $\delta$ such that
\begin{equation}  cos\ \delta(w) = \frac{w_o^2-w^2}{w_o^2} \end{equation} 
\begin{equation}  sin\ \delta(w) = \frac{w_o\Gamma(w)}{w_o^2} 
\label{eqn:sindel}
\end{equation} 
therefore
\begin{equation}  e^{-i\delta} = cos \delta -i sin \delta = \frac{ w_o^2-w^2-iw_o\Gamma(w)}{w_o^2} 
\label{eqn:edel}
\end{equation} 
\begin{equation}  cot \delta = \frac{cos\delta}{sin\delta}=\frac{w_o^2-w^2}{w_o\Gamma(w)} \end{equation} 
and using equations (\ref{eqn:sindel}) and (\ref{eqn:edel}) into equation (\ref{eqn:BWE}):
\begin{equation}  \boxed{ \Psi(w) = \frac{\Gamma_o}{\Gamma(w)} e^{i\delta}sin \delta . } \end{equation}

Resonance shapes are also influenced by {\it centrifugal-barrier effects} caused by the angular (spin) factors on the potentials. We will follow the discussion and use the results obtained by Von Hippel and Quigg \cite{Hippel}.

The radial (semiclassical) component of the wave ($A_{l}$) equation has the form \cite{Leader}
\begin{equation}  \frac{\partial^2}{\partial \rho^2}A_{l}(\rho) \cong \begin{pmatrix} \frac{b^2}{r^2}-1\end{pmatrix}A_{l}(\rho) \end{equation} 
where
\begin{equation}  \rho = q r \end{equation} 
and $b$ is the impact parameter
\begin{equation}  \pi b^{2} \sim  \frac{l(l+1)}{q^2} \end{equation} 
$q$ is the breakup momentum of the decay products in the rest frame of the decaying particle. 
The solution of this equation for large $r$ (outgoing wave) is given in terms of the spherical Hankel functions ($h_{l}^{1}$), in the form
\begin{equation}  A_{l}(\rho) \cong i q r  h_{l}^{1}(qr) .\end{equation} 

Let's assume that the centrifugal-barrier will act up to some {\it interaction radius} $R$ and consider that $R << b$, then, that radial probability density will grow rapidly with $r$. The probability to reach the barrier at $R$ will be

\begin{equation}  (qR)^2\begin{vmatrix}h_{l}^{1}(qr)\end{vmatrix}^{2} . \end{equation} 

The {\it transmission coefficient} (inverse probability) through the barrier could be defined as

\begin{equation}  T_{l}(R/b) \cong (qR)^{-2}\begin{vmatrix}h_{l}^{1}(qr)\end{vmatrix}^{-2} \end{equation} 
Therefore, the time associated with the resonance (or the width, $\Gamma_o$) needs to be weighted by this probability (of {\it escape}). Using equation (\ref{eqn:gamma}) weighted by the transmission coefficient we have
\begin{equation}   \Gamma(w) = (q/w) T_{l}(R/b) \end{equation} 
and normalizing
\begin{equation}   \Gamma(q,w,R,b) = \Gamma_{o} \frac{(q/w) T_{l}(R/b)}{(q_{o}/w_{o}) T_{l}(R/b_{o})} \end{equation} 
where $w_{o}$,  $q_o$ and $b_o$ are the values of the mass, momentum and impact parameter at the resonance, and $\Gamma_o$ the energy-independent resonance's width.

Changing the notation to call $F^{2}_{l}=T_{l}$, taking $R$ about one fermi and recalling that $b \sim \frac{1}{q}$; we have

\begin{equation}  \boxed{ \Gamma(w,q)=\Gamma_{0}\frac{w_{0}qF^{2}_{l}(q)}{wq_{0}F^{2}_{l}(q_{0})}. } \end{equation} 

The Hankel functions are related to the spherical Bessel functions, these are calculated by a power-series expansion~\cite{Morse}

\begin{equation}  h_{l}^{(1)}(x)=(-i/x)e^{i(x-l\pi/2)} \sum_{k=0}^{l} (-1)^{k} \frac{(l+k)!}{k!(l-k)!}(2ix)^{-k} \end{equation} 
and
\begin{equation}  F_{l}(q) = \sqrt{q^{-2} \begin{vmatrix}h_{l}^{1}(q)\end{vmatrix}^{-2}} .\end{equation} 

The $F$'s are called the Blatt-Weisskopf centrifugal-barrier factors \cite{Blatt}. The first four are given by

\begin{equation}  F_{0}(q)=1 \end{equation} 

\begin{equation}  F_{1}(q)=\sqrt{\frac{2z}{z+1}} \end{equation} 

\begin{equation}  F_{2}(q)=\sqrt{\frac{13z^{2}}{(z-3)^{2}+9z}} \end{equation} 

\begin{equation}  F_{3}(q)=\sqrt{\frac{277z^{3}}{z(z-15)^{2}+9(2z-5)^{2}}} \end{equation} 
with
\begin{equation}  z=(q/0.1973)^{2}\ in\ GeV \end{equation}

The value of $q_{R}=0.1973 GeV/c$ corresponds to a centrifugal barrier at 1 fermi.

The mass dependent amplitude must also be modified to account for the centrifugal barrier, and the new amplitude can be written as
\begin{equation}  V_l(w) \equiv e^{i\phi_l} \Psi(w) F_l(q) \end{equation} 
where we have introduced a $\phi_{l}$ production phase for the wave $l$ (which is independent of the mass).

\begin{equation}  \boxed{ V_l(w)=e^{i(\phi_l+\delta(w))}  \frac{\Gamma_o}{\Gamma(w)} sin\delta F_l(q) }
\label{eqn:massd}
 \end{equation} 

The measured (cross section) distribution is
\begin{equation}  \frac{d\sigma}{dw} = \begin{vmatrix} V_{l}(w)\end{vmatrix}^{2}pq \end{equation} 
therefore
\begin{equation}  \frac{d\sigma}{dw} = e^{2i\delta(w)}  \frac{\Gamma_o^2}{\Gamma(w)^2} sin^2\delta F^2_l(q) \end{equation} 
or using
\begin{equation}  e^{2i\delta} =\frac{w_o^4}{(w_o^2-w^2)^2+w_o^2\Gamma^2} \end{equation} 
and 
\begin{equation}  sin^2 \delta= \frac{ \Gamma^2}{w_0^2} \end{equation} 
we obtain

\begin{equation}  \boxed{ \frac{d\sigma}{dw} =  \frac{1}{(w_o^2-w^2)^2+w_o^2\Gamma^2} w_o^2 \Gamma_o^2 F^2_l(q) .} 
\label{eqn:masscross}
\end{equation} 

The observed mass distribution can be hard to fit through this function. Sometimes we can find a better fit to the data just phenomenologically adjusting the observed distribution by a polynomial function of $(w_o-w)$:

\begin{equation}  \frac{d\sigma}{dw} =  \frac{1}{(w_o^2-w^2)^2+w_o^2\Gamma^2} w_o^2 \Gamma_o^2 F^2_l(q) \mathbb{P}(w_o-w) \end{equation} 

For example, BNL-E852 \cite{SUC97}, has taken the following form in one of their analyses

\begin{equation}  \frac{d\sigma}{dw} =  \frac{1}{(w_o^2-w^2)^2+w_o^2\Gamma^2} w_o^2 \Gamma_o^2 F^2_l(q)[a_l+b_l(w-w^o_l)+c_l(w-w^o_l)^2] .\end{equation} 
Fitting this distribution to a region with an identified mass distribution enhancement , we obtain values for $w_{o}$ and $\Gamma_{o}$. However, as discussed earlier, a mass peak (enhancement) in a mass distribution is not akin of a resonance. A phase motion study must also be done and we come to this next.

\section{Phase Motion}
\label{sect:Phase}

From the mass-independent fit we obtained complex amplitudes for each wave in each mass bin. We have used their magnitudes (intensities) to calculate predicted counts and distributions, but we can also use their complex phase to look for resonant behavior. A single wave phase is arbitrary and thus ambiguous, therefore, only the difference of phases between two waves contains physical information. We normally examine the behavior of the phase difference (phase motion) of the wave under study against a well established resonant wave.

The structure of the phase-motion is determined by the mass-dependent cross section. After performing a mass-dependent fit, we have a form for the mass distribution for each ($\epsilon, k, b$) wave. These mass distributions will have the form described in the previous section
\begin{equation}  \frac{d\sigma}{dw} =  \frac{1}{(w_o^2-w^2)^2+w_o^2\Gamma^2} w_o^2 \Gamma_o^2 F^2_l(q) \end{equation} 
or
\begin{equation}  \frac{d\sigma}{dw} = e^{2i\delta(w)}  \frac{\Gamma_o^2}{\Gamma(w)^2} sin^2\delta F^2_l(q) . \end{equation}

The values of $\Gamma_o$ and $w_o$ are determined by fitting the above form to the mass distribution. Knowing those values, the mass dependent amplitudes may be determined up to a phase and a normalization constant. These are
\begin{equation}  A_{l}(w)=A_{l_{o}} e^{i(\phi_{l}+\delta_{l}(w))}  \frac{\Gamma_{o_{l}}}{\Gamma_{l}(w)} sin\delta_{l}(w) F_{l}(q) 
\label{eqn:ampmass}
\end{equation} 
where we use the symbol $l$ for the collective identification of the ($\epsilon, k, b$) waves, $A_{l_o}$ is a normalization constant and $\phi_{l}$ is an arbitrary phase (independent of the mass, $w$).

The total phase of the amplitude is then
\begin{equation}  \Phi_l = \phi_l + \delta_{l} (w) \end{equation} 
and the phase difference between two partial waves, $l_{1}$ and $l_{2}$ is
\begin{equation}  \Delta \Phi_{l_1,l_2} (w)= \Phi_{l_{1}} - \Phi_{l_{2}} = \phi_{l_{1}} -\phi_{l_{2}} + \delta_{l_{1}}(w) - \delta_{l_{2}} (w) .\end{equation} 
Given the calculated values of $A_{l_1}$ and $A_{l_2}$ (using equation \ref{eqn:ampmass}), we obtain
\begin{equation}  A_{l_{1}}(w)=A_{l_{o_{1}}} e^{i(\phi_{l_{1}}+\delta_{l_{1}}(w))}  \frac{\Gamma_{o_{l_{1}}}}{\Gamma_{l_{1}}(w)} sin\delta_{l_{1}}(w) F_{l_{1}}(q) \end{equation} 
\begin{equation}  A_{l_{2}}(w)=A_{l_{o_{2}}} e^{i(\phi_{l_{2}}+\delta_{l_{2}}(w))}  \frac{\Gamma_{o_{l_{2}}}}{\Gamma_{l_{2}}(w)} sin\delta_{l_{2}}(w) F_{l_{2}}(q) . \end{equation} 

Therefore
\begin{equation*}  A_{l_{1}}(w)A^{*}_{l_{2}}(w) =A_{l_{o_{1}}}  \frac{\Gamma_{o_{l_{1}}}}{\Gamma_{l_{1}}(w)} sin\delta_{l_{1}}(w) F_{l_{1}}(q) A_{l_{o_{2}}}  \frac{\Gamma_{o_{l_{2}}}}{\Gamma_{l_{2}}(w)} \end{equation*} \begin{equation} \times\  sin\delta_{l_{2}}(w) F_{l_{2}}(q) e^{i(\phi_{l_{1}}-\phi_{l_{2}} +\delta_{l_{1}}(w)-\delta_{l_{2}}(w))}  \end{equation} 
then
\begin{equation}  A_{l_{1}}(w)A^{*}_{l_{2}}(w) = C(w) e^{i\Delta \Phi_{l_1,l_2} (w)} \end{equation}
\begin{equation}  \boxed{ \Delta \Phi_{l_1,l_2} (w) = arctan \begin{bmatrix} \frac{\it{Im}(A_{l_{1}}(w)A^{*}_{l_{2}}(w))}{\it{Re}(A_{l_{1}}(w)A^{*}_{l_{2}}(w))}  \end{bmatrix}  }
\label{eqn:delphi}
\end{equation}
where we can obtain the phase difference between two waves from their calculated (fitted) mass dependent amplitudes.

\begin{figure}[!htp]
\centerline{\includegraphics[width=12cm]{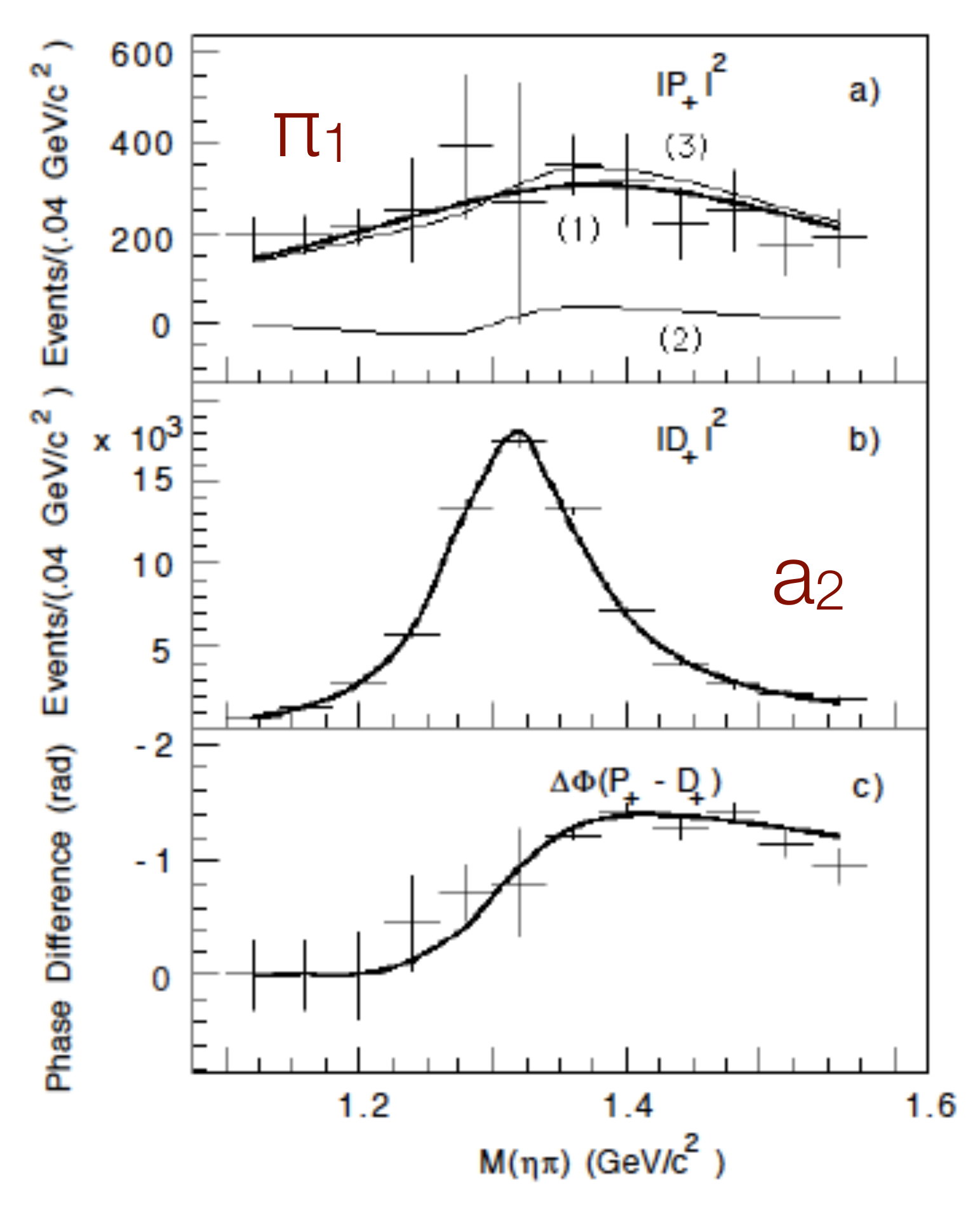}}
\caption{\label{fig:852phase}An example of phase motion in the $\eta \pi$ system from E852 data~\cite{865phase}. The plot shows the fit results of the mass-dependent PWA (curves) to the mass independent PWA (crosses) fit results. Plot a) shows the $P_{+}$ results and b) the $D_{+}$ results. Plot c) shows their relative phase $\Delta \Phi(P_{+}-D_{+})$. The curve is calculated using equation (\ref{eqn:delphi}).}
\end{figure}

Since:
\begin{equation}  cot (\delta) = \frac{w_o^{2}-w^2}{w_o\Gamma} \end{equation} 
we expect the phase difference to change from $-\pi/2$ to $\pi/2$ with rapid {\it phase motion} around the value $w=w_o$. However, because of the difference $\Delta \Phi_{l_1,l_2}$ differs from ($\delta_{l_{1}}(w) - \delta_{l_{2}} (w)$) by a constant phase ($\Delta \phi$), this change might not happen exactly at the resonance mass, that is where the intensity is maximum. The {\it sharpness} of this phase motion will also depend on the $\delta(w)$ function of the reference wave to which the phase motion is measured. $\delta(w)$ depends on the BW parameters, i.e. the width and mass of the resonances. Therefore, the presence of a resonance must be associated with structure on the $\Delta \Phi_{l_1,l_2} (w)$ distribution. We can perform a $\chi^{2}$ fit to the combine mass and phase distributions using formulas (\ref{eqn:massd}) and (\ref{eqn:delphi}). For example, in  figure~\ref{fig:852phase}, we can see an example of a phase motion associated with a resonant peak found by the E852 collaboration at Brookhaven. The figure shows the phase motion of the $P_{+}$ wave versus the $D_{+}$ wave in the $\eta \pi$ system~\cite{865phase}. It is important to notice that the phase motion is related to the complex behavior of the amplitudes (phase) and not to the intensities, therefore, phase motion is a good tool for the study of resonances with small cross sections.

Let's discuss, in a little more detail, how we can extract (from data) the phase differences between two waves and their errors. Equation (\ref{eqn:yield}) gives the number of predicted events in each mass bin for a given wave, we can write

\begin{equation}  N_{\alpha} = \big{|} V_{\alpha} (\Psi_{\alpha,\alpha}^{r})^{1/2} \big{|}^{2} 
\end{equation} 

%or

%\begin{equation}  N_{b} = {^{1} V_{b}} \  {^{1}V^{*}_{b}}\  {^{1,1}} \Psi_{b,b}^{r} +
%{^{-1} V_{b}} \  {^{1}V^{*}_{b}}\  {^{-1,1}} \Psi_{b,b}^{r}  + {^{-1} V_{b}} \  {^{-1}V^{*}_{b}}\  {^{-1,-1}} %\Psi_{b,b}^{r}  + {^{1} V_{b}} \  {^{-1}V^{*}_{b}}\  {^{1,-1}} \Psi_{b,b}^{r} 
%\end{equation} 

therefore, the predicted number of events can be considered as the square of a complex amplitude ${\cal{A}}_{b}$
such as

\begin{equation} {\cal{A}}_{b} =  V_{\alpha} (\Psi_{\alpha,\alpha}^{r})^{1/2}  .
\end{equation} 

 The phase difference between two waves is then obtained from

\begin{equation}
\Delta \phi = (\phi_{1} - \phi_{2}).
\end{equation}

We will write ${\cal{A}}_{1} = A_{1}e^{i (\phi_{1} + \delta)}$ and ${\cal{A}}_{2} = A_{2}e^{i \phi_{2}}$ , with an arbitrary phase ($\delta$) since only the magnitude is derived from data. Therefore (with $C=e^{\delta}$ a constant),

\begin{equation}
{\cal{A}}_{1}{\cal{A}}^{*}_{2} = C A_{1}A_{2}e^{i(\phi_{1}-\phi_{2})} = C  A_{1}A_{2}e^{i \Delta \phi}
\end{equation}

since

\begin{equation}
{\cal{A}}_{1}{\cal{A}}^{*}_{2} =  V_{1} V^{*}_{2} (\Psi_{1,1}^{r} \Psi_{2,2}^{r})^{1/2} 
\end{equation}

we have

\begin{equation}
C  A_{1}A_{2}e^{i \Delta \phi} = V_{1} V^{*}_{2} (\Psi_{1,1}^{r} \Psi_{2,2}^{r})^{1/2} 
\end{equation}

and therefore

\begin{equation} \boxed{ 
\Delta \phi =arctan \bigg{(} \frac{Im(V_{1} V^{*}_{2})}{Re(V_{1} V^{*}_{2})} \bigg{)}. }
\end{equation}

A similar equation to (\ref{eqn:delphi}) but using the fitted production amplitudes from the mass-independent analysis.

The errors in the phase differences are calculated by (error propagation) taken $\Delta \phi$ as a function of the $V_{1}$ and  $V^{*}_{2}$ values. Calling $a=Re(V_{1} V^{*}_{2})$ and $b=Im(V_{1} V^{*}_{2})$ we have

\begin{equation}
{\sigma^{2}_{\Delta \phi}} = \bigg{(}\frac{\partial {\Delta \phi}}{\partial {a}} \bigg{)}^{2} {\sigma^{2}_{a}} + \bigg{(}\frac{\partial {\Delta \phi}}{\partial {b}} \bigg{)}^{2} {\sigma^{2}_{b}} + 2 \bigg{(}\frac{\partial {\Delta \phi}}{\partial {a}} \bigg{)} \bigg{(}\frac{\partial {\Delta \phi}}{\partial {b}} \bigg{)} {\sigma^{2}_{ab}}
\end{equation}

where

\begin{equation}
\frac{\partial {\Delta \phi}}{\partial {a}} = \frac{a}{a^{2}+b^{2}}
\
;
\  
\frac{\partial {\Delta \phi}}{\partial {b}} = \frac{-b}{a^{2}+b^{2}}.
\end{equation}

The errors in $a$ and $b$ are derived from the error matrix, $\mathscr{C}$, of equation (\ref{eqn:errormat}). We have

\begin{equation}
\sigma^{2}_{a(b)} = \mathscr{J}_{a(b)} \cdot \mathscr{C} \cdot \mathscr{J}^{T}_{a(b)}
\end{equation}

and

\begin{equation}
\sigma^{2}_{ab} = \mathscr{J}_{a} \cdot \mathscr{C} \cdot \mathscr{J}^{T}_{b}.
\end{equation}

We use the notation $V_{1} = v_{1R} + iv_{1I}$ and $V^{*}_{2} = v_{2R} - iv_{2I}$, therefore

\begin{equation}
V_{1} V^{*}_{2} = (v_{1R}v_{2R} + v_{1I}v_{2I}) + i(v_{1I}v_{2R} - v_{1R}v_{2I})
\end{equation}

from which we found

\begin{equation}
\mathscr{J}_{a} = (v_{2R},v_{2I},v_{1R},v_{1I})
%\end{equation}
\
and
\
%\begin{equation}
\mathscr{J}_{b} = (-v_{2I},v_{2R},v_{1I},-v_{1R}).
\end{equation}

\section{Moments Analysis}

We have seen that, after we take care of the dynamic quantities by binning the data in $(M,t)$ bins, the intensity, I($\tau$), depends mainly on angular quantities. This is exactly true in the case of two particles final states. Therefore, it is natural to try a multipole expansion of the intensity distribution~\cite{Harmonics}.

Let's consider the case of two spinless particles in the final state, produced by a definite spin particle or unpolarized photon beam.
We have seen that in this case the amplitudes are
\begin{equation}  A_{lm}(\tau) = A_{lm}(\phi,\theta) = \sqrt{\frac{2l+1}{4\pi}} { D^{l\ *}_{m0}(\phi,\theta,0)} \end{equation} 
and the intensity is
\begin{equation}  I(\tau) =I(\phi,\theta) = \sum_{l.m,l',m'} A_{l,m}(\phi,\theta)\ \rho_{l,m,l',m'} A^{*}_{l',m'}(\phi,\theta) \end{equation} 
where $\rho_{l,m,l',m'}$ are the resonance spin density matrices, therefore
\begin{equation}  I(\phi,\theta) = \sum_{l.m,l',m'} \sqrt{\frac{2l+1}{4\pi}}\sqrt{\frac{2l'+1}{4\pi}}\ \rho_{l,m,l',m'} { D^{l\ *}_{m0}(\phi,\theta,0)}{ D^{l'}_{m'0}(\phi,\theta,0)} .
\label{eqn:Inorm}
\end{equation} 

A multipole expansion of the intensity function on spherical harmonics will have the form \cite{Harmonics}
\begin{equation}  I(\phi,\theta) = \sum_{L,M} H(LM) {Y^{M}_{L}(\phi,\theta)} \end{equation} 
where the coefficients $H(LM)$ are called the {\it moments}.
But since
\begin{equation}   Y^{M}_{L}(\phi,\theta)  = \sqrt{\frac{2L+1}{4\pi}} D^{L\ *}_{M0}(\phi,\theta,0) \end{equation} 
we can write
\begin{equation}  I(\phi,\theta)=\sum_{LM} \begin{pmatrix} \frac{2L+1}{4\pi} \end{pmatrix} H(LM) D^{L\ *}_{M0}(\phi,\theta,0) .
\label{eqn:mom1}
\end{equation} 

Using $L=l+l'$ and $M=m+m'$, and the following relation between Wigner-D functions (see Appendix C)
\begin{equation}  D^{l\ *}_{m0}D^{l'}_{m'0} = \sum_{LM} \frac{2L+1}{2l+1} (l'm'LM|lm)(l'0L0|l0)D^{L\ *}_{M0}
\label{eqn:mulD}
 \end{equation} 
where $ (l'm'LM|lm)$ and $(l'0L0|l0)$ are Clebsch-Gordan coefficients, we can write equation (\ref{eqn:Inorm}) as
\begin{equation}  I(\phi,\theta) = \sum_{l.m,l',m'} \sqrt{\frac{2l+1}{4\pi}}\sqrt{\frac{2l'+1}{4\pi}}\ \rho_{l,m,l',m'} \sum_{LM} \frac{2L+1}{2l+1} (l'm'LM|lm)(l'0L0|l0)D^{L\ *}_{M0}\end{equation} 
re-arranging terms
\begin{equation}  I(\phi,\theta) = \sum_{LM}   \frac{2L+1}{4\pi} \sum_{l.m,l',m'} {(\frac{2l'+1}{2l+1})}^{1/2}\ \rho_{l,m,l',m'} (l'm'LM|lm)(l'0L0|l0)D^{L\ *}_{M0}
\label{eqn:mom2}
\end{equation} 
comparing equations (\ref{eqn:mom1}) and (\ref{eqn:mom2}), we see that
\begin{equation}  \boxed{ H(LM) = \sum_{l.m,l',m'} {(\frac{2l'+1}{2l+1})}^{1/2}\ \rho_{l,m,l',m'} (l'm'LM|lm)(l'0L0|l0) .} 
\label{eqn:momB}
\end{equation} 

This equation provides a relation between the {\it partial waves} and the {\it moments}. This important relation allows a check of PWA results. We can use a full PWA to obtain the relevant waves contributing to the intensity, and then use the obtained waves to {\it predict moments}. We can then compare the measured moments against the calculated moments.

Let's consider the extraction of the moments from the data. 
Using the following relation between Wigner-D functions

\begin{equation}  \int d{\Omega} D^{l\ *}_{m0}D^{l'}_{m'0}D^{L}_{M0} = \frac{8\pi^{2}}{2l+1} (l'm'LM|lm)(l'0L0|l0) \end{equation} 
into equation (\ref{eqn:momB})
\begin{equation}  H(LM) = \sum_{l.m,l',m'} {(\frac{2l'+1}{2l+1})}^{1/2}\ \rho_{l,m,l',m'} \frac{2l+1}{8\pi^{2}} \int d{\Omega} D^{l\ *}_{m0}D^{l'}_{m'0}D^{L}_{M0}   \end{equation} 
and re-arranging terms, we have
\begin{equation}  H(LM) = \int d{\Omega} \bigg{[} \sum_{l.m,l',m'} \sqrt{\frac{2l+1}{4\pi}}\sqrt{\frac{2l'+1}{4\pi}} \ \rho_{l,m,l',m'}    D^{l\ *}_{m0}D^{l'}_{m'0} \bigg{]} D^{L}_{M0}   \end{equation} 
or 
\begin{equation}  H(LM) = \int d{\Omega} I(\Omega) D^{L}_{M0}(\phi,\theta,0) .  \end{equation} 

We find, then,  that the moments are the intensity-averaged Wigner-D functions in a $(M,t)$ bin
\begin{equation}  \boxed{ H(LM) = <D^{L}_{M0}(\phi,\theta,0)>  } \end{equation} 
we will call these, the {\it normalized moments}.

In particular
\begin{equation}  H(00) = \int d{\Omega} I(\Omega) = \mbox{number of events in bin (M,t)}  . \end{equation} 

Let's call $H_{exp}$, the {\it experimental moments}, those moments are affected by detection acceptance.  We introduce an acceptance function, $\eta(\Omega)$, such that $\eta = 1$ for accepted events and $\eta =0$ for unaccepted events.
Therefore
\begin{equation}  H_{exp}(LM) = \int d\Omega\ \eta(\Omega) I(\Omega) D^{L}_{M0}(\phi,\theta,0)    \end{equation} 
The intensity is
\begin{equation}  I(\Omega)=\sum_{L'M'} \begin{pmatrix} \frac{2L'+1}{4\pi} \end{pmatrix} H(L'M') D^{L'\ *}_{M'0}(\phi,\theta,0) \end{equation} 
where $H(L'M')$ are the {\it true moments} (those calculated from the fitted waves).

Therefore
\begin{equation}  H_{exp}(LM) = \int d\Omega\ \eta(\Omega) \sum_{L'M'} \begin{pmatrix} \frac{2L'+1}{4\pi} \end{pmatrix} H(L'M') D^{L'\ *}_{M'0}(\phi,\theta,0)  D^{L}_{M0}(\phi,\theta,0) \end{equation} 
and re-arranging
\begin{equation}  H_{exp}(LM) =  \sum_{L'M'} H(L'M') \int d\Omega\ \eta(\Omega)  \begin{pmatrix} \frac{2L'+1}{4\pi} \end{pmatrix}  D^{L'\ *}_{M'0}(\phi,\theta,0)  D^{L}_{M0}(\phi,\theta,0) . \end{equation} 
 
Let's call
\begin{equation}  \Phi(LML'M') = \int d\Omega\ \eta(\Omega)  \begin{pmatrix} \frac{2L'+1}{4\pi} \end{pmatrix}  D^{L'\ *}_{M'0}(\phi,\theta,0)  D^{L}_{M0}(\phi,\theta,0) \end{equation} 
this function acts as a normalization integral, we have
\begin{equation}  H_{exp}(LM) =  \sum_{L'M'} H(L'M') \Phi(LML'M') . \end{equation} 

The normalization integral can be calculated by MC, generating $N_{g}$ events, of which $N_{a}$ are accepted, we have

\begin{equation}  \Phi(LML'M') = \frac{1}{N_{g}}  \begin{pmatrix} \frac{2L'+1}{4\pi} \end{pmatrix}  \sum^{N_{a}}_{i} D^{L'\ *}_{M'0}(\phi_{i},\theta_{i},0)  D^{L}_{M0}(\phi_{i},\theta_{i},0) .\end{equation} 

As discussed before (section \ref{sect:reflect}), we calculate the intensity, amplitudes and density matrix in the reflectivity basis to take into account selection rules based on parity conservation. 

Let's consider an unpolarized photon beam, where $\epsilon = \epsilon'$, and therefore the resonance spin density matrix is block diagonal. We will now calculate moments on the reflectivity basis starting from
\begin{equation}  I(\phi,\theta) = \sum_{\epsilon} \sum_{l.m,l',m'} \sqrt{\frac{2l+1}{4\pi}}\sqrt{\frac{2l'+1}{4\pi}}\ {^\epsilon \rho_{l,m,l',m'}} {^\epsilon D^{l\ *}_{m0}(\phi,\theta,0)}{^\epsilon D^{l'}_{m'0}(\phi,\theta,0)}. \end{equation} 

The Wigner-D function in the reflectivity basis are related to the canonical basis through equation (\ref{eqn:dofref})

\begin{equation}   ^\epsilon D _{m0}^{l *}(\phi,\theta,0)=\theta(m)\begin{bmatrix} D _{m0}^{l *}(\phi,\theta,0)-\epsilon D _{m0}^{l}(\phi,\theta,0)\end{bmatrix} .\end{equation} 
Therefore
\begin{equation*}  ^\epsilon D^{l\ *}_{m0} {^\epsilon D^{l'}_{m'0}} =  \theta(m)\begin{bmatrix} D _{m0}^{l *}-\epsilon D _{m0}^{l}\end{bmatrix}\theta(m')\begin{bmatrix} D _{m'0}^{l'}-\epsilon D _{m'0}^{l' *}\end{bmatrix} = \end{equation*} 
\begin{equation}   =  \theta(m')\theta(m)\begin{bmatrix} D _{m0}^{l *}D _{m'0}^{l'}-\epsilon D _{m0}^{l}D _{m'0}^{l'}-\epsilon D _{m0}^{l *}D _{m'0}^{l' *}+\epsilon^{2} D _{m'0}^{l' *}D _{m0}^{l}\end{bmatrix}. \end{equation} 
Using that $\epsilon^{2}=1$ and that : $D _{m0}^{l *}=(-1)^{m}D _{-m0}^{l}$, we have
\begin{equation*}   ^\epsilon D^{l\ *}_{m0} {^\epsilon D^{l'}_{m'0}}  =  \theta(m')\theta(m)\lbrack D _{m0}^{l *}D _{m'0}^{l'}-\epsilon(-1)^{m} D _{-m0}^{l *}D _{m'0}^{l'} \end{equation*} 
\begin{equation} -\epsilon (-1)^{m'}D _{m0}^{l *}D _{-m'0}^{l'}+(-1)^{m+m'}D _{m'0}^{l'}D _{m0}^{l *}\rbrack  \end{equation} 
and using
\begin{equation}  D^{l\ *}_{m0} D^{l'}_{m'0} = \sum_{LM} \frac{2L+1}{2l+1} (l'm'LM|lm)(l'0L0|l0)D^{L\ *}_{M0} \end{equation} 
we have (with $M=m+m'$)
\begin{equation*}  ^\epsilon D^{l\ *}_{m0} {^\epsilon D^{l'}_{m'0}}  = \end{equation*} 
\begin{equation*}  =  \sum_{LM} \frac{2L+1}{2l+1} \theta(m')\theta(m) \lbrack (l'm'LM|lm)-\epsilon(-1)^{m} (l'm'LM|l-m)  \end{equation*}
\begin{equation} -\epsilon (-1)^{m'}(l'-m'LM|lm)+(-1)^{M}(l'm'L-M|lm) \rbrack (l'0L0|l0)D^{L\ *}_{M0} . \end{equation} 
Let's introduce the notation (a {\it combination of Clebsch-Gordan coefficients})
\begin{equation*}  ^\epsilon B [l'm'LMlm]= \theta(m')\theta(m) \lbrack (l'm'LM|lm)-\epsilon(-1)^{m} (l'm'LM|l-m)  \end{equation*}
\begin{equation} -\epsilon (-1)^{m'}(l'-m'LM|lm) +(-1)^{M}(l'm'L-M|lm) \rbrack \end{equation} 
then
\begin{equation}  ^\epsilon D^{l\ *}_{m0} {^\epsilon D^{l'}_{m'0}}  = \end{equation} 
\begin{equation}  =  \sum_{LM} \frac{2L+1}{2l+1} {^\epsilon B} [l'm'LMlm] (l'0L0|l0)D^{L\ *}_{M0} \end{equation} 
similar to our previous expression, in equation (\ref{eqn:mulD}). Therefore, following the same previous steps, we found:
\begin{equation}  \boxed{ H(LM) = \sum_{\epsilon} \sum_{l.m,l',m'} {(\frac{2l'+1}{2l+1})}^{1/2} \ {^\epsilon\rho_{l,m,l',m'}} {^\epsilon B} [l'm'LMlm](l'0L0|l0) } 
\label{eqn:mom3}
\end{equation} 

This formula shows how to calculate moments as a function of partial waves (included in the density matrix).
The general form (for more than two final particles) is rather complicated, and then not very useful in practice \cite{SUC71}.

Let's calculate a rather simple example: let's take $k=1$ (rank one) and for each wave $l$ (S,P,D...) only consider the values $m=0$ and $m= 1$ ($ {^\epsilon V_{lmk}}$=0 if $m > 1$).  We also will consider the case of an unpolarized beam, therefore the spin density matrix of the photon can be factorized out. Since
\begin{equation}  {^\epsilon\rho_{l,m,l',m'}} = \sum_{k} {^\epsilon V_{lmk}}{^\epsilon V^{*}_{l'm'k}} =  {^\epsilon V_{lm}}{^\epsilon V^{*}_{l'm'}} \end{equation} 
Remembering that for $m=0$ the only available states are for negative reflectivity, for each value of $l$, we have
\begin{equation}  {^{(-)} V_{l0}}\ ; {^{(-)} V_{l1}}\ and\  {^{(+)} V_{l1}} \end{equation} 
It is a standard notation to call:\\
For $l=0$
\begin{equation}  S_{0} = {^{(-)} V_{00}}\ \end{equation} 
For $l=1$
\begin{equation}  P_{0} = {^{(-)} V_{10}}\ ; P_{-} = {^{(-)} V_{11}}\ and\ P_{+} =  {^{(+)} V_{11}} \end{equation} 
For $l=2$
\begin{equation}  D_{0} = {^{(-)} V_{20}}\ ; D_{-} = {^{(-)} V_{21}}\ and\ D_{+} = {^{(+)} V_{21}} \end{equation} 
and so on, such that for a wave $l$, we can use $[l]_{0}$, $[l]_{-}$ and $[l]_{+}$, where $[l]=S,P,D, \cdots$
Using relation (\ref{eqn:mom3}), we obtain the H(L0) moments such that
\begin{equation}  H(L0) = \sum_{\epsilon} \sum_{l,m,l',m'}  {(\frac{2l'+1}{2l+1})}^{1/2} \ {^\epsilon\rho_{l,m,l',m'}} {^\epsilon B} [l'm'L0lm](l'0L0|l0) \end{equation} 
or
\begin{equation}  H(L0) =  \sum_{l.l'}  {(\frac{2l'+1}{2l+1})}^{1/2} (l'0L0|l0) \sum_{\epsilon=-,+} \sum_{m,m'=0,1}  {^\epsilon V_{lm}}{^\epsilon V^{*}_{l'm'}} {^\epsilon B} [l'm'L0lm] \end{equation} 

Let's look at the last term of this relation

\begin{equation*}  \sum_{\epsilon=-,+} \sum_{m,m'=0,1}  {^\epsilon V_{lm}}{^\epsilon V^{*}_{l'm'}} {^\epsilon B} [l'm'LMlm] = \end{equation*}
\begin{equation}
\sum_{m,m'=0,1} \begin{bmatrix}  {^{(-)} V_{lm}}{^{(-)} V^{*}_{l'm'}} {^{(-)} B} [l'm'L0lm] + {^{(+)} V_{lm}}{^{(+)} V^{*}_{l'm'}} {^{(+)} B} [l'm'L0lm] \end{bmatrix}  . \end{equation} 
Lat's use
\begin{equation*}  ^\epsilon B [l'm'L0lm]= \theta(m')\theta(m) \lbrack (l'm'L0|lm)-\epsilon(-1)^{m} (l'm'L0|l-m) \end{equation*}
\begin{equation} -\epsilon (-1)^{m'}(l'-m'L0|lm)+(l'm'L0|lm)\rbrack \end{equation} 
We can see that there are a few possible combinations that are not zero
\begin{equation}  ^\epsilon B [l'0L0l0]= \begin{bmatrix} (l'0L0|l0)-\epsilon (l'0L0|l0) \end{bmatrix} \end{equation} 
therefore
\begin{equation}  ^{(-)} B [l'0L0l0]= (l'0L0|l0) \end{equation} 
\begin{equation}  ^{(+)} B [l'0L0l0]= 0 \end{equation} 
and
\begin{equation}  ^\epsilon B [l'1L0l1]= (l'1L0|l1) . \end{equation} 

Therefore
\begin{equation*}  H(L0) =  \sum_{l.l'}  {(\frac{2l'+1}{2l+1})}^{1/2} (l'0L0|l0) \times  \end{equation*} 
\begin{equation} \begin{bmatrix} [l]_{0}[l']^{*}_{0}(l'0L0|l0) + [l]_{-}[l']^{*}_{-}(l'1L0|l1) + [l]_{+}[l']^{*}_{+}(l'1L0|l1) \end{bmatrix} \end{equation} 
In a similar fashion, it is possible to prove that
\begin{equation}  H(L1) =  \sum_{l.l'}  {(\frac{2l'+1}{2l+1})}^{1/2} (l'0L0|l0) \begin{bmatrix} [l]_{-}[l']^{*}_{0}(l'0L1|l1) - [l]_{0}[l']^{*}_{-}(l'-1L1|l0)  \end{bmatrix} \end{equation} 
\begin{equation}  H(L2) =  \sum_{l.l'}  {(\frac{2l'+1}{2l+1})}^{1/2} (l'0L0|l0) \begin{bmatrix} -[l]_{-}[l']^{*}_{-}(l'-1L2|l1) + [l]_{+}[l']^{*}_{+}(l'-1L2|l1)  \end{bmatrix} \end{equation} 
For example:
\begin{equation}  H(00)=  \sum_{l.l'}  \begin{bmatrix} [l]_{0}[l']^{*}_{0} + [l]_{-}[l']^{*}_{-} + [l]_{+}[l']^{*}_{+}  \end{bmatrix} = |S_{0}|^{2}+ |P_{0}|^{2} + |P_{-}|^{2}+|P_{+}|^{2} \end{equation} 
as we should expect, this H(00) moment is the total probability, or the sum of all the waves probabilities entering the analysis.

How useful are moments experimentally?
We have seen that the moments include information about the spin density matrix. From the definition
\begin{equation}  H(LM) = \int d{\Omega} I(\Omega) D^{L}_{M0}(\phi,\theta,0)   \end{equation} 
If we assume that $I(\Omega)$ is uniform in the (M,t) bin under study, we can write
\begin{equation}  H(LM) =  I(\Omega) \sum^{N}_{i} D^{L}_{M0}(\phi_{i},\theta_{i},0)   \end{equation} 
We can obtain, then,  directly from data the (unnormalized) moments
\begin{equation}  H(LM) =  \sum^{N}_{i} D^{L}_{M0}(\phi_{i},\theta_{i},0)   \end{equation} 

These are sums of Wigner-D function directly calculable from data, completely model independent.
Including the values of the Wigner-D functions, the first few moments (for two spinless particles final states) have the following forms (all angles are calculated in the Gottfried-Jackson frame):

 \begin{equation}      H(00) = 1 \end{equation} 
 \begin{equation}     H(10) = cos(\theta) \end{equation} 
 \begin{equation}      H(11) = \frac{-1}{\sqrt{2}} sin(\theta) cos(\phi) \end{equation} 
  \begin{equation}     H(20) = \frac{1}{2}*(3 cos^{2}(\theta)-1) \end{equation} 
  \begin{equation}     H(21) = \frac{-\sqrt{3}}{2} sin(\theta) cos(\theta) cos(\phi) \end{equation} 
  \begin{equation}     H(22) = \frac{\sqrt{6}}{4} (1-cos^{2}(\theta)) cos(2\phi) \end{equation} 
   \begin{equation}     H(30) = \frac{1}{2} (5 cos^{3}(\theta)-3cos(\theta)) \end{equation} 
  \begin{equation}     H(31) = \frac{-\sqrt{3}}{4} sin(\theta) (5 cos^{2}(\theta)-1) cos(\phi) \end{equation} 
 \begin{equation}      H(32) = \sqrt{\frac{15}{8}} (1-cos^{2}(\theta)) cos(\theta) cos(2\phi) \end{equation} 
  \begin{equation}     H(33) = \frac{-\sqrt{5}}{4} (1-cos^{2}(\theta))^{\frac{3}{2}} cos(3\phi) \end{equation} 
  \begin{equation}     H(40) = \frac{1}{8} (35cos^{4}(\theta)-30 cos^{2}(\theta)+3) \end{equation} 
   \begin{equation}    H(41) = \frac{-\sqrt{5}}{4} sin(\theta) (7 cos^{3}(\theta)-3cos(\theta)) cos(\phi) \end{equation} 
   \begin{equation}    H(42) = \sqrt{\frac{5}{32}} (1-cos^{2}(\theta)) (7 cos^{2}(\theta)-1) cos(2\phi). \end{equation} 

If we now weight our mass histogram (number of events for each bin) by these quantities and plot versus mass, the first moment (H(00)) will give us the mass distribution. The others will give us some information on the spin density matrix versus mass. However, as we have discussed, the relation of those moments with partial waves is complex. Furthermore, since the acceptance corrected moments contain interference between different moments, the comparison also requires several important assumption and it is model dependent.
Moments analysis has been used in limited cases to analyze data, as when small number of waves are necessary against final two-pseudoscalar systems~\cite{MOM1}. 
In general, even if the moments are directly calculable from the data, the association of structure with spin and other quantum numbers is not direct, and perhaps less enlighten that performing a formal PWA of the data.

\section{Target Polarization}

In equation (\ref{eqn:densdef}), we defined the operator $|in\rangle \langle in | $ as the initial spin density matrix operator, $\widehat{\rho_{in}}$,

\begin{equation}  \widehat{\rho_{in}} \equiv |in\rangle \langle in | = \sum^{2}_{i,j=1} \rho_{i,j}\end{equation} 

We had considered the case where we prepare or measure the polarization of the incoming photons. The average over the polarization states is described by the photon spin-density matrix, as it was described in detail in section \ref{sect:SpinDen}.

Then we had
\begin{equation}  I(\tau) =  \sum_{k} \sum_{i,j} \langle out | \widehat{T} \rho_{i,j} \widehat{T}^{\dagger} |out \rangle \end{equation} 
where $k$ (the rank of the resonance spin-density matrix) is the number of different spin orientations of the incoming (target) and outgoing (recoiling) states. For example, in the case of fermions of spin $1/2$ (protons or neutrons), we have $k=2 \times 2 = 4$.

However, there is an experimental possibility of polarizing the target nucleon. In this case, the initial state spin density matrix will also include a target spin density factor that we can write as 

\begin{equation}  \widehat{\rho_{in}} \equiv |in\rangle \langle in | =\widehat{\rho_{\gamma}} \widehat{ \rho_{target}} 
\label{eqn:tar1}
\end{equation} 

The rank of the fit will be reduced by two. Since the spin density matrix of the target multiplies that of the beam, it does not introduce any extra experimental leverage in the PWA fitting or the naturality of the exchange particle (we obtain same advantages by knowing the polarization of the beam {\it or} the target). To obtain valuable information about the exchanged particle we need to measure the spin of the outgoing nucleon. To determine the spin flip or non-flip of the nucleon we need information on the recoiling nucleon.  However, if the beam polarization is known, adding the target polarization might introduce extra analyzing power by limiting the possible reactions at the baryon vertex.  If our purpose is to obtain the full analytic form of the cross section, a {\it complete experiment} ~\cite{sand11} with polarized beams and targets is required.
 \section{Restricting Exchanged Naturality Using  Nuclear Targets}

Consider an experiment with an unpolarized photon beam where the recoil particle is measured. A first generation of this experiment has already been performed at JLab-CLAS~\cite{Stepan}. As an example, consider coherent scattering from a spin zero target producing a two-particle final state (i.e. $\eta\pi$ or $\eta'\pi$). Using equation (\ref{eqn:totall}) for two final state particles (\ref{eqn:twopart})

\begin{equation}   I(\phi,\theta) = \sum_{\epsilon=-,+} \sum_{l.m,l',m'} \sqrt{\frac{2l+1}{4\pi}}\sqrt{\frac{2l'+1}{4\pi}}\  {^{\epsilon}\rho_{l,m,l',m'}} {^{\epsilon} { D^{l\ *}_{m0}(\phi,\theta,0)}} {^{\epsilon} { D^{l'}_{m'0}(\phi,\theta,0)}} . \end{equation} 

Let's take $k=1$ (rank one), and for each wave $l$ ($S,P,D \dots $) consider only the values $m=0$ and $m=1$ ($ {^\epsilon V_{lmk}}$=0 if $m > 1$).  We consider the case of an unpolarized beam (polarization does not add to the problem since the naturality of the exchange will be already restricted), therefore, the spin density matrix of the photon can be factored out. Then,

\begin{equation}  {^\epsilon\rho_{l,m,l',m'}} = \sum_{k} {^\epsilon V_{lmk}}{^\epsilon V^{*}_{l'm'k}} =  {^\epsilon V_{lm}}{^\epsilon V^{*}_{l'm'}} .\end{equation} 

Remembering that for $m=0$ the only available states have negative reflectivity, for each value of $l$ we have ($\epsilon= \pm$), then
\begin{equation}  {^{(-)} V_{l0}}\ ; {^{(-)} V_{l1}}\ and\  {^{(+)} V_{l1}} \end{equation} 
Recalling our standard notation, for $l=0$
\begin{equation}  S_{0} = {^{(-)} V_{00}}\ \end{equation} 
and for $l=1$
\begin{equation}  P_{0} = {^{(-)} V_{10}}\ ; P_{-} = {^{(-)} V_{11}}\ and\ P_{+} =  {^{(+)} V_{11}} \end{equation} 
and for $l=2$
\begin{equation}  D_{0} = {^{(-)} V_{20}}\ ; D_{-} = {^{(-)} V_{21}}\ and\ D_{+} = {^{(+)} V_{21}} \end{equation} 
and so on, such that for a wave $l$, we can use $[l]_{0}$, $[l]_{-}$ and $[l]_{+}$, where [$l$]=$S,P,D,F\dots$

Considering coherent scattering from a spin zero target (i.e. ${^{4}H_{e}}$) \cite{Aznauryan,Asatryan}, ($J_{ex}=0$), only natural parity exchange (+) is allowed (only $\rho$ or $\omega$ exchange). Also, we assume that a state with $l=0$ is forbidden due to s-channel helicity conservation (SCHC) and the spin of the produced system have to be  that of the incoming photon, $J=1$ ($m=1$). Therefore

\begin{equation}   I(\phi,\theta) = \sum_{l.m,l',m'} \sqrt{\frac{2l+1}{4\pi}}\sqrt{\frac{2l'+1}{4\pi}}\  {^{+} \rho_{l,m,l',m'}} {^{+} { D^{l\ *}_{m0}(\phi,\theta,0)}} {^{+} { D^{l'}_{m'0}(\phi,\theta,0)}} . \end{equation} 

But, with our assumptions $[l]_{0}$, $[l]_{-}$ are zero, therefore
\begin{equation}  {^{+} \rho_{l,m,l',m'}} =  [l]_{+} [l']^{*}_{+} \end{equation}
and with $m=m'=1$, we can then write
\begin{equation}   I(\phi,\theta) = \sum^{l_{max}}_{l=1} \begin{vmatrix} \sqrt{\frac{2l+1}{4\pi}} [l]_{+} {^{+} { D^{l\ *}_{10}(\phi,\theta,0)}} \end{vmatrix}^{2}  \end{equation} 
but
\begin{equation}   {^{+} { D^{l\ *}_{10}(\phi,\theta,0)}} = \frac{1}{\sqrt{2}} \begin{bmatrix} D _{10}^{l *}(\phi,\theta,0)- D _{10}^{l}(\phi,\theta,0)\end{bmatrix} \end{equation} 
and
\begin{equation}  D _{10}^{l *}(\phi,\theta,0) = e^{i\phi} d _{10}^{l}(\theta) \end{equation} 
%5
then
\begin{equation}   {^{+} { D^{l\ *}_{10}(\phi,\theta,0)}} = i\frac{2}{\sqrt{2}}  d _{10}^{l}(\theta) sin(\phi). \end{equation} 

The intensity is then
\begin{equation}   I(\phi,\theta) = \sum^{l_{max}}_{l=1} \begin{vmatrix} \sqrt{\frac{2l+1}{\pi}}d _{10}^{l}(\theta) [l]_{+} \end{vmatrix}^{2} sin^{2}{\phi} \end{equation} 
and since
\begin{equation}  d^{1}_{10}=-\frac{1}{\sqrt{2}}sin(\theta) \end{equation} 
\begin{equation}  d^{2}_{10}=-\sqrt{\frac{3}{2}} sin(\theta)cos(\theta) \end{equation} 
\begin{equation}  d^{3}_{10}=-\frac{\sqrt{3}}{4} (5cos^{2}(\theta)-1) \end{equation} 
We can write
\begin{equation}   I(\phi,\theta) = sin^{2}{\phi} \begin{vmatrix} -\sqrt{3} P_{+} sin(\theta)-\sqrt{15}D_{+} sin(\theta)cos(\theta)-\frac{\sqrt{15}}{2} F_{+}(5cos^{2}(\theta)-1) \end{vmatrix}^{2} \end{equation} 
where we have only three amplitudes: $P_{+}$, $D_{+}$ and $F_{+}$ to be extracted from the data.

\section{Improvements to the Model}

To illustrate the directions for future improvements, we discuss in the next sections some enhancements to the PWA method which are not yet fully implemented.

\subsection{Kinematic Effects (Deck Effect)}

More than resonances contribute to the production of multiparticle final states. For a full description of the data we will need to account for all possible effects, specially those that might mimic resonant behavior. Deck~\cite{Deck,Dudek2} showed that in the peripheral production of three-particle final states ($\pi \pi \pi$), it is possible to obtain a substantial enhancement of the cross section at low-mass over the phase-space in two-particle subchannels.
The {\it Deck effect} is now used to name an array of different kinematic effects that can distort the phase- space of multiparticle final states. These are contributions from intermediate diagrams through the baryon vertex that contribute to partial waves through modification to the mass terms of the assumed isobars. They are essentially of kinematic origin, and should appear in any amplitude of the partial wave decomposition and be independent of dynamical resonance poles.  

\begin{figure}[!htp]
\centerline{\includegraphics[width=12cm]{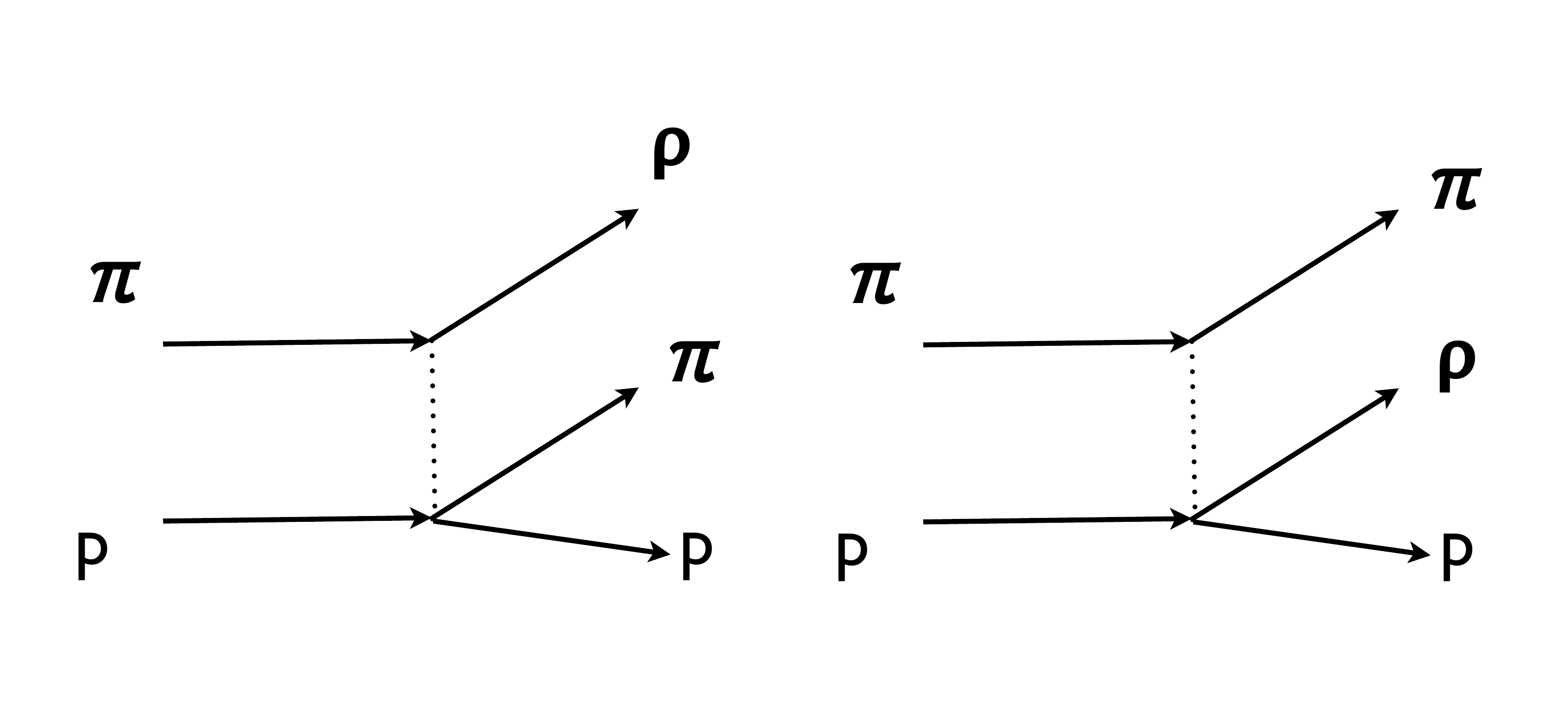}}
\caption{\label{fig:deckdia}Diagrams giving rise to a kinematical peak in the $\pi- \rho$ mass spectrum (Deck effect)~\cite{Deck}.}
\end{figure}

\begin{figure}[!htp]
\centerline{\includegraphics[width=10cm]{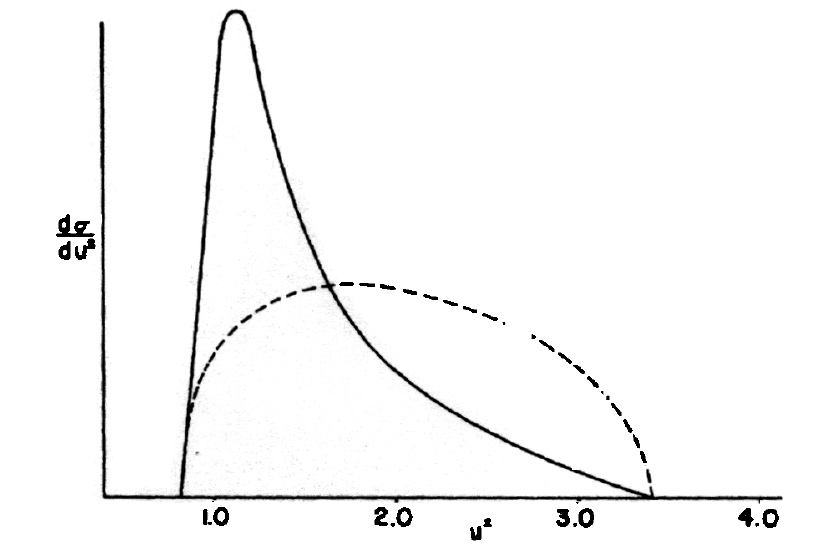}}
\caption{\label{fig:deck}Mass distribution (cross section)  versus masses, showing the Deck Effect at low masses (solid curve) compared to the.~\cite{Deck} phase-space (dashed curve).}
\end{figure}

Deck studied pion production of charged $3 \pi$ final states, calculating the cross section for the complementary diagram with one of the pions coming from the nucleon vertex. Diagrams are shown in figure~\ref{fig:deckdia}. The cross section  for the $\rho$ -$ \pi$ system is given by

\begin{equation*} \frac{d \sigma}{du^{2}} = \frac{g^{2}}{4 \pi} \frac{(m^{2}_{2}-4m^{2})} {4F^{2}_{I}} \begin{bmatrix} \frac{d \sigma}{d \Omega} \end{bmatrix}_{0} \end{equation*}
\begin{equation} \times\int^{{-t^{2}}_{max}}_{{-t^{2}}_{min}} d(-t^{2}) \int^{\omega^{2}_{max}}_{\omega^{2}_{0}} d \omega^{2} e^{-\delta (-t^{2})} \frac{1}{B} \frac{q}{u}
\frac{(-\omega^{2})} {[a-b \/ B(A-\omega^{2})]^{2}} \end{equation}
where $a,b,A$ and $B$ are functions of the Mandelstam variables.
This cross section is plotted in figure~\ref{fig:deck} and compared with the phase-space. A peak is observed at low $\rho - \pi$ invariant mass. This excess in the phase-space, a broad threshold kinematic enhancement is produced because the $\rho$ and the $\pi$ are boosted forward. This {\it peak} is produced in the $a_{1}$ region where a normal $P$ state of the quarkonium is also expected.

The Deck effect might interfere with the $a_{1}$ state and distort the mass distribution of other observed resonances. A similar effect has also predicted to affect the $\pi_{2}$~\cite{Dudek2}.

There are two possible ways to include the Deck effect into the PWA. Firstly, one may calculate the Deck cross section and add it to the total intensity, fitting an overall normalization constant for the Deck process. A better method is to expand the Deck diagrams into partial waves and then incorporate those partial waves into the intensity distribution, as the Deck process interferes with meson resonance production. Deck type diagrams favor forward peaked final multiparticle system, therefore, suggesting some possibilities for reducing the effects in the sample. In the analysis of large number of final particles the momenta of each particle will be reduced and we will need to include more complex and important terms from Deck effects.

\subsection{Threshold Effects (Flatt\'{e} Formula)}

Doing mass-dependent fits, the mass distribution must be corrected by effects produced when production energies come close to threshold energies. In 1976 S. M. Flatt\'{e} ~\cite{Flatte} analyzed the $\pi\eta$ and the $KK$ coupled channel systems and proposed the following parametrization of the S-wave production amplitudes $A_{i}$ when analyzed near the $KK$ threshold. The BW distribution needs to be modified such that

\begin{equation}  A_{i} \sim \frac{M_{R} \sqrt{\Gamma_{0} \Gamma_{i}}} { {M_{R}}^{2} - E^{2} - i M_{R} (\Gamma_{1}+\Gamma_{2})} .\end{equation}  with $ i=1,2$

Here $E$ is the effective mass (c.m. energy), $M_{R}$ is a resonance mass (the $a_{0}(980)$ mass in this particular case), and the first channel width is

\begin{equation}  \Gamma_{1} = g_{1} \times \frac{1}{2E}  \sqrt{ [E^{2}-(m_{\eta} + m_{\pi})^{2}] [E^{2}-(m_{\eta} - m_{\pi})^{2}]  }. \end{equation} 

Above the $KK$ threshold a second channel opens, the width  is
\begin{equation}  \Gamma_{2} = g_{2} \times \sqrt{ \frac{E^{2}}{4}-m^{2}_{K} } .\end{equation} 
Below the threshold we have
\begin{equation}  \Gamma_{2}=i g_{2} \times \sqrt{m^{2}_{K}-E^{2}} .\end{equation} 
At the threshold energy 
\begin{equation}   E_{0}=2m_{K}  \end{equation} 
and 
\begin{equation}  \Gamma_{0} = g_{1}  \sqrt{ [E^{2}_{0}-(m_{\eta} + m_{\pi})^{2}] [E^{2}_{0}-(m_{\eta} - m_{\pi})^{2}]  } .\end{equation}  

The Flatt\'{e}  production amplitudes depend on three real parameters: the
resonance mass $M_{R }$ and the two coupling constants $g_{1}$ and $g_{2}$. More discussions related to the Flatt\'{e}  parametrization can be found in reference ~\cite{Baru}.
 \section{Other Variations}
\label{sect:varia}
\subsection{Flat Background}

For most final multiparticle states, the resonant signal overlaps a non-resonant {\it background}. This non-resonant background can be taken as an isotropic (infinite number of non-resonant waves), {\it flat} overlapping noise.  The {\it flat amplitude} is an incoherent contribution to the intensity, that is isotropic in phase-space. It is allowed in the fit in order to absorb any contributions that are not projected onto the partial waves. These are experimental {\it broken} events (misidentified particles, out-of-time particles, etc.) These are random in nature, a flat phase-space distribution means equal probability density over all phase-space elements, this results in a flat angular distribution. 
It can be then associated to a new set of (flat) amplitudes, $\{c\}$, such that
\begin{equation} {^{\epsilon}A_{c}(\tau)} =  {^{\epsilon}} C_{c} (M,t) \end{equation}
a new set of waves (independent of the angles ($\tau$)) to be added to the intensity. 

We can add them to the rest of the waves to form the intensity, $I(M,t,\tau)$, in a coherent or incoherent (non-interfering) manner.  If added incoherently
\begin{equation} I_{Total}(M,t,\tau) = I(M,t,\tau) + I_{FLAT}(M,t) \end{equation}
where
\begin{equation}  I(M,t,\tau) = \sum_{k} \sum_{\epsilon,\epsilon'}  \sum_{b,b'} \  {^{\epsilon}V^{k}_{b}}(M,t)\  {^{\epsilon}A_{b}(\tau)}\  \rho_{\epsilon,\epsilon'} {^{\epsilon'}V^{k*}_{b'}}(M,t)\  {^{\epsilon'}A^{*}_{b'}(\tau)} \end{equation} 
and
\begin{equation}  I_{FLAT}(M,t) = \sum_{k,\epsilon}  \sum_{c,c'} \  {^{\epsilon}V^{k}_{c}}(M,t) {^{\epsilon}} C_{c} (M,t)\   \widehat{\rho_{\gamma}} {^{\epsilon}V^{k*}_{c}}(M,t)  {^{\epsilon}} C_{c'} (M,t) .\end{equation}
Binning in $M$ and $t$,
\begin{equation}  I_{FLAT} = \sum_{k} \sum_{\epsilon,\epsilon'}  \sum_{c,c'}  \  {^{\epsilon}V^{k}_{c}} {^{\epsilon}} C_{c} \  \rho_{\epsilon,\epsilon'} {^{\epsilon'}V^{k*}_{c}} \ {^{\epsilon'}} C_{c'} . \end{equation}

We normally take just one flat background amplitude (one complex number) such that
\begin{equation} {A} =  C (M,t) \end{equation}
as an overall constant. Therefore, we add just one {\it flat background intensity} to our set
\begin{equation}  I_{FLAT} = \sum_{k} \sum_{\epsilon,\epsilon'}  \  {^{\epsilon}V^{k}}  \   \rho_{\epsilon,\epsilon'} {^{\epsilon'}V^{k*}} .\end{equation}

In general, we might increase the number of $V$'s to our fit by adding a new term to each reflectivity and rank. In practice, we take the $V$'s from background to be independent of the reflectivity and the rank, therefore, we have only one amplitude,  ${^{\epsilon}V^{k}} = V_{FLAT}$, and the intensity for the fit is

\begin{equation}  I(M,t,\tau) = \sum_{k} \sum_{\epsilon,\epsilon'}  \sum_{b,b'} \  {^{\epsilon}V^{k}_{b}}(M,t)\  {^{\epsilon}A_{b}(\tau)} \  \rho_{\epsilon,\epsilon'} {^{\epsilon'}V^{k*}_{b'}}(M,t)\  {^{\epsilon'}A^{*}_{b'}(\tau)} +|V_{FLAT}|^2 . \end{equation}

\subsection{t-Dependence}

We indicated (equation~\ref{eqn:fermi}) that the data needs to be binned in mass ($M$) and Mandelstam-$t$. However, for data sets with limited statistics is normal practice to integrate over a large range of values of  Mandelstam-$t$. We can assume no dependence on $t$, or include a modeled $t$ dependence. We can write

\begin{equation}  I(M,t,\tau) = \sum_{k} \sum_{\epsilon,\epsilon'}  \sum_{b,b'} \  {^{\epsilon}V^{k}_{b}}(M,t)\  {^{\epsilon}A_{b}(\tau)}\ \rho_{\epsilon,\epsilon'} {^{\epsilon'}V^{k*}_{b'}}(M,t)\  {^{\epsilon'}A^{*}_{b'}(\tau)} \end{equation} 
\begin{equation} {^{\epsilon}V^{k}_{b}}(M,t) ={^{\epsilon}Z^{k}_{b}}(M)e^{-\beta_{b}t} . \end{equation}
Binning only in $M$, we have then
\begin{equation}  I(\tau) = \sum_{k} \sum_{\epsilon,\epsilon'}  \sum_{b,b'} \  {^{\epsilon}Z^{k}_{b}}(M)e^{-\beta_{b}t}\  {^{\epsilon}A_{b}(\tau)}\ \rho_{\epsilon,\epsilon'} {^{\epsilon'}Z^{k*}_{b'}}(M)e^{-\beta_{b'}t}\  {^{\epsilon'}A^{*}_{b'}(\tau)} \end{equation} 
where we have introduced a new parameter to fit, $\beta_{b}$, one new real parameter for each wave.
\section{Wave  Ambiguities, Leakages and Baryon Contamination}
\label{sect:Baryon}

It is difficult to address ambiguities and leakages in general terms. Their significance in PWA depends very much on the channel under study and the detector characteristics, but all analyses need to address the effects. There are a few important ideas we wish to convey:

\begin{itemize}
\item More final state particles in the channel implies fewer ambiguities.

\item Better known or larger acceptance implies less leakage.

\item Better resolution implies less leakage.

\end{itemize}

Ambiguities are present in all parameter estimation problems. They are generated when there is not a one to one mapping between the modeled space and the measurement space. Our mass-independent fit is essentially a fit to the angular distribution of the final particles. Unfortunately, a given angular distribution may not map univocally to one model. Different partial waves could produce very similar angular distributions (wave ambiguities), the detector acceptance and resolution can mimic the angular distribution of one wave into another (leakage) and a baryon vertex decay can reproduce the same or part of the final state angular distribution (baryon contamination). 

Ambiguities mostly relate to the uniqueness of the final state. For few particles in the final state, ambiguities are inherent of the PWA method. Generally the more particles in the final state less ambiguous is the wave set. As an example, the decay of a resonance into two pseudoscalar mesons has an exact eight-fold ambiguity up to $D$-wave as it has been shown in the following references~\cite{Sadovsky,ChungAmbig}. 

Strong (large cross section) resonances produced in a given wave, are more likely to produce leakage into other waves. Leakage is the term used when events from one wave can be found in another wave. The fit misrepresented the quantum numbers of those events. This is normally due to experimental effects (acceptance and resolution), however it could also be caused by weak waves to close (in angular space) to a large, predominant wave.

\begin{figure}[!htp]
\centerline{\includegraphics[width=12cm]{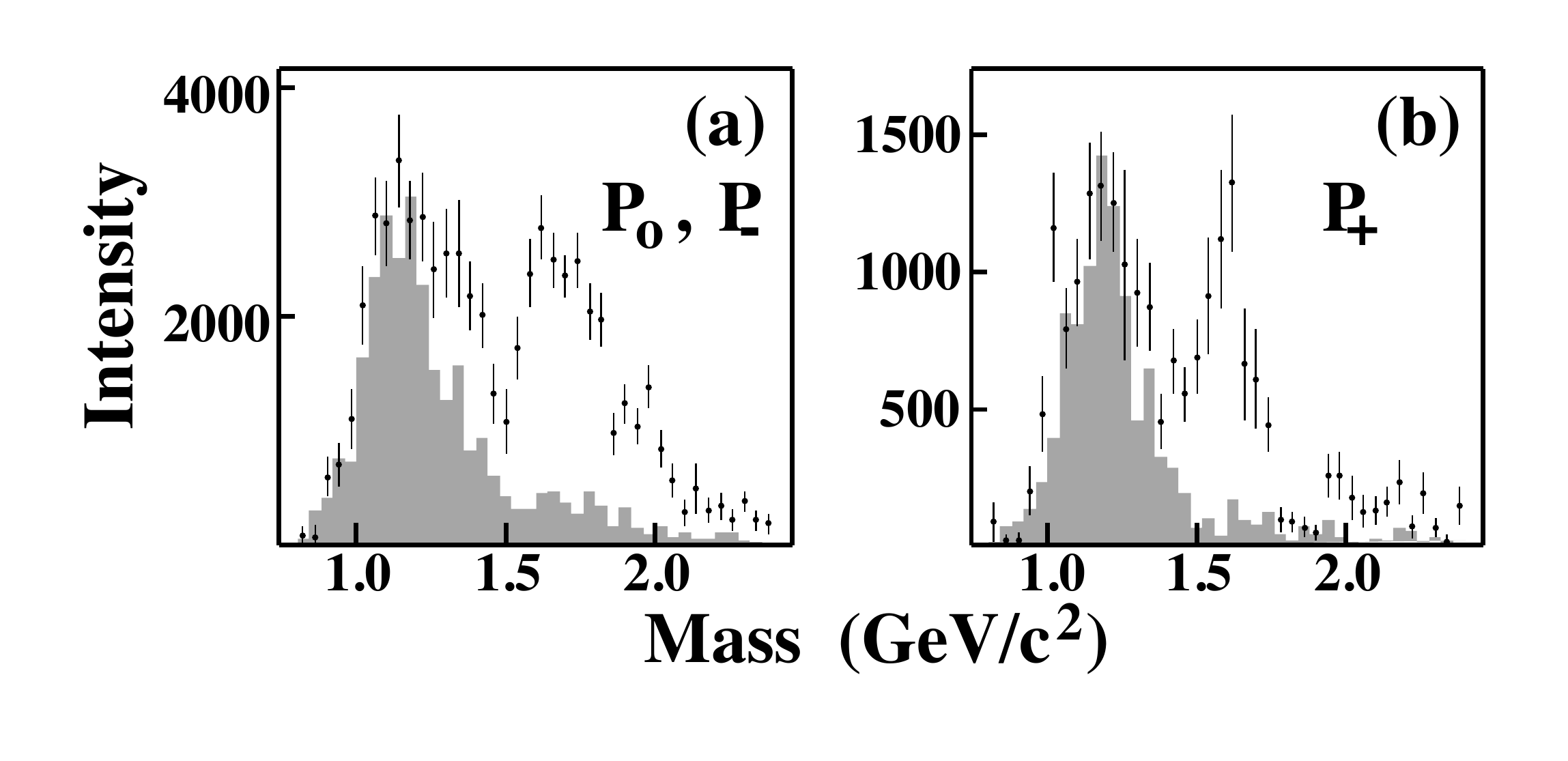}}
\caption{\label{fig:E852_example}An example of Leakage from BNL-e852 experiment. The final state $p\pi+\pi-\pi-$ is studied. The plots show wave intensities of the $1^{-+}[\rho(770)P$ exotic waves. The left plots fo $m^{\epsilon}=0^{-},1^{-}$ combined and the right plot for $m^{\epsilon}=1^{+}$ waves. The PWA fits to the data are shown as points with error bars. The shaded histograms show estimated contributions from all non-exotic waves to this waves due to leakages.~\cite{AdamsEx}.}
\end{figure}

Leakages are normally studied through Monte Carlo methods. We generate a phase-space sample in a specific multiparticle final state. Then, a set of resonances with an assumed relative strength and correspondent partial waves are assumed (normally extracted from data). An intensity is then generated with our model that will be used  to weight the previously generated events. These events are then processed through the detector simulation and analyzed through the standard reconstruction software. A PWA is then performed on this simulated data . With this method we can then recognize if a given input wave will be recognized by our PWA as another output wave. We can even trace the acceptance or resolution of the detector, the relative strengths and specific significant resonances that will spill leakages over the signal waves. For example, figure~\ref{fig:E852_example} shows a leakage study performed by the BNL-E852 collaboration in their study of the exotic $1^{-+}$ states~\cite{AdamsEx}. The shaded histograms are Monte Carlo events generated using a spin-density matrix extracted from the data, that includes all non-exotic waves excepts the matrix elements corresponding to the $1^{-+}$ waves that were set to zero. A PWA is done, and as observed in the figure, considerable leakage (non $P$ waves contributing to fitted $P$ waves) are observed. These events are mask into other waves by detector acceptances and we can work the Monte Carlo to know what elements contribute the leakages. In the case on BNL-E852, there is leakage but they concluded that is not affecting the signal that is observed at the right of the plot.

\begin{figure}[!htp]
\centerline{\includegraphics[width=10cm]{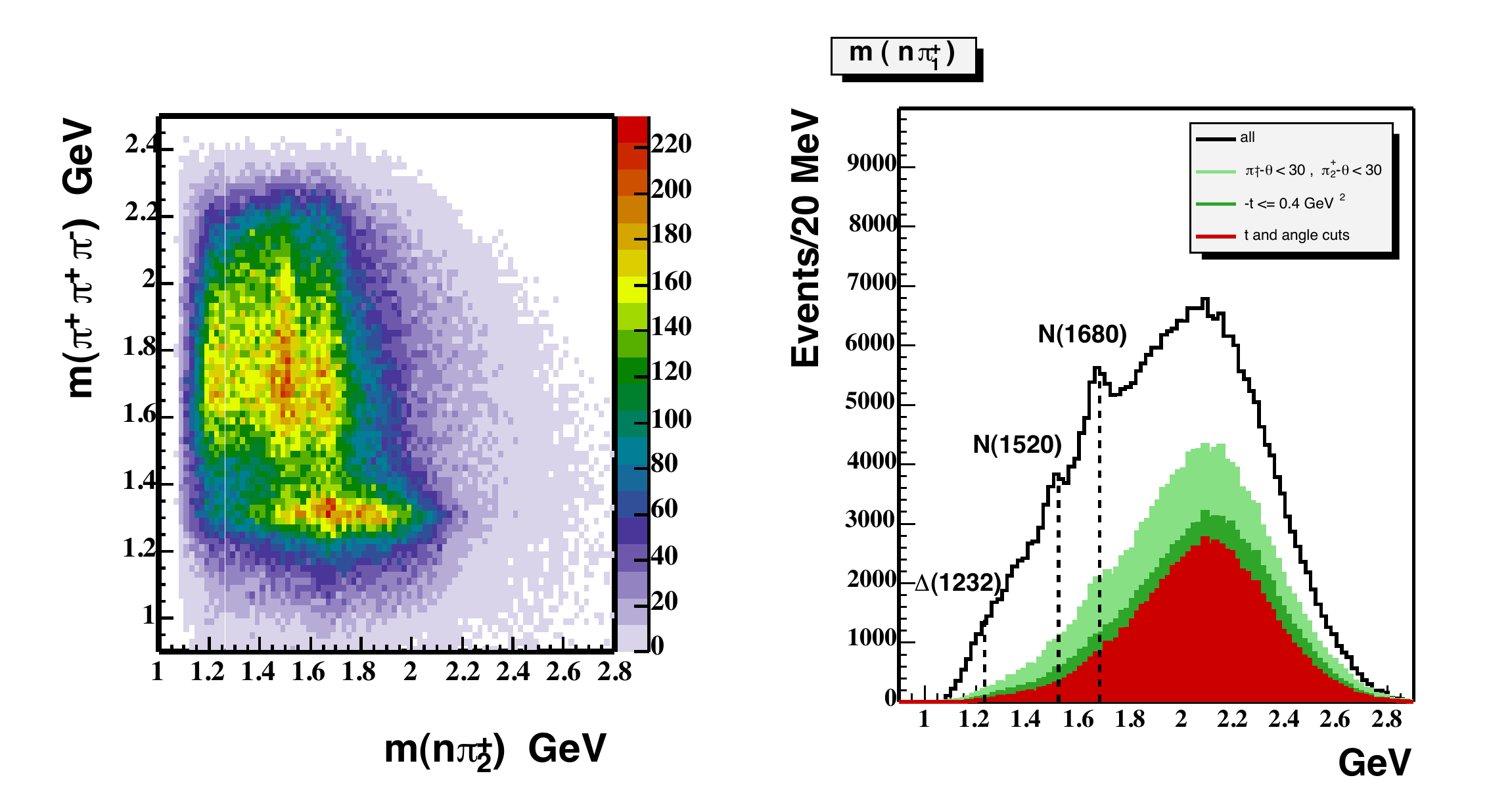}}
\caption{\label{fig:g6c_example}An example of baryon contamination from CLAS-g6c photoproduction experiment. The final state $n\pi+\pi+\pi-$ is studied, shown in the left plot is the $\pi+\pi+\pi-$ mass versus the $n\pi+_{1}$ mass. The plot shows a clear overlap of meson (3${\pi}$) and baryon (n${\pi}$) resonances in the high mass region. The sub-index 1 indicates the largest momenta $\pi+$. The right hand plot shows the $n\pi+$ mass distribution, where several known baryon resonances can be observed (different cuts applied to the data to diminish the baryon contribution are shown in colors).~\cite{Nozar2}.}
\end{figure}

Contamination due to the production of hadronic baryon resonances poses another difficulty, particularly for experiments at intermediate energies where there is significant kinematic overlap between mesonic resonances and baryon resonances. Figure~\ref{fig:g6c_example} shows an example for the reaction $\gamma p \rightarrow n \pi^{+} \pi^{+} \pi^{-}$ in the CLAS-g6c experiment~\cite{Nozar2}. The right plot of the figure shows several baryon resonances observed decaying to $n \pi^{+}$. The left side of the figure shows a two dimensional plot of the $n \pi^{+}$ mass versus the $\pi^{+} \pi^{+} \pi^{-}$ mass (where we expect to observe the meson resonances). As shown in the figure, we see the $a_{2}$ at about 1.3 GeV and then a complete overlap of mesonic and baryonic states at higher ($\pi^{+} \pi^{+} \pi^{-}$) masses. One way to reduce the number of baryonic events in the sample and thus increase the signal-to-noise ratio is by using kinematical cuts (as seen at the right plot in the figure). This method has been used in analyses by the CLAS collaboration at Jefferson Lab. These analyses used cuts in t (momentum transfer) to enhance t-channel, peripheral production, and in laboratory angles of the final states~\cite{Nozar, Craig}. The assumption is that meson resonances are more likely to be produced forward in the Lab than baryon decays.
Another possibility is to correct for the baryon contamination by accounting for the baryon production in the model itself,  adding amplitudes for baryon decays at the nucleon vertex. This has been explored at CLAS but resulted with very limited success. We are currently working to improve this technique.

We have to consider that in most cases mesons will be a small signal on top of baryon decay backgrounds, at least at 6 GeV, therefore the use a phenomenological model based on fitting the data for the PWA (Breit-Wigner shapes) produced by baryons could introduce large systematic errors. Studies of the systematic effects caused by the kinematical cuts must be done comparing different models of the baryonic decays.

\section{Future Directions}
\label{sect:New}

We expect that future experiments will obtain data with such better systematic and statistical errors that the limitations of the current method could be exposed.  Several groups (including the authors) are working on refining the current analysis methods to confront the challenges that will be imposed by the new data sets. The new refinements will require a close collaboration between theorists and experimentalists and between baryon and meson spectroscopy groups. Models and data analysis tools will need to be jointly revised. We give in this section a very brief description on future directions for the field as they are being discussed in ongoing meetings, workshops, and mini-schools.

Quantum Field Theories (QFT) provide the modern description of the fundamental interactions~\cite{Weinberg}.  QFT based numerical calculations, in particular lattice QCD, have recently made great advances in calculating the hadron spectrum, and it is the hope of the community that, in a near future, it will provide the theoretical basis for the study of resonances at intermediate energies. Understanding confinement must come from the application of QFT to the study of resonances at the intermediate energy regime where confinement is realized. However, this program of analysis is still incomplete but to incorporate LQCD into the fabric of PWA is one the future challenges.

Most of the current methods of analysis in the intermediate energy regime has been done using the S-matrix formalism~\cite{Eden}. The S-matrix formalism, introduced long ago by Heisenberg~\cite{Heisen}, adapts itself very well to short-range interactions as it is the case with the strong interactions. Only the incoming and outgoing properties of the interaction are considered; it could be said that is a more experimental approach to the problem since these are the properties directly measured. The method uses general principles of relativity and quantum mechanics to restrict the properties of the scattering matrix. These properties include the superposition principle, the requirements of special relativity, conservation of probability, and causality. The S-matrix is broken into a coherent sum of amplitudes, and the conservation of probability and causality translate to conditions of unitarity and analyticity on those complex amplitudes. Analyticity and causality play a very important role in making it possible to write the amplitudes using dispersion relations on the complex plane.
 The transition amplitudes are the real-boundary values of these complex analytic amplitudes. Different approaches have been used to obtain the amplitudes and most of the current discussions are on how to obtain a better set of amplitudes. For example, Regge theory introduces spin and angular momentum into the model, treating angular momenta as complex and it is able to make prediction on the energy and angular behavior of higher mass resonances. By application of complex analysis relations, resonances becomes poles in the Riemann surfaces of the complex amplitudes. This treatment of scattering, developed initially in the 1960s and 1970s, has been very successful describing hadron interactions and it provides a better representation for interference effects when several and overlapping resonances are present. 

The isobar model, as used in this report, does not respect unitarity and analyticity, as it breaks probability conservation. This, of course, has been known for a long time, and several approaches has been used to improve on this problem~\cite{AIT}. The use of the isobar model by experimentalists was mostly a practical approach, dictated by the available amount of data and computer power, that it has seemed adequate by previous analysis results. As data quality, statistics and computer power are now greatly improving and the study concentrate in wider and less defined resonances, the model may need refinement, and the spectroscopy community is making efforts to go beyond the isobar model.  Future directions in meson spectroscopy are directed in improving the model (theory) and the computational tools. The computer power is naturally improving as technology advances, but efforts are also being pursued to include parallelism and more flexible code for the inclusion of different models, and more user friendly software packages for the fitting of data.

As discussed in section~\ref{sect:Baryon}, at intermediate energies, we expect baryon contamination in our sample. This contamination comes from s-channel processes, at the low energy end of the energy spectrum, and most important for future photoproduction experiments, by the splitting of the target nucleon into other baryons (into $\Delta$'s,  higher mass $N^*$'s and $Y^{*}$s). We discussed an experimental approach to deal with this problem in section~\ref{sect:Baryon},  another possible solution of this problem will involve the inclusion of new baryon vertices into diagrams as in  figure~\ref{fig:decay1}. The baryon amplitudes may be extracted using the SAID~\cite{SAID}, MAID~\cite{MAID} or Bonn-Gatchina~\cite{Bonn} formalisms. These amplitudes will be then included (coherently) in the intensities to be fitted to the data. There will be an increase in the number of parameters of the fits that will greatly increase the complexity of the fit (and the need for more computer power). Therefore, it is important that improvements in the formalism are followed by improvements in computer tools and hardware. A solution for combination of s- and t-channel processes was used by the CLAS collaboration on the PWA of $\omega$ photoproduction~\cite{CLASW} (at lower energies than those considered here). In that analysis, covariant helicity-coupling amplitudes were constructed using the Zemach operator expansion method~\cite{Zemach,CHUNG2001}. However, s-channel contamination will be a minor problem in future photoproduction experiments using photon energies of about 12 GeV.

Advances to the theory have been centered in using LQCD and refining calculations based in the S-matrix formalism. Amplitudes are being calculated for specific reactions in the Regge formalism, using dispersion relations~\cite{AdamP} and including baryon vertices. This treatment brings unitarity and analyticity into the model. There are some indications that the isobar formalism can misrepresent phase differences, and therefore misrepresent resonance properties in high statistics experiments~\cite{EBAC}. 

Another important direction for improvement is the incorporation of the latest computing advances to make PWA faster, therefore, allowing for more iterations and larger wave sets. Parallelism or vectorization using GPU's~\cite{Wilt}, the GRID~\cite{grid} or the new Intel Xeon-Phi processors~\cite{Jeffers}  are being worked out as near future alternatives.

Most of these directions of work are not yet implemented into full PWA codes. They are mentioned here for completeness, with the expectation that these topics will be the subject of future reports.

\section{Concluding Remarks}

Our goal is to study strong interactions within the framework of Quantum Chromodynamics (QCD), and in particular, to understand hadronic matter as the aggregates of quarks and gluons interacting via the QCD Lagrangian. Meson spectroscopy may shed some light about QCD at intermediate baryon-light-meson production energies, and on the phenomenological models that have been so successful in this regime (i.e. Regge theory, Quark constituent models, etc.).  In the absence of perturbative calculations, lattice QCD is our best model for comparison. The comparison of theory and data has been normally done through the extraction of resonance's properties. However, the definition of hadron observables is not easy or precise, as even their definition  (i.e. mass and width) are in discussion. 
The analyses in meson spectroscopy have been done using different methods and tools, differentiated by both beams and geographical-lab borders. Two main analysis tools have been developed, the one presented in this report based on the Jacob and Wick spin formalism, has been used mainly by pion, kaon, and photon beam experiments (SLAC and Brookhaven, JLab, and a different version at VES), and the covariant tensor formalism (going back to Rarita-Schwinger)~\cite{Zemach} used by $p \bar{p}$  and $J/ \psi$ decay experiments (Crystal Ball and now at BES~\cite{Crystal,BES}).

We have described a particular method to search for strong interaction resonances, as well as to determine their properties through the measurement of multiparticle final states. Several assumptions have been introduced to make this method possible (i.e. truncation of the partial wave expansion, selection of specific interactions, decay modes, \dots), and we discussed how these assumptions affect the predictability of the systematic errors. Furthermore, in cases with more than two final state particles, we use a simple cascade model for the decays, the so-called isobar model.  In this model the unitary of the S-matrix is not assured,  i.e. we are not taken into account {\it all possible mechanisms} able to produce the observed final state. Moreover, this simplification does not include the consideration of the inter-particle interactions beyond what is purposely included in the model. In doing so, this method run into the danger of masking the experimentally extracted properties of the resonances. Clearly, the nature of multi-particle final states is more complicated than the isobar model decay predicts. However, we use this model to examine data structures, and it provides a particular infrastructure for a {\it data driven} approach to search for resonances. Currently the most interesting topics in spectroscopy, gluonia and hybrid-quarkonia searches, will mostly come from the analysis of small, wide, and overlapping resonances decaying to multi-particle final states. These structures will be more difficult to identify and better control over the systematic effects will be required. To make progress we will need to improve on the isobar model. However, the method presented here has worked on extracting the most obvious resonances and should still be considered as the starting point. For practical reasons, it is clear that experimentalists will always have to select {\it a particular model and associated wave set} that adequately describes the observed data, and accordingly, they will need to understand the source of systematic error associated with that selection. We will also need to understand that a more complex model, which will include more parameters to be fit, is not a warranty for a better isolation of the signal. It is unlikely that a single signal or experiment will be able to determine unambiguously the nature and properties of a resonance. The determination of resonance's properties, as well as their existence, requires the accumulation of data from many channels, detectors, combined data and different methods and models of analysis.

Several improvements to the presented method are obvious and, somehow, already being pursued. Amplitudes for three body final states are being calculated and can be added to the present model~\cite{EBAC}. The baryon vertex and final state interactions diagrams need to be included to the model. New computational methods needs to be explored to make the calculation of normalization integrals faster and more flexible, and to be able to determine fit quality and the best set of partial waves in a more automatic way. 

In concluding, the method we described needs to be considered in the context of its utility to identify resonances and measure their properties.  The method, including the model of the intensity,  is not intended to provide a theoretical description of the data from basic principles; it is not our intent to validate a theoretical model. There are obviously several shortcuts in the method when examined against well establish physical principles that may introduce spurious solutions to our search.   The method has been proven to work well for the identification and measurement of well separated and relatively narrow resonances. The data is driving the model such that the assumptions could be justifiable. Will this method work to obtain information from less defined, overlapping and wide resonances? We will need to compare this model with more complex or sophisticated models and found what data tells us about the improvements, systematic uncertainties, and predicting powers. The final answer will remain with the analysis of data of future experiments with high statistics, coupled-channels analysis and multi-particle final states.

 \section{Appendix A: Frames of Reference}
We have used  four different frames of reference to write particle properties: the laboratory frame (Lab), the center-of-mass frame (CM), the Gottfried-Jackson frame (GJ) and the helicity frame (HEL). We will here define these frames and write the transformations used to relate the values of the four-momenta between those frames.
Let's call $p^{\mu}=(E;\overrightarrow{p})$ the four-momentum of a particle. We will consider the beam ($p^{\mu}_{B}$), the target ($p^{\mu}_{T}$), the recoil ($p^{\mu}_{R}$), the resonance ($p^{\mu}_{X}$) and the final products ($p^{\mu}_{i}$).

\subsection{The Lab frame}
The laboratory (Lab) frame is the frame at rest with respect to the target ($\overrightarrow{p}_{T}=0$).
The $z$ direction is in the direction of the beam (pointing downstream), the $y$ is vertical (pointing up) and $x$ is given to produce a right-handed system. The origin of the frame is situated at the center of the target. Particle's four-momenta are experimentally measured on this frame. By definition then

\begin{equation}  p^{\mu}_{T} = (m_{T};0,0,0) \end{equation} 
\begin{equation}  p^{\mu}_{B} = (E_{B};0,0,p_{zB}) \end{equation} 

\subsection{The Center-of-mass frame}
The center-of-mass frame (CM) is defined as the frame where
\begin{equation}  \overrightarrow{p}_{B}+\overrightarrow{p}_{T}=0 \end{equation} 
and by momentum conservation we also have
\begin{equation}  \overrightarrow{p}_{X}+\overrightarrow{p}_{R}=0 \end{equation} 
all those momenta are in a plane called, the {\it production plane}.
Since the relation between Lab and CM is a boost in $z$, the directions of the all axes in the CM frame remain unchanged with respect to the Lab frame (i.e., $z$ is in the beam direction).

\subsection{The Gottfried-Jackson frame}
The Gottfried-Jackson frame (GJ) is a frame where the resonance ($X$) is at rest. $z$ is in the direction of the beam and $y$ is perpendicular to the production plane, such that $\overrightarrow{y}=\overrightarrow{p}_{B} \times \overrightarrow{p}_{X}$ and $x$ is given to produce a right-handed system. The GJ angles, $(\theta_{GJ},\phi_{GJ})$, are the standard polar coordinates in the GJ frame. Sometimes the $\phi_{TY}$ (Trieman-Yang angle) is used. This angle is defined by projecting the GJ particle's momentum in the $xy$ plane and determining the angle between the projection and the $y$ axis.

\subsection{The Helicity frame}
In the isobar model, the helicity frame (HEL) is a frame where the isobar particle is at rest. $z$ is in the direction of the isobar GJ frame momentum and $y$ is such that $\overrightarrow{y}=\overrightarrow{p}_{B} \times \overrightarrow{p}_{isobar}$ and $x$ is given to produce a right-handed system. The HEL angles, $(\theta_{hJ},\phi_{h})$, are the standard polar coordinates in the helicity frame.

\subsection{Transformation between Frames}

Lorentz's boosts and space rotations (Poincare group transformations) are used to transform particle's four-momenta among frames~\cite{Hagedorn}.
The rotation of a space vector $\overrightarrow{V}$, by an angle $\theta$, around an axis given by a unit vector $\hat{n}=(n_{x},n_{y},n_{z})$ (i.e. n's are the directional cosines) is given by
\begin{equation}  \overrightarrow{V'} = \mathbb{R} \cdot \overrightarrow{V} \end{equation} 
where $\mathbb{R}$ is the matrix
\begin{equation}  \mathbb{R} = \begin{bmatrix} cos\theta+n^{2}_{x}(1-cos\theta) & n_{x}n_{y}(1-cos\theta) -n_{z}sin\theta & n_{x}n_{z}(1-cos\theta)+n_{y}sin\theta \\
 n_{y}n_{x}(1-cos\theta) +n_{z}sin\theta
 &  cos\theta+n^{2}_{y}(1-cos\theta)& n_{y}n_{z}(1-cos\theta)-n_{x}sin\theta \\
n_{z}n_{x}(1-cos\theta) -n_{y}sin\theta  & n_{z}n_{y}(1-cos\theta)+n_{x}sin\theta &  cos\theta+n^{2}_{z}(1-cos\theta)\end{bmatrix} \end{equation} 
We define the boost given by a momentum $\overrightarrow{p}$ in the following way~\cite{Hagedorn}.
Since 
\begin{equation}  \overrightarrow{\beta} = \frac{\overrightarrow{p}}{E} \end{equation} 
\begin{equation}  \gamma = \frac{1}{\sqrt{1-\beta^{2}}} = \frac{E}{m} \end{equation} 
\begin{equation}  \gamma\beta = \frac{p}{m} \end{equation} 
and the boost (in the $z$ direction) $\mathbb{L}$ is
\begin{equation}  \mathbb{L} = \begin{bmatrix} \gamma &  0 & 0 & -\gamma\beta \\ 0 & 1 & 0 & 0 \\ 0 & 0 & 1 & 0 \\
-\gamma\beta & 0 & 0 & \gamma \end{bmatrix} \end{equation} 
or
\begin{equation}  \mathbb{L} = \begin{bmatrix} \frac{E}{m} &  0 & 0 & -\frac{p}{m} \\ 0 & 1 & 0 & 0 \\ 0 & 0 & 1 & 0 \\
-\frac{p}{m} & 0 & 0 & \frac{E}{m} \end{bmatrix} \end{equation} 
and
\begin{equation}  p^{'\mu} = \mathbb{L} \cdot p^{\mu} \end{equation} 

To transform from the Lab to the CM we need to perform a $-\overrightarrow{p}_{B}$ boost.
To transform from the CM to the GJ we need first to rotate the axis $z$, such that the axis $y$ becomes perpendicular to the production plane. Then, boost to the frame where the resonance ($X$) is at rest and then rotate the $z$ axis to coincide again with the beam direction.
Similarly, we boost to the isobar rest frame to obtain the helicity frame quantities.

 \section{Appendix B: Rotations:Wigner-D functions}
The Wigner-D functions define the rotations on the spin space \cite{Rose}. Given a set of Euler angles $(\alpha,\beta,\gamma)$ (we follow the definition on reference \cite{Rose}), a rotation of an state of total spin $J$, called $|Jm \rangle$, can be expressed as an expansion in a complete  space basis
as
\begin{equation}  \mathbb{R}|Jm \rangle =\sum_{m'} D^{J}_{m'm}(\alpha \beta \gamma) |Jm' \rangle \end{equation} 
the Wigner-D functions are defined as the coefficients on this expansion.
Since we can also express
\begin{equation}  \mathbb{R} = e^{-i\alpha J_{z}}e^{-i\beta J_{y}}e^{-i\gamma J_{z}} \end{equation} 
\begin{equation}  D^{J}_{m'm}(\alpha \beta \gamma) =  \langle Jm'|e^{-i\alpha J_{z}}e^{-i\beta J_{y}}e^{-i\gamma J_{z}}|Jm \rangle \end{equation} 
or
\begin{equation}  D^{J}_{m'm}(\alpha \beta \gamma) = e^{-im'\alpha} \langle Jm'|e^{-i\beta J_{y}}|Jm \rangle e^{-im\gamma} \end{equation} 
 There are two angles (polar coordinates) to define direction, we use $\alpha = \phi$ and $\beta = \theta$. The $m'$ indexes, that are summed over, do not have physical meaning, and therefore the angle $\gamma$. The choice of $\gamma$ is conventional, reference \cite{Wick} uses $\gamma = -\phi$, we adopt reference \cite{SUC71} convention, taken $\gamma = 0$.
Calling (Wigner "small d" functions)
\begin{equation}  d^{J}_{m'm}(\beta) =  \langle Jm'|e^{-i\beta J_{y}}|Jm \rangle \end{equation} 
we write
\begin{equation}  D^{J}_{m'm}(\alpha \beta \gamma) = e^{-im'\alpha} d^{J}_{m'm}(\beta) e^{-im\gamma} \end{equation} 

The $d^{J}_{m'm}(\beta)$ were calculated by Wigner (see Rose Appendix II~\cite{Rose}), they have the form
\begin{equation*}  d^{J}_{m'm}(\beta) = \begin{bmatrix}(J+m)!(J-m)!(J+m')!(J-m')! \end{bmatrix}^{\frac{1}{2}} \end{equation*} 
\begin{equation*}  \times \sum_{S} \frac{(-1)^{S}}{ (J-m'-S)!(J+m-S)!(S+m'-m)!S!} \end{equation*} 
\begin{equation}  \times\begin{bmatrix} cos(\frac{\beta}{2})\end{bmatrix}^{2J+m-m'-2S}\begin{bmatrix}-sin(\frac{\beta}{2})\end{bmatrix}^{m'-m+2S} \end{equation} 
where the sum is over the values of the integer S for which the factorials are greater than or equal to zero.
The Wigner-D functions are normalized such that \cite{Richman}
\begin{equation}  \int d\Omega D^{J_{1}\ *}_{m'_{1}m_{1}}D^{J_{2}}_{m'_{2}m_{2}} = \frac{8\pi^{2}}{2J_{1}+1} \delta_{J_{1}J_{2}}\delta_{m'_{1}m'_{2}}\delta_{m_{1}m_{2}} 
\label{eqn:DNorm}
\end{equation} 

We have the following properties \cite{Rose}
\begin{equation}  D^{J\ *}_{m'm}(\alpha \beta \gamma) = (-1)^{m'-m}D^{J}_{-m'-m}(\alpha \beta \gamma) \end{equation} 
\begin{equation}  D^{J_{1}}_{m'_{1}m_{1}}D^{J_{2}}_{m'_{2}m_{2}} = \sum_{J_{3},m'_{3},m_{3}} (J_{1}m'_{1}J_{2}m'_{2}|J_{3}m'_{3})(J_{1}m_{1}J_{2}m_{2}|J_{3}m_{3})D^{J_{3}}_{m'_{3}m_{3}} \end{equation} 
or 
\begin{equation}  D^{J_{1}}_{m'_{1}m_{1}}D^{J_{3\ *}}_{m'_{3}m_{3}} = \sum_{J_{3},m'_{3},m_{3}} (J_{1}m'_{1}J_{2}m'_{2}|J_{3}m'_{3})(J_{1}m_{1}J_{2}m_{2}|J_{3}m_{3})D^{J_{2\ *}}_{m'_{2}m_{2}} \end{equation} 
where the parentheses are the Clebsch-Gordan coefficients (see Appendix C). The normalization relation is

\begin{equation}  \int d\Omega D^{J_{1}}_{m'_{1}m_{1}}D^{J_{2}}_{m'_{2}m_{2}}D^{J_{3\ *}}_{m'_{3}m_{3}} = \frac{8\pi^{2}}{2J_{3}+1} (J_{1}m'_{1}J_{2}m'_{2}|J_{3}m'_{3})(J_{1}m_{1}J_{2}m_{2}|J_{3}m_{3}) \end{equation} 
 \section{Appendix C: Clebsch-Gordan coefficients}

Consider two angular momentum (spin) states $|J_{1}m_{1} \rangle$ and $|J_{2}m_{2} \rangle$.We want to add those states to form an state of angular momentum (spin), $|Jm \rangle$, such that

\begin{equation}  J = J_{1} \oplus J_{2} \end{equation} 

The states are related by a linear relation given by
\begin{equation}  |Jm \rangle =\sum_{m_{1}m_{2}} (J_{1}m_{1}J_{2}m_{2}|Jm)|J_{1}m_{1} \rangle |J_{2}m_{2} \rangle \end{equation} 
where the coefficients, $(J_{1}m_{1}J_{2}m_{2}|Jm)$, are called {\it Clebsch-Gordan coefficients} (CG). We have that $m=m_{1}+m_{2}$.

An expression for the CG coefficients was also derived by Wigner~\cite{Rose}
\begin{equation}  (J_{1}m_{1}J_{2}m_{2}|Jm) =\delta_{m,m_{1}+m_{2}} \end{equation} 
\begin{equation}  \times \begin{bmatrix} (2J+1) \frac{(J+J_{1}-J_{2})! (J-J_{1}+J_{2})! (J_{1}+J_{2}-J)! (J+m)! (J-m)!}{(J+J_{1}+J_{2}+1)! (J_{1}-m_{1})! (J_{1}+m_{1})! (J_{2}-m_{2})! (J_{2}+m_{2})! } \end{bmatrix}^{\frac{1}{2}} \end{equation} 
\begin{equation}  \times \sum_{S} \frac{(-1)^{S+J+m}}{S!}  \frac{(J+m_{1}+J_{2}-S)! (J_{1}-m_{1}+S)!}{J-(J_{1}+J_{2}-S)! (J+m-S)! (S+J_{1}-J_{2}-m)! } \end{equation} 
Using the orthogonality relations, we can write the reciprocal relations
\begin{equation}  |J_{1}m_{1} \rangle|J_{2}m_{2} \rangle =\sum_{J} (J_{1}m_{1}J_{2}m_{2}|Jm) |Jm \rangle \end{equation} 

Some important symmetry properties can be extracted from these formulas (See reference \cite{Rose}).

\begin{equation}  (J_{1}m_{1}J_{2}m_{2}|Jm) = (-1)^{J_{1}+J_{2}-J} (J_{1}-m_{1}J_{2}-m_{2}|J-m) \end{equation} 

\begin{equation}  (J_{1}m_{1}J_{2}m_{2}|Jm) = (-1)^{J_{1}+J_{2}-J} (J_{2}m_{2}J_{1}m_{1}|Jm) \end{equation} 

\begin{equation}  (J_{1}m_{1}J_{2}m_{2}|Jm) = (-1)^{J_{1}-m_{1}} \begin{bmatrix}\frac{2J+1}{2J_{2}+1}\end{bmatrix}^{\frac{1}{2}} (J_{1}m_{1}J-m|J_{2}-m_{2}) \end{equation}

 \section{Appendix D: Data Simulation.}

We want to generate simulated data from a set of $\alpha = 1, \cdots, N$ waves coming from $R=1, \cdots , n$ resonances.

Let's assume a resonance, $R$, with a relativistic Breit-Wigner mass distribution (we might, of course, assume any other distribution) such that (equation \ref{eqn:masscross})

\begin{equation}  \frac{d\sigma}{dw} =  \frac{1}{(w_o^2-w^2)^2+w_o^2\Gamma^2} w_o^2 \Gamma_o^2 F^2_l(q) . \end{equation} 

Let's take $\Gamma = \Gamma_{o}$ and the Blatt-Weisskopf coefficients $F^2_l(q) =1$. The number of events for a given mass, $w$, are then
\begin{equation}  N_{R}(w) =  C_{R} \frac{1}{(w_o^2-w^2)^2+w_o^2\Gamma_{o}^2} w_o^2 \Gamma_o^2  . \end{equation} 

where $C_{R}$ is a normalization coefficient that correspond to the number of events expected at $w=w_{o}$ (maximum). The value of these coefficients needs to be assumed to obtain the desired relation of cross sections among the resonances. A wave $\alpha$ will contribute to this number of events with a weight $W_{R, \alpha}$, that will also need to be assumed, such that

\begin{equation}  N_{R,\alpha} (w)=  W_{R,\alpha} C_{R} \frac{1}{(w_o^2-w^2)^2+w_o^2\Gamma_{o}^2} w_o^2 \Gamma_o^2  .
\label{eqn:EQ1}
 \end{equation} 
By equation (\ref{eqn:yield}), this number of events is also given by

\begin{equation}  N_{R,\alpha } (w) =  \big{|} V_{R, \alpha} (w)\big{|}^{2} \Psi_{\alpha,\alpha}^{r} (w) 
\label{eqn:EQ2}
\end{equation} 

therefore, from equations (\ref{eqn:EQ1}) and (\ref{eqn:EQ2}) we have

\begin{equation}  \big{|} V_{R, \alpha} (w)\big{|} = \sqrt{ \frac{1}{ \Psi_{\alpha,\alpha}^{r} (w)} W_{R,\alpha} C_{R} \frac{1}{(w_o^2-w^2)^2+w_o^2\Gamma_{o}^2} w_o^2 \Gamma_o^2} 
\end{equation} 

the magnitude of the contribution of the wave $\alpha$ to the resonance $R$ at the value of the mass $w$. Values for the magnitude and angle of polarization ($\mathscr{P}$, $\alpha$) will also need to be assumed to calculate the photon spin density matrix, $\rho_{\epsilon,\epsilon'}$, and the normalization integrals, $\Psi_{\alpha,\alpha}^{r} (w)$.

Recalling equation (\ref{eqn:BWE}), the amplitude given by the Breit-Wigner formula is

\begin{equation}  V_{R,\alpha}(w)=\frac{ \sqrt{C_{R}W_{R,\alpha}}w_o\Gamma_o}{(w_o^2-w^2-iw_o\Gamma(w))} =   \big{|} V_{R, \alpha} (w)\big{|} e^{i \Phi_{R,\alpha}}
\end{equation}

where  $\Phi_{R,\alpha}$ is the BW phase of the contribution of the wave $\alpha$ to the resonance $R$. Therefore

\begin{equation}  e^{i \Phi_{R,\alpha}} =\frac{ \sqrt{((w_o^2-w^2)^{2}+w^{2}_o\Gamma^{2}_{o})}}{(w_o^2-w^2-iw_o\Gamma(w))} \sqrt{ \Psi_{\alpha,\alpha}^{r} (w) }
\end{equation}

from which we can obtain $\Phi_{R,\alpha}$ and, therefore,  the complex structure of the $V_{R,\alpha}(w)$ coefficients.
Using equation (\ref{eqn:trueN}), the simulated mass spectrum is

\begin{equation}  N (w) =  \sum^{n}_{R,R'} \sum^{N}_{\alpha,\alpha'}   V_{R,\alpha} (w) V^{*}_{R',\alpha'} (w) \Psi_{\alpha,\alpha'}^{r} (w)
\end{equation} 

where we include all quantum numbers in one index defining a wave, $\alpha = (b, \epsilon, k)$. Notice that this calculation accounts for the interference between waves.

Suppose that we want to simulate a set of events produced by the previous set of waves and resonances. The probability that an event with characteristics given by $\tau$ is produced, in a given mass bin $w$,  is given by

\begin{equation}  I(\tau, w) = \sum_{R,R'} \sum_{\alpha,\alpha'} \  V_{R,\alpha}(w)  A_{b}(\tau) \rho_{\epsilon,\epsilon'} V^{*}_{R',\alpha'} (w) A^{*}_{\alpha'}(\tau). 
\label{eqn:wte}
\end{equation}

We start by generating events from a phase-space distribution.  We then calculate for each generated event the value
of $I(\tau, w)$ using (\ref{eqn:wte}) and take the maximum value of this quantity in the full event set, $I_{max}$. We calculate the normalized probability for each event, $w_{n} = I(\tau, w)/I_{max}$. For each event, we then generate a random number ($RAN$) between 0 and 1, if $ w_{n} \leq RAN$ the event is kept in the sample, otherwise, the event is discarded. The final sample will then mirror a sample produced by the waves and resonances assumed.
When the $V$ values used in (\ref{eqn:wte}) are the fitted values, this is the method used to generate a predicted sample of events.

 \section{Acknowledgements}

C. W. Salgado was partially supported by National Science Foundation grants \# 0855338 and \# 1205763. D. P. Weygand is supported by the United States Department of Energy under contract DE-AC05-84ER40150. We wish to thank Veronique Ziegler for reading of the manuscript and useful comments.

%\section{Bibliography}

%\noindent 

\end{document}